\documentclass[11pt]{article}
\usepackage[left=2cm, right=2cm, top=1in, bottom=1in]{geometry}
\newcommand\blfootnote[1]{%
  \begingroup
  \renewcommand\thefootnote{}\footnote{#1}%
  \addtocounter{footnote}{-1}%
  \endgroup
}

\usepackage[hyphens]{url}
\usepackage{xcolor}
\usepackage{algorithm} 
\usepackage{algorithmic}
\usepackage{booktabs} 
\usepackage{graphicx} 
\usepackage{subcaption}
\usepackage{amsmath}
\usepackage{amsthm}
\usepackage{amsfonts}
\usepackage{microtype}
\usepackage{amssymb,bbm,mathrsfs}
\usepackage{pgfplots}
\usepackage{tikz}
\usetikzlibrary{graphs}
\usetikzlibrary{pgfplots.groupplots}
\pgfplotsset{compat=1.3}

\def\MA{\textrm{MA}}
\def\MVP{\textrm{MVP}}
\def\MVM{\textrm{MVM}}
\def\EVM{\textrm{EVM}}
\def\EVP{\textrm{EVP}}
\usepackage{enumitem}
\usepackage{soul} 

\theoremstyle{definition}
\newtheorem{definition}{Definition}[section]

\usepackage{biblatex}
\title{A Variance-aware Multiobjective Louvain-like Method for Community Detection in Multiplex Networks}
\author{Sara Venturini\thanks{Department of Mathematics “Tullio Levi-Civita”, University of Padova, 35121 Padova, Italy (sara.venturini@math.unipd.it)} , 
Andrea Cristofari\thanks{Department of Civil Engineering and Computer Science Engineering, University of Rome ``Tor Vergata'', 00133 Rome, Italy (andrea.cristofari@uniroma2.it)},
Francesco Rinaldi\thanks{Department of Mathematics “Tullio Levi-Civita”, University of Padova, 35121 Padova, Italy (rinaldi@math.unipd.it)},
Francesco Tudisco\thanks{School of Mathematics, Gran Sasso Science Institute, Italy, (francesco.tudisco@gssi.it)}}
\date{\vspace{-5ex}}

\begin{document}

\maketitle
\blfootnote{A. Cristofari, F. Rinaldi and F. Tudisco contributed equally to this work.}
\textbf{Abstract.}
In this paper, we focus on the community detection problem in multiplex networks, i.e., networks with multiple layers having same node sets and no inter-layer connections. In particular, we look for groups of nodes that can be recognized as communities consistently across the layers. To this end, we propose a new approach that generalizes the Louvain method by (a) simultaneously updating 
average and variance of the modularity scores across the layers, and (b) reformulating the greedy search procedure in terms of a filter-based multiobjective optimization scheme. Unlike many previous modularity maximization strategies, which rely on some form of aggregation of the various layers, our multiobjective approach aims at maximizing the individual modularities on each layer simultaneously. 
We report experiments on synthetic and real-world networks, showing the effectiveness and the robustness of the proposed strategies both in the informative case, where all layers show the same community structure, and in the noisy case, where some layers represent only noise.\\

\textbf{Key words.}
community detection, multilayer networks, multiplex networks.\\

\section{Introduction}
Networks represented as graphs with nodes and edges have emerged as effective tools for modelling and analysing complex systems of interacting entities. In fact, graphs arise naturally in many disciplines, such as social networks, information networks, infrastructure networks, biological networks (see, e.g., \cite{estrada2012structure, newman2018networks}).

One of the most relevant issues in the analysis of graphs representing real systems is the identification of communities, i.e.,  groups of nodes that are densely connected to each other and loosely connected to the nodes in the other communities. 
While many community detection and clustering algorithms have been developed over the recent years,  most of these are designed for standard single-layer graphs. On the other hand, having one single graph is  often an oversimplifying assumption, which can lead to misleading models and results. Fortunato conducted a survey on this topic \cite{Fortunato}.

Advances in the study of networked systems have shown that the interconnected world is often composed of  networks that are coupled to one another through different layers, where each layer represents one of many possible types of interactions \cite{kivela2014multilayer}. Multi-layer networks arise naturally in diverse applications such as transportation networks \cite{gallotti2015}, financial-asset markets \cite{bazzi2016}, temporal dynamics \cite{taylor2017, taylor2016}, semantic world clustering \cite{sedoc2017}, multivideo face analysis \cite{cao2015}, mobile phone networks \cite{kiukkonen2010}, social balance \cite{cartwright1956}, citation analysis \cite{tang2009}, and many others. 

As for standard (single-layer) models, community detection is one key problem in the analysis of multi-layer networks and the  presence of multiple layers poses several additional challenges. For instance, networks may have different types of multi-layer structures and  communities may or may not be consistent across the layers.

Here, we focus on multiplex networks, modeled by a sequence of graphs (the layers) with a common set of nodes and no edges between nodes of different layers. We also assume that each layer is  undirected and simple. 
Moreover, with the terminology introduced in~\cite{magnani2021community}, 
our aim is to find a set of communities that is \textit{total} (i.e., every node belongs to at least one community),
\textit{node-disjoint} (i.e., no node belongs to more than one cluster on a single layer), and \textit{pillar} (i.e., each node belongs to the same community across the layers).

Several community detection algorithms for multiplex networks
have been proposed in recent years.
Among the successful approaches, we find methods that suitably reduce the multiplex to a single-layer graph,  methods that are based on adaptations of single-layer consensus and spectral clustering, methods based on information-theoretic flow diffusion, and methods that infer communities by fitting suitable planted partition models. We attempt to summarize the related literature in \S\ref{sec:related_work} and refer to the survey~\cite{magnani2021community} for further details.

In this paper, we propose a new strategy that directly approaches the multiobjective optimization problem of maximizing the modularity score of each individual layer. To this end, we adapt the popular Louvain heuristic method for single-layer networks \cite{Blondel}, a locally greedy modularity-increasing sampling process over the set of node partitions. A natural extension of this method to the multiplex 
case, already studied in the literature, e.g. in \cite{mucha2010,bazzi2016}, is to locally maximize a weighted average $M_Q$ of the modularity of the layers, instead of the modularity of a single layer. 
One of the advantages of using a linear combination of the layer modularities is that the increment  of $M_Q$ can be directly computed using the increment of the modularity of each layer, whose computation is efficiently handled by the original Louvain technique.

In a similar spirit, in \S\ref{Methods} we consider a variant of the Louvain strategy which directly aims at maximizing the vector-valued function of the modularities of all the layers. To address the resulting multiobjective optimization problem, we propose a technique to memorize and dynamically update a list of several solutions, according to a suitably developed Pareto search. The size of the list as well as the final community assignment are controlled by means of a scalar cost function which takes into account for the variance of the layer modularities, in addition to their average. 
The use of either a positive or a negative variance regularization term allows us to better control the amount of variability of  the modularity scores across the layers and allows us to effectively handle both purely informative layers as well as the presence of noise. Moreover, while the resulting scalar cost function is a nonlinear combination of the layer modularities, we show that iteratively computing the modularity of each layer allows us to efficiently update their variance as well, resulting in an efficient variance-aware multiobjective Louvain-like scheme.

As such, an important feature of the proposed approach is that, unlike many methods presented in the literature, it does not need to preassign the number of classes at the beginning. This is particularly useful as we usually do not have information about the community structure of the multiplex and one must make some a-priori (possibly unjustified) assumption about the number of clusters otherwise.

To verify the performance and robustness of our technique, in our experimental set-up we study multiplex graphs in two settings: in the first one all layers are informative, i.e., all layers contain information about the community structure, whereas in the second setting at least one of the layers is a \textit{noisy layer}, which accounts for corrupted measurements.
In \S\ref{SBM} and \S\ref{LFR}, we compare our method with nine baselines on
 synthetic multilayer networks generated via the  Stochastic Block Model (SBM) \cite{mercado2018} and the Lancichinetti-Fortunato-Radicchi (LFR) Benchmark \cite{lancichinetti2008benchmark}, 
properly extended to the multiplex setting. Then, in \S\ref{real}, we compare the performance  on several real-world multiplex graphs. Our results show that directly taking into account the variance across the layers may lead to much better performance, especially in the presence of noisy data. In particular, the proposed filter-based algorithm often leads to the best or second-best classification results, thus confirming the added value  of the  multiobjective approach.

\section{Related work}\label{sec:related_work}
In the following, we attempt to review and summarize some recent approaches to community detection for multi-layers. 
As in \cite{magnani2021community}, we focus on algorithms explicitly designed to discover communities in multiplex
networks.

Flattening methods reduce the multiplex network into a single-layer unweighted network and then apply a traditional community detection algorithm.
The simplest algorithm of this type constructs a single-layer graph where two nodes are adjacent if they are neighbours in at least one layer \cite{berlingerio2011finding}.
An alternative is to create a weighted single-layer graph, where weights reflect some structural properties of the multiplex \cite{berlingerio2011finding,kim2016differential}.

Layer flattening coincides with summing the adjacency matrices of each layer into an aggregated adjacency matrix. More powerful community detection algorithms can be obtained, i.e.,  by merging  modularity matrices or specific node embeddings.  A study of these types of extensions is done in \cite{tang2012community}, with a novel method that integrates the layer-based node structural features.

Another popular approach is to perform aggregation at the cost function level, by extending single-layer community detection cost functions to the multilayer setting. The Generalized Louvain (GL) method proposed by Mucha et al.~\cite{mucha2010} is used to maximize a multilayer version of Newman's modularity \cite{Newman3}. 
Bazzi et al.~\cite{bazzi2016} propose a related approach with a Louvain-based modularity-maximization method for community detection in multilayer temporal networks. 
Both these approaches use a cross-layer modularity function that is essentially based on a weighted arithmetic mean of the modularities of the single layers. As we will discuss in the next section, our proposed approach defines a new variance-aware Louvain-like method,  which leverages this idea as starting point. 
A related approach is proposed by Pramanik et al. \cite{pramanik2017}, where a weighted linear combination of the modularities of each layer is used to define a multilayer modularity index. Their approach focuses on the case of two-layer networks with both inter- and intra-layer connections and  aims at detecting inter- and intra-layer communities simultaneously. 

A variety of alternative strategies has been developed in relatively recent years.
Pizzuti and Socievole \cite{pizzuti2017} develop a  genetic algorithm for community detection in multilayer networks that makes use of a  multiobjective optimization formulation. In particular,  the proposed algorithmic scheme exploits the concept of Pareto dominance when creating new populations at a given iteration, and returns a family of solutions that represent different trade-offs between the objectives at the end of the optimization process (the best solution is finally chosen using some tailored strategies).

De Domenico et al.~\cite{de2015identifying} extend the famous information-theoretic approach of \cite{rosvall2008maps}, by proposing a method that generalizes the so-called map equation for single graphs and identifies communities as groups of nodes that capture flow dynamics within and across layers for a long time.

De Bacco et al. \cite{de2017community} proposed a method based on likelihood maximization. They define a mixed-membership multilayer stochastic block model and propose a method that infers the communities by fitting this model to a given multilayer dataset via log-likelihood maximization.

Wilson et al.~\cite{Wilson} propose a technique for multilayer data that aims at identifying densely connected sets of vertex-layer pairs via a significance-based score that quantifies the connectivity of such sets as compared to a suitable fixed-degree random graph model.

Methods inspired by data clustering techniques  are another popular line of work.
Zeng et al. \cite{Zeng} proposed a pattern mining algorithm for finding closed quasi-cliques that appear on multiple layers with a frequency above a given threshold. A cross-graph quasi-clique is defined as a set of vertices belonging to a maximal quasi-clique that appears on all layers \cite{Pei}.

Tang et al. \cite{tang2009} and  Dong et al. \cite{Dong} proposed graph clustering algorithms for multilayer graphs based on matrix factorization. The key point is to extract common factors from multiple graphs to be used for  various clustering methods. Tang et al. \cite{tang2009} factorize adjacency matrices while Dong et al. \cite{Dong} factorize graph Laplacian matrices.
Liu et al. \cite{3sources} proposed a nonnegative matrix factorization based multiview clustering algorithm, where the factors representing clustering structures from multiple views are regularized toward a common consensus.

Another popular line of research tries to  extend spectral clustering to multilayer graphs. In general, these algorithms aim to define a graph operator that contains all the information of the multilayer graph such that the eigenvectors corresponding to the smallest eigenvalues are informative about the clustering structure. These methods usually rely on some sort of ``mean operator'', e.g., the Laplacian of the average adjacency matrix or the average Laplacian matrix \cite{Paul}. Further examples are the work of Zhou and Burges \cite{Zhou2}, which developed a multiview spectral clustering via generalizing the usual single view normalized cut to the multi-view case and tried to find a cut which is close to optimal on each layer, and the algorithm designed by Chen and Hero \cite{Chen2}, which performs convex aggregation of layers based on signal-plus-noise models.

Alternative approaches are proposed for instance by Dong et al.~\cite{dong2013clustering}, where  spectral clustering is extended by merging the informative Laplacian eigenspaces of different layers via a subspace optimization analysis on Grassmann manifolds.
Zhan et al. \cite{zhan2017}, \cite{zhan2018b}, \cite{zhan2018} developed several multiview graph learning approaches which  merge multiple graphs into a unified graph with the desired number of connected components. 
Other multiview clustering approaches exploit the idea of maximizing clustering agreement. 
Zong et al. \cite{zong2018} introduced Weighted Multi-View Spectral Clustering, where the largest canonical angle is used to measure the difference between spectral clustering results of different views. 
Nie et al. \cite{nie2018} proposed a self-weighted scheme for fusing multiple graphs with the importance of each view considered, called Procrustes Analysis technique.

A common limitation of the proposed multiview clustering methods is that they do not consider to deal with possibly noisy or corrupted data, because they focus on the consistency of multiple layers and do not consider the inconsistency.  
To address this issue, Xia et al.\ \cite{xia2014} proposed  Robust Multi-view Spectral Clustering, a Markov chain method that aims to learn an intrinsic transition matrix from multiple views by restricting the transition matrix to be low-rank.
This aspect has also been considered by Mercado et al.\ \cite{mercado2018}, \cite{Mercado}, where they propose a Laplacian operator obtained by merging the Laplacians from different layers via a   one-parameter family of nonlinear matrix power means. Recently, Liang at al.\ \cite{liang2020} proposed a multiview graph learning framework, which simultaneously models
multi-view consistency and multiview inconsistency in a unified objective function.

Another line of work  adopts Bayesian inference \cite{Winkler}, in which certain hypotheses about connections between nodes are made to find the best fit of a model to the graph through the optimization of a suitable likelihood \cite{Peixoto}.

Bickel and Scheffer \cite{ bickel2004} extended the semi-supervised co-training approach \cite{ blum1998} to multi-view clustering. The basic idea of co-training is to iterate over all views and optimize the objective function in the next view using the result obtained from the latter one. 
A co-training approach is proposed by  Kumar and Daum\'e \cite{Kumar}, where the algorithm aims to find a consistent clustering that
agrees across the views under the main assumption that all layers can be independently used for clustering. 
Under the same assumption, Kumar et al. \cite{Kumar2} proposed Co-regularized Spectral Clustering, where they concentrated on this approach under the notion of co-regularization, maximizing the agreement between different views.

\section{Multiobjective Louvain-like method for multiplex networks}
\label{Methods}
In this section, we present our method for community detection in multiplex networks which, based on the popular 
 Louvain heuristic method for single-layer networks  \cite{Blondel}, aims at maximizing the modularity of all layers simultaneously.
 Unlike many alternative strategies, where either the multiplex or the cost function are aggregated into a single-layer representative of the original multiple layers, our approach directly takes into account the multiobjective nature of the problem under analysis, i.e., the existence of more than one objective to optimize. To this end, we maintain and update 
a list of suitable community assignments during the algorithm, each of them being preferable over the others with respect to a specific criterion.
 
More formally, consider a multiplex with $k$ layers $G_1,\dots,G_k$, where $G_s=(V,E_s)$ is the graph forming the $s$-th layer. Thus, consider the vector of layer modularities $Q=(Q_1,\dots,Q_k)$. Here and everywhere in the text, we shall always assume vectors are column vectors, unless otherwise specified. 
The modularity score of the $s$-th layer is defined as
\begin{equation}
\label{modularity}
Q_s = \frac 1 {2m_s} \sum_{ij} \Big(A^{(s)}_{ij}-\frac{d^{(s)}_i d^{(s)}_j}{2m_s}\Big)\delta(C_i,C_j)\, ,
\end{equation}
where the sum runs over all pairs of vertices, $A^{(s)}$ is the adjacency matrix of $G_s$, $m_s$ the total number of edges in  $E_s$ (or the sum of all their weights, in the case of weighted graphs), 
$d_i^{(s)}$ is the degree (or weighted degree) of the node $i$ in $G_s$, $C_i$ is the community node $i$ belongs to, and the function $\delta$ yields one if vertices $i$ and $j$ are in the same community (i.e.,\ $C_i=C_j$) and zero otherwise.

We aim at maximizing all entries of $Q$ simultaneously, i.e., we consider the following multiobjective optimization problem:
\begin{equation}\label{eq:problem}
\max_{\{\text{partitions of }V\}} \, (Q_1,\dots, Q_k)
\end{equation}

In multiobjective optimization, there is not a unique way to define optimality, since there is no a-priori  total order for $\mathbb R^k$ and each partial order leads to different strategies. Here, we consider the well-established definition of optimality according to Pareto \cite{Pareto}: 
\begin{definition}
Given two vectors $z^1,z^2\in \mathbb R^k$, we write $z^1 \succeq_P z^2$ if  $z^1$ dominates $z^2$   according to Pareto, that is:
\begin{align*}
&z^1_i \geq z^2_i \quad \text{for each index } i=1,..,k \text{ and}\\
&z^1_j > z^2_j \quad \textrm{for at least one index } j=1,..,k\, .
\end{align*}
A vector $z^*\in \mathbb R^k$ is Pareto optimal if there is no other vector $z \in \mathbb R^k$  such that $z \succeq_P z^*$. 
Moreover, the Pareto front is the set of all Pareto optimal points.
\end{definition}

To address \eqref{eq:problem}, we proceed by adapting the standard Louvain method with the aim of approaching a suitable candidate solution on the Pareto front of the modularity vector. To this end, we start with an initial partition where each node represents a community, to which corresponds the initial modularity vector $Q$. Then, we proceed with a two-phase scheme which generates a list $L$ of community assignments and corresponding modularity vectors such that no one is Pareto-dominated by the others. The final approximate solution is then chosen in terms of  a scalar function $F$ used to asses the ``quality'' of a partition across the multiple layers. The choice of $F$ is discussed in \S\ref{sec:choice_of_F}. To start with, the list $L$ consists of the initial vector $Q=(Q_s)_s$, the initial community partitioning and the corresponding value of $F$.

In the first phase, the algorithm picks one node at a time, following a given initial node ordering. 
For each node $i$, for every layer $s$, and for every neighbor $j$ of node $i$ (among the $j$s that have not  been  considered yet), we compute the change of modularity $\Delta Q_s^{(i\to j)}$ as the sum of the change in modularity on layer $s$ obtained by removing $i$ from its community $C_i$ and the variation of modularity on layer $s$ obtained by including $i$ in the community $C_j$.
For each such change of community assignment we evaluate the new modularity vector $Q^{(i\to j)} = (Q_1 +\Delta Q_1^{(i\to j)},\dots,Q_k +\Delta Q_k^{(i\to j)})$ by efficiently updating the previous modularity scores as in the original Louvain scheme. If $Q^{(i\to j)}$ is not Pareto-dominated by any of the modularity vectors in $L$, $Q^{(i\to j)}$ is a good candidate to be added to $L$. However, as in the original Louvain method, we want to consider only new modularity vectors that yield a ``strict improvement''. To this end, we use the modularity updates $\Delta Q_s^{(i\to j)}$ to efficiently  evaluate the variation $\Delta F^{(i\to j)}$ of the scalar function $F$. If additionally the new modularity vector $Q^{(i\to j)}$ corresponds to a positive increment $\Delta F^{(i\to j)}$ of the quality function $F$, we add $Q^{(i\to j)}$, the new value of the quality function $ F + \Delta F^{(i\to j)}$ and the corresponding partition to $L$. Thus,  we remove from $L$ all the partitions whose modularity vectors are dominated by the newly inserted  one. Moreover, in order to avoid the list computed this way to grow exponentially in size, we add a final control on the length of $L$ by filtering out the elements of $L$ that have small value of $F$, maintaining only the $h$ partitions that achieve the largest value.  

We proceed this way until the list stops changing and contains only vectors that do not dominate each other. At this point, the method selects from $L$ the best partition with respect to $F$ and uses this as the new starting point. We then move to the second phase, which consists of an aggregation step where the communities in the selected partition are merged into single vertices, forming a smaller graph. The whole procedure is then repeated iteratively, until no further improvement in the Pareto sense is possible, in  the same spirit of the original Louvain method for single-layer graphs. 
The overall scheme is summarized in Algorithm~\ref{LMM}. 

Clearly, the choice of the function $F$ may significantly affect the performance of the proposed strategy, both in terms of the final community assignment and in terms of computational time, as evaluating $F$ may be an expensive operation. 
We argue below that a reasonable choice of $F$ takes into account both average and variance of the layer modularities, and we show how these can be evaluated via an inexpensive iterative update.

\begin{algorithm}[t]
\caption{Louvain Multiobjective Method 
}\label{LMM}
\hspace*{\algorithmicindent} \!\!\!\! \textbf{Input} $G$ multiplex graph, $F$ scalar quality function\\
\hspace*{\algorithmicindent} \textbf{Output} final partition
\begin{algorithmic}
\STATE $L$ initialized with node-based partition, the corresponding modularity vector $Q$ and the value of $F$
\STATE Set {\tt terminate} = false

\REPEAT 

\STATE Set {\tt updateL} = true

\REPEAT

\FORALL{node $i$ of $G$}
\FORALL{partition $C$ in list $L$}
\STATE place $i$ in every neighboring community which yields a positive increment of $F$. 
If the corresponding modularity vector $Q$ is not Pareto-dominated by any of the modularity vectors in $L$, insert in $L$ the vector $Q$, the corresponding partition and the $F$ value. Delete from $L$ all terms corresponding to modularity vectors that are dominated by $Q$.
\ENDFOR
\ENDFOR
\STATE If $L$ is longer than $h$, cut it to length $h$ using $F$

\IF{$L$ does not change}
\STATE  {\tt updateL} = false
\ENDIF

\UNTIL( {\tt updateL} == false)

\STATE Consider the partition of the list $L$  which maximizes the function $F$ gain 
\IF{$L$ has changed}
\STATE $G$ =
reduced graph where each community of the selected partition is a node
\ELSE
\STATE Set {\tt terminate} = true
\ENDIF
\UNTIL({\tt terminate} == true)
\end{algorithmic}
\end{algorithm}

\subsection{Variance-aware cross-layer modularity function}\label{sec:choice_of_F}
One possibility to quantify the quality of a partition into communities for a multiplex is to measure the average of the corresponding modularity functions across all the layers. This corresponds to choosing $F=M_Q$, with  
\begin{equation}
\label{Media}
M_Q = \frac 1 k \sum_{s=1}^k Q_s\, .
\end{equation}
 The idea of considering a linear (possibly weighted) combination of the modularity functions of single layers is relatively natural and has been considered, for instance, in~\cite{bazzi2016,mucha2010,pramanik2017}.
 A key advantage of this choice is that the gain  $\Delta F^{(i\to j)}$, measuring how the chosen function $F$ changes when node $i$ is moved from $C_i$ to $C_j$ during phase one of the algorithm, can be straightforwardly computed due to the linearity of $F$ with respect to $Q_s$:
\begin{equation}
\label{average}
\Delta M_Q^{(i\to j)} = \frac 1 k \sum_{s=1}^k \Delta Q_s^{(i\to j)}
\end{equation}
This observation is at the basis of the GL method,
proposed in~\cite{mucha2010}, see also~\cite{magnani2021community,bazzi2016}.

While the average yields a reasonable overview of the graph community structure across the layers,  in many situations this may lead to an oversimplification \cite{mercado2018,Mercado}. For instance, in the presence of noisy layers, linear averages over the multiple layers perform poorly \cite{mercado2018}. To overcome this issue, we consider two functions that embed the sampled variance of the modularity of the layers:
\begin{align}
    \label{F_pm} F_- = (1-\gamma)M_Q-\gamma V_Q \qquad \text{and} \qquad
     F_+ = (1-\gamma)M_Q+\gamma V_Q
\end{align}
where $\gamma \in (0,1)$ is a parameter and $V_Q$ is the sampled variance of the modularity of the layers, which we compute~as
\begin{equation}
V_Q = \frac 1 {k-1} \sum_{s=1}^k(Q_s-M_Q)^2\, .
\end{equation}
While $F_\pm$ is now quadratic in $Q_s$, we observe below that, as for the linear choice $F=M_Q$, the increment $\Delta F_\pm^{(i\to j)}$ of both $F_-$ and $F_+$ can be computed in an efficient way during the algorithm. In fact, a direct computation shows that the following formula holds: 
$$
\Delta F_{\pm}^{(i\to j)} = (1-\gamma) \Delta M_Q^{(i\to j)} \pm \gamma \, R_Q^{(i\to j)},
$$
where $R_Q^{(i\to j)}$ is the coefficient
$$
R_Q^{(i\to j)} = V_{\Delta Q}^{(i\to j)}+\frac 2 {k-1}(Q-M_Q\mathbf 1)^\top (\Delta Q^{(i\to j)} -\Delta M_Q^{(i\to j)}\mathbf 1)\, ,
$$
$\mathbf 1=(1,\dots,1)$ is the vector of all ones, $Q=(Q_1,\dots,Q_k)$ is the vector of the layer modularities,   $\Delta Q^{(i\to j)}$ is the column vector  whose $s$-th component is the gain $\Delta Q_s^{(i\to j)}$,  
and $V_{\Delta Q}^{(i\to j)}$  is the sampled variance of $\Delta Q_s^{(i\to j)}$, which we compute as follows:
\begin{equation}
 V_{\Delta Q}^{(i\to j)} = \frac 1 {k-1} {\sum_{s=1}^{k}  ( \Delta Q_s^{(i\to j)}  -  \Delta M_Q^{(i\to j)}  )^2 }\, .
\end{equation}

These formulas show that, as for $M_Q$, iteratively updating $\Delta Q_s^{(i\to j)}$ allows us to iteratively update the nonlinear quality functions $F_\pm$ during Algorithm~\ref{LMM}  in a computationally  efficient way, using the  increment  $\Delta F_{\pm}^{(i\to j)}$.  This allows us to efficiently compute the quality of the new community assignments  keeping simultaneously track of both mean and variance of the layer modularities.


\subsection{Positive vs negative variance regularization}
The quality functions $F_+$ and $F_-$ allow us to consider different types of variability of modularity across the layers. In particular, in the setting where all the layers have consistent community structure, we use $F_-$. The rationale behind this choice  is that we want  to obtain a
 trade-off between large modularity and sufficiently small  variability in the layers  (the larger $\gamma$, the smaller the variance in the final solution), which is what an ideal solution would look like if the community structure of all layers is in agreement. On the other hand, in the presence of some noisy layers, i.e., layers with no community structure, $F_+$ is a better choice. In fact,   in this case we want to favor solutions which have at the same time a good level of variability across the layers and a large enough modularity (the larger $\gamma$, the greater the level of allowed variability).  

Overall, we study three  variants of the proposed Louvain Multiobjective Method in Algorithm~\ref{LMM}, which correspond to the following three different choices of the function $F$: the  modularity average $M_Q$, defined in \eqref{Media},  and the functions $F_-$ and $F_+$ defined in \eqref{F_pm}. We refer to the corresponding algorithms respectively as \textit{Louvain Multiobjective Average} (\textbf{\MA}), \textit{Louvain Multiobjective Variance Minus} (\textbf{\MVM}) and \textit{Louvain Multiobjective Variance Plus} (\textbf{\MVP}). 

All these methods require to choose the length of the list~$L$, that we called $h$. The longer such list is, the better is the way we approach the Pareto front and explore the space related to the layer modularities. At the same time, this comes at a higher computational cost, which grows exponentially with $h$. 
Our experiments show that already  $h=2,3$ leads to remarkable performance in terms of accuracy and NMI, as compared to $h=1$. 
We emphasize that, when we limit the length to $h=1$, the method boils down to a form of aggregation strategy where the multiobjective approach is completely ignored and, instead, we aim at maximizing the chosen scalar function $F$ by means of a Louvain-like greedy strategy. Thus, for example, \MA{} with $h=1$ essentially corresponds to the GL method~\cite{mucha2010}. From this point of view, the methods \MVP{} and \MVM{} for $h=1$ are particularly interesting as they yield a variance-aware extension of the popular GL approach, whereas \MA{} for $h>1$ provides a form of multiobjective GL. Moreover, when $h=1$, the whole procedure of the method significantly simplifies. For these reasons, 
we use a separate notation for the case $h=1$,  referring to the method that uses the function $F_-$  as \textit{Louvain Expansion Variance Minus} (\textbf{\EVM}) and to the method that uses the function $F_+$  as \textit{Louvain Expansion Variance Plus} (\textbf{\EVP}).


 \section{Experiments}
\label{Experiments}
We implemented the methods described in \S\ref{Methods} using \texttt{Matlab}. Our codes are all available at the GitHub page: 
\url{https://github.com/saraventurini/A-Variance-aware-Multiobjective-Louvain-like-Method-for-Community-Detection-in-Multiplex-Networks}.

We considered both synthetic and real-world networks, performing extensive experiments to compare the proposed methods against nine multilayer community detection baselines (see \S\ref{sec:related_work} for details), namely:

\begin{itemize}[leftmargin=*]
\item \textbf{GL}: Generalized Louvain  
\cite{bazzi2016,mucha2010,pramanik2017};
\item \textbf{CoReg}: Co-Regularized spectral clustering, with parameter $\lambda = 0.01$~\cite{Kumar2};
\item \textbf{AWP}: Multi-view clustering via Adaptively Weighted Procrustes \cite{nie2018};
\item \textbf{MCGC}: Multi-view Consensus Graph Clustering, with parameter $\beta = 0.6$~\cite{zhan2018};
\item \textbf{PM}: Power mean Laplacian multilayer clustering, with parameter $p = -10$~\cite{mercado2018};
\item \textbf{MT}: Multitensor expectation maximization~\cite{de2017community};
\item \textbf{SCML}: Subspace Analysis on Grassmann Manifolds, with parameter $\alpha = 0.5$~\cite{dong2013clustering};
\item \textbf{PMM}: Principal Modularity Maximization, with parameters $l = 10$ and $\text{maxKmeans} = 5$~\cite{tang2009uncoverning,tang2012community};
\item \textbf{IM}: Information-theoretic generalized map equation~\cite{de2015identifying}.
\end{itemize}
%
%

Notably,  all these algorithms, except for GL and IM, require the user to specify  the number of communities we look for a-priori. This is a potential drawback in practice, as we usually do not have information about the community structure of the graph and would have to  make some (possibly unjustified) assumptions about the number of clusters.

Methods' performance is evaluated using two metrics: the accuracy, measured as the percentage of nodes assigned to the correct community~\cite{zhan2018}, and the Normalized Mutual Information (NMI)~\cite{strehl2002}.

For all experiments, we considered two settings: the informative case, i.e.,
all the layers carry useful information about the underlying clusters, and the noisy case, where some of the layers are just randomly generated noise.    As discussed in \S\ref{sec:choice_of_F}, a negative variance contribution as in $F_-$ is suitable for the informative setting, while a positive variance term as in $F_+$  may help in the presence of noisy layers. Thus, we test \EVM{} and \MVM{} for the informative case, while we use \EVP{}   and \MVP{} in the noisy setting. Moreover, in order to confirm the advantages of the variance term in $F_\pm$, we further report the performance of \MA, which only accounts for the sum of modularities across the layers (and thus provides a form of multiobjective version of GL).   
For the Louvain Multiobjective model, we consider the two list lengths   $h = 2, 3$ which we indicate by adding a number to the method's acronym (e.g., \MA{}2 stands for the Louvain Multiobjective Average method with list length $h=2$).

In order to evaluate the performance of the method as the variance regularizing parameter changes, we let $\gamma \in \{0.1, 0.3, 0.5, 0.7, 0.9\}$ in the definition of $F_-$ and $F_+$. In Figures 1-4 and Tables 2-4, we report the scores obtained with the  parameter  achieving the highest NMI on each dataset.
However, a comparable performance is achieved for a large number of parameters. 
In particular, analyzing the performance of the methods as the parameter $\gamma$ varies, provides an experimental guiding principle on the choice of $\gamma$, suggesting that a balanced contribution $\gamma\approx 0.5$ is most appropriate in the informative setting, while a larger $\gamma\approx 0.9$ seems to perform best in the presence of noise.

%


Being locally-greedy algorithms, the initial ordering of the nodes in phase one of all our methods may affect the final performance, just like the standard Louvain method. Based on our computational experience, it seems that choosing a specific ordering has a minor effect on the cost function itself, while it may have an impact on the computational time. Choosing the appropriate initial ordering is a  nontrivial question and a well-known issue  of this type of greedy strategies. In our experiments, we choose the initial ordering depending on the specific network setting we deal with.
More specifically, we order the nodes according to the community size in the informative setting, while we list the nodes in random order in the noisy setting. This is mainly due to computational reasons: sorting the nodes according to the size of the corresponding community is  more expensive than assigning a random order. The sorting cost is reasonable in 
the experiments for the informative case, while it becomes prohibitive for the noisy case.

\subsection{Synthetic Networks via SBM}
\label{SBM}
We consider here networks generated via the (multi-layer) Stochastic Block Model (SBM),  a generative model for graphs with planted communities generated through the parameters $p$ and $q$. These parameters represent the edge probabilities: given nodes $i$ and $j$, the probability of observing an edge between them is $p$ (resp.\ $q$), if $i$ and $j$ belong to the same (resp. different) cluster. 

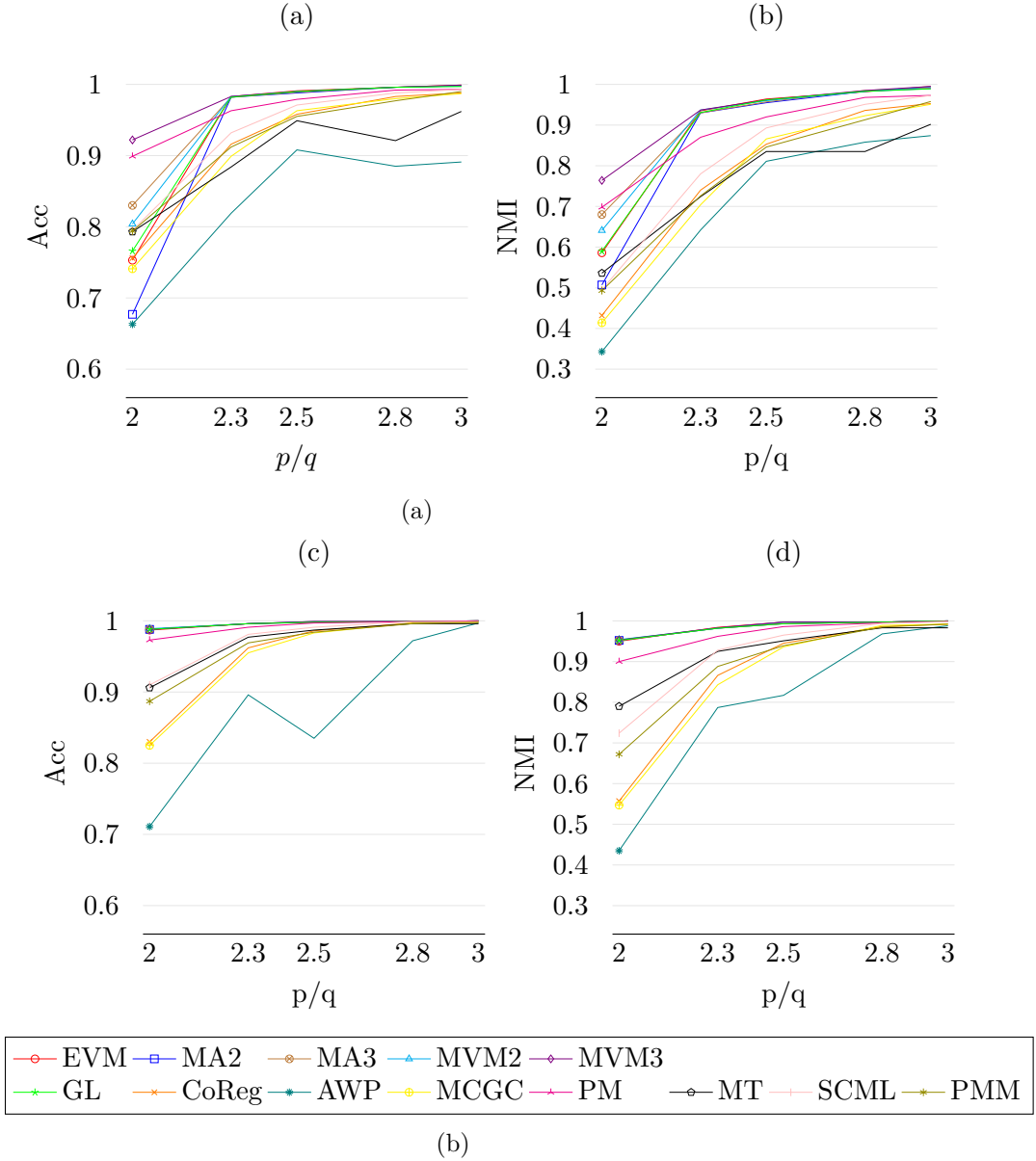
\begin{figure*}[p]
\centering
\captionsetup{oneside,margin={-0.53cm,0.6cm}}
    \hspace*{-2cm}
    \subfloat[]
    {
\pgfplotstableread{
Methods p/q Acc NMI 
AWP 3 0.891 0.874
AWP 2.8 0.885 0.858
AWP 2.5 0.908 0.811
AWP 2.3 0.819 0.642
AWP 2 0.663 0.343
}{\AWP}

\pgfplotstableread{
Methods p/q Acc NMI 
CoReg 3 0.988 0.953
CoReg 2.8 0.983 0.936
CoReg 2.5 0.958 0.853
CoReg 2.3 0.916 0.74
CoReg 2 0.756 0.432
}{\CoReg}

\pgfplotstableread{
Methods p/q Acc NMI 
EVM 3 0.999 0.995
EVM 2.8 0.996 0.982
EVM 2.5 0.991 0.964
EVM 2.3 0.983 0.935
EVM 2 0.753 0.586
}{\EVM}

\pgfplotstableread{
Methods p/q Acc NMI 
GL 3 0.997 0.989
GL 2.8 0.996 0.983
GL 2.5 0.99 0.962
GL 2.3 0.982 0.931
GL 2 0.766 0.59
}{\GL}

\pgfplotstableread{
Methods p/q Acc NMI 
MA2 3 0.998 0.99
MA2 2.8 0.996 0.983
MA2 2.5 0.988 0.955
MA2 2.3 0.982 0.93
MA2 2 0.677 0.507
}{\MA}

\pgfplotstableread{
Methods p/q Acc NMI 
MA3 3 0.998 0.992
MA3 2.8 0.996 0.985
MA3 2.5 0.989 0.957
MA3 2.3 0.982 0.93
MA3 2 0.83 0.68
}{\MAt}

\pgfplotstableread{
Methods p/q Acc NMI 
MCGC 3 0.987 0.951
MCGC 2.8 0.98 0.923
MCGC 2.5 0.963 0.866
MCGC 2.3 0.899 0.704
MCGC 2 0.741 0.414
}{\MCGC}

\pgfplotstableread{
Methods p/q Acc NMI 
MT 3 0.962 0.902
MT 2.8 0.921 0.835
MT 2.5 0.949 0.835
MT 2.3 0.884 0.724
MT 2 0.793 0.536
}{\MT}

\pgfplotstableread{
Methods p/q Acc NMI 
MVM2 3 0.998 0.994
MVM2 2.8 0.996 0.983
MVM2 2.5 0.99 0.96
MVM2 2.3 0.983 0.937
MVM2 2 0.804 0.641
}{\MVM}

\pgfplotstableread{
Methods p/q Acc NMI 
MVM3 3 0.999 0.995
MVM3 2.8 0.996 0.985
MVM3 2.5 0.99 0.962
MVM3 2.3 0.983 0.937
MVM3 2 0.922 0.764
}{\MVMt}

\pgfplotstableread{
Methods p/q Acc NMI 
PM 3 0.993 0.973
PM 2.8 0.992 0.968
PM 2.5 0.979 0.92
PM 2.3 0.963 0.87
PM 2 0.899 0.698
}{\PM}

\pgfplotstableread{
Methods p/q Acc NMI 
PMM 3 0.99 0.958
PMM 2.8 0.977 0.913
PMM 2.5 0.955 0.846
PMM 2.3 0.912 0.726
PMM 2 0.794 0.493
}{\PMM}

\pgfplotstableread{
Methods p/q Acc NMI 
SCML 3 0.993 0.973
SCML 2.8 0.988 0.951
SCML 2.5 0.971 0.893
SCML 2.3 0.932 0.78
SCML 2 0.795 0.495
}{\SCML}

\begin{tikzpicture}
    \begin{groupplot}[
        group style={
        group name=art_info, 
        group size= 2 by 1,
        horizontal sep=50pt
        },
        width=6.3cm,
        height=6.3cm,
        enlarge y limits,
        axis y line*=left,
        axis x line*=bottom, 
        xtick=data,
        ymajorgrids,
		grid style={line width=.2pt,draw=gray!20},
        y axis line style={draw=none},
        xtick style={draw=none},
		ytick style={draw=none},
        yticklabel style={xshift=-0.5em},
    ]
     
     \nextgroupplot[title=(a),xmin=1.98, xmax=3.02, ymin=0.6, ymax = 1,extra y ticks={0.7,0.9}, xlabel={$p/q$}, ylabel={Acc}]
     
       \addplot [line width=0.05mm, 
     mark=o, color=red, mark size=1.5, mark options={fill=red!20!red,mark indices=5}, 
     ] table[x=p/q,y=Acc]{\EVM}; 
     
     \addplot [line width=0.05mm, 
     mark=square, color=blue, mark size=1.5, mark options={fill=blue!20!blue,mark indices=5}, 
     ] table[x=p/q,y=Acc]{\MA}; 
     
      \addplot [line width=0.05mm, 
     mark=otimes, color=brown, mark size=1.5, mark options={fill=brown!20!brown,mark indices=5}, 
     ] table[x=p/q,y=Acc]{\MAt}; 
     
       \addplot [line width=0.05mm, 
     mark=triangle, color=cyan, mark size=1.5, mark options={fill=cyan!20!cyan,mark indices=5}, 
     ] table[x=p/q,y=Acc]{\MVM}; 
     
       \addplot [line width=0.05mm, 
     mark=diamond, color=violet, mark size=1.5, mark options={fill=violet!20!violet,mark indices=5}, 
     ] table[x=p/q,y=Acc]{\MVMt}; 
     
       \addplot [line width=0.05mm, 
     mark=star, color=green, mark size=1.5, mark options={fill=green!20!green,mark indices=5}, 
     ] table[x=p/q,y=Acc]{\GL}; 
     
       \addplot [line width=0.05mm, 
     mark=x, color=orange, mark size=1.5, mark options={fill=orange!20!orange,mark indices=5}, 
     ] table[x=p/q,y=Acc]{\CoReg}; 
     
       \addplot [line width=0.05mm, 
     mark=10-pointed star, color=teal, mark size=1.5, mark options={fill=teal!20!teal,mark indices=5}, 
     ] table[x=p/q,y=Acc]{\AWP};  
     
       \addplot [line width=0.05mm, 
     mark=oplus, color=yellow, mark size=1.5, mark options={fill=yellow!20!yellow,mark indices=5}, 
     ] table[x=p/q,y=Acc]{\MCGC}; 
     
       \addplot [line width=0.05mm,
     mark=Mercedes star, color=magenta, mark size=1.5, mark options={fill=magenta!20!magenta,mark indices=5}, 
     ] table[x=p/q,y=Acc]{\PM}; 
     
       \addplot [line width=0.05mm,
     mark=pentagon, color=black, mark size=1.5, mark options={fill=black!20!black,mark indices=5}, 
     ] table[x=p/q,y=Acc]{\MT}; 
     
       \addplot [line width=0.05mm,
     mark=|, color=pink, mark size=1.5, mark options={fill=pink!20!pink,mark indices=5}, 
     ] table[x=p/q,y=Acc]{\SCML}; 
     
       \addplot [line width=0.05mm, 
     mark=asterisk, color=olive, mark size=1.5, mark options={fill=olive!20!olive,mark indices=5}, 
     ] table[x=p/q,y=Acc]{\PMM}; 
     
     
     \nextgroupplot[title=(b),xmin=1.98, xmax=3.02, ymin=0.3, ymax = 1,extra y ticks={0.3,0.5,0.7,0.9},xlabel={p/q}, ylabel={NMI}]
     
       \addplot [line width=0.05mm, 
     mark=o, color=red, mark size=1.5, mark options={fill=red!20!red,mark indices=5}, 
     ] table[x=p/q,y=NMI]{\EVM}; 
     
     \addplot [line width=0.05mm, 
     mark=square, color=blue, mark size=1.5, mark options={fill=blue!20!blue,mark indices=5}, 
     ] table[x=p/q,y=NMI]{\MA}; 
     
      \addplot [line width=0.05mm, 
     mark=otimes, color=brown, mark size=1.5, mark options={fill=brown!20!brown,mark indices=5}, 
     ] table[x=p/q,y=NMI]{\MAt}; 
     
       \addplot [line width=0.05mm, 
     mark=triangle, color=cyan, mark size=1.5, mark options={fill=cyan!20!cyan,mark indices=5}, 
     ] table[x=p/q,y=NMI]{\MVM}; 
     
       \addplot [line width=0.05mm, 
     mark=diamond, color=violet, mark size=1.5, mark options={fill=violet!20!violet,mark indices=5}, 
     ] table[x=p/q,y=NMI]{\MVMt}; 
     
       \addplot [line width=0.05mm, 
     mark=star, color=green, mark size=1.5, mark options={fill=green!20!green,mark indices=5}, 
     ] table[x=p/q,y=NMI]{\GL}; 
     
       \addplot [line width=0.05mm, 
     mark=x, color=orange, mark size=1.5, mark options={fill=orange!20!orange,mark indices=5}, 
     ] table[x=p/q,y=NMI]{\CoReg}; 
     
       \addplot [line width=0.05mm, 
     mark=10-pointed star, color=teal, mark size=1.5, mark options={fill=teal!20!teal,mark indices=5}, 
     ] table[x=p/q,y=NMI]{\AWP};  
     
       \addplot [line width=0.05mm, 
     mark=oplus, color=yellow, mark size=1.5, mark options={fill=yellow!20!yellow,mark indices=5}, 
     ] table[x=p/q,y=NMI]{\MCGC}; 
     
       \addplot [line width=0.05mm,
     mark=Mercedes star, color=magenta, mark size=1.5, mark options={fill=magenta!20!magenta,mark indices=5}, 
     ] table[x=p/q,y=NMI]{\PM}; 
     
       \addplot [line width=0.05mm,
     mark=pentagon, color=black, mark size=1.5, mark options={fill=black!20!black,mark indices=5}, 
     ] table[x=p/q,y=NMI]{\MT}; 
     
       \addplot [line width=0.05mm,
     mark=|, color=pink, mark size=1.5, mark options={fill=pink!20!pink,mark indices=5}, 
     ] table[x=p/q,y=NMI]{\SCML}; 
     
       \addplot [line width=0.05mm, 
     mark=asterisk, color=olive, mark size=1.5, mark options={fill=olive!20!olive,mark indices=5}, 
     ] table[x=p/q,y=NMI]{\PMM}; 

    \end{groupplot}
\end{tikzpicture}}
    \hfill
    \hspace*{-1cm}
    \subfloat[]
    {

\pgfplotstableread{
Methods p/q Acc NMI 
AWP 3 0.997 0.988
AWP 2.8 0.972 0.968
AWP 2.5 0.835 0.817
AWP 2.3 0.896 0.787
AWP 2 0.711 0.435
}{\AWP}

\pgfplotstableread{
Methods p/q Acc NMI 
CoReg 3 0.998 0.992
CoReg 2.8 0.997 0.986
CoReg 2.5 0.986 0.945
CoReg 2.3 0.962 0.866
CoReg 2 0.83 0.557
}{\CoReg}

\pgfplotstableread{
Methods p/q Acc NMI 
EVM 3 1 1
EVM 2.8 0.999 0.997
EVM 2.5 0.999 0.997
EVM 2.3 0.996 0.984
EVM 2 0.987 0.95
}{\EVM}

\pgfplotstableread{
Methods p/q Acc NMI 
GL 3 1 1
GL 2.8 0.999 0.997
GL 2.5 0.998 0.993
GL 2.3 0.996 0.982
GL 2 0.988 0.952
}{\GL}

\pgfplotstableread{
Methods p/q Acc NMI 
MA2 3 1 1
MA2 2.8 0.999 0.997
MA2 2.5 0.998 0.993
MA2 2.3 0.996 0.982
MA2 2 0.988 0.952
}{\MA}

\pgfplotstableread{
Methods p/q Acc NMI 
MA3 3 1 1
MA3 2.8 0.999 0.997
MA3 2.5 0.998 0.993
MA3 2.3 0.996 0.982
MA3 2 0.988 0.952
}{\MAt}

\pgfplotstableread{
Methods p/q Acc NMI 
MCGC 3 0.998 0.992
MCGC 2.8 0.997 0.989
MCGC 2.5 0.983 0.936
MCGC 2.3 0.955 0.843
MCGC 2 0.825 0.547
}{\MCGC}

\pgfplotstableread{
Methods p/q Acc NMI 
MT 3 0.996 0.983
MT 2.8 0.996 0.983
MT 2.5 0.987 0.951
MT 2.3 0.977 0.925
MT 2 0.906 0.79
}{\MT}

\pgfplotstableread{
Methods p/q Acc NMI 
MVM2 3 1 1
MVM2 2.8 0.999 0.997
MVM2 2.5 0.999 0.997
MVM2 2.3 0.996 0.982
MVM2 2 0.989 0.954
}{\MVM}

\pgfplotstableread{
Methods p/q Acc NMI 
MVM3 3 1 1
MVM3 2.8 0.999 0.997
MVM3 2.5 0.999 0.997
MVM3 2.3 0.996 0.983
MVM3 2 0.988 0.953
}{\MVMt}

\pgfplotstableread{
Methods p/q Acc NMI 
PM 3 1 0.999
PM 2.8 0.999 0.995
PM 2.5 0.997 0.986
PM 2.3 0.991 0.962
PM 2 0.973 0.9
}{\PM}

\pgfplotstableread{
Methods p/q Acc NMI 
PMM 3 0.998 0.992
PMM 2.8 0.996 0.985
PMM 2.5 0.984 0.939
PMM 2.3 0.969 0.888
PMM 2 0.887 0.672
}{\PMM}

\pgfplotstableread{
Methods p/q Acc NMI 
SCML 3 0.999 0.998
SCML 2.8 0.999 0.994
SCML 2.5 0.991 0.965
SCML 2.3 0.981 0.928
SCML 2 0.91 0.724
}{\SCML}

\begin{tikzpicture}
    \begin{groupplot}[
        group style={
        group name=art_info, 
        group size= 2 by 1,
        horizontal sep=50pt
        },
        width=6.3cm,
        height=6.3cm,
        enlarge y limits,
        axis y line*=left,
        axis x line*=bottom, 
        xtick=data,
        ymajorgrids,
		grid style={draw=gray!20},
        y axis line style={draw=none},
        xtick style={draw=none},
		ytick style={draw=none},
        yticklabel style={xshift=-0.5em},
        legend style={
        at={(1.1,-0.30)},
        legend columns=8,
        legend cell align=left,
        anchor=north},
        legend image post style={mark options={scale=1}}
    ]
     
     \nextgroupplot[title=(c),xmin=1.98, xmax=3.02, ymin=0.6, ymax = 1,extra y ticks={0.7,0.9},xlabel={p/q}, ylabel={Acc}]
     
      \addplot [line width=0.05mm, 
     mark=o, color=red, mark size=1.5, mark options={fill=red!20!red,mark indices=5}, 
     ] table[x=p/q,y=Acc]{\EVM}; \addlegendentry{EVM};
     
     \addplot [line width=0.05mm, 
     mark=square, color=blue, mark size=1.5, mark options={fill=blue!20!blue,mark indices=5}, 
     ] table[x=p/q,y=Acc]{\MA}; \addlegendentry{MA2};
     
      \addplot [line width=0.05mm, 
     mark=otimes, color=brown, mark size=1.5, mark options={fill=brown!20!brown,mark indices=5}, 
     ] table[x=p/q,y=Acc]{\MAt}; \addlegendentry{MA3};
     
       \addplot [line width=0.05mm, 
     mark=triangle, color=cyan, mark size=1.5, mark options={fill=cyan!20!cyan,mark indices=5}, 
     ] table[x=p/q,y=Acc]{\MVM}; \addlegendentry{MVM2};
     
       \addplot [line width=0.05mm, 
     mark=diamond, color=violet, mark size=1.5, mark options={fill=violet!20!violet,mark indices=5}, 
     ] table[x=p/q,y=Acc]{\MVMt}; \addlegendentry{MVM3};
     
      \addlegendimage{empty legend}
            \addlegendentry{}
            
     \addlegendimage{empty legend}
            \addlegendentry{}
            
     \addlegendimage{empty legend}
            \addlegendentry{}
            
       \addplot [line width=0.05mm, 
     mark=star, color=green, mark size=1.5, mark options={fill=green!20!green,mark indices=5}, 
     ] table[x=p/q,y=Acc]{\GL}; \addlegendentry{GL};
     
       \addplot [line width=0.05mm, 
     mark=x, color=orange, mark size=1.5, mark options={fill=orange!20!orange,mark indices=5}, 
     ] table[x=p/q,y=Acc]{\CoReg}; \addlegendentry{CoReg};
     
       \addplot [line width=0.05mm, 
     mark=10-pointed star, color=teal, mark size=1.5, mark options={fill=teal!20!teal,mark indices=5}, 
     ] table[x=p/q,y=Acc]{\AWP};  \addlegendentry{AWP};
     
       \addplot [line width=0.05mm, 
     mark=oplus, color=yellow, mark size=1.5, mark options={fill=yellow!20!yellow,mark indices=5}, 
     ] table[x=p/q,y=Acc]{\MCGC}; \addlegendentry{MCGC};
     
       \addplot [line width=0.05mm,
     mark=Mercedes star, color=magenta, mark size=1.5, mark options={fill=magenta!20!magenta,mark indices=5}, 
     ] table[x=p/q,y=Acc]{\PM}; \addlegendentry{PM};
     
       \addplot [line width=0.05mm,
     mark=pentagon, color=black, mark size=1.5, mark options={fill=black!20!black,mark indices=5}, 
     ] table[x=p/q,y=Acc]{\MT}; \addlegendentry{MT};
     
       \addplot [line width=0.05mm,
     mark=|, color=pink, mark size=1.5, mark options={fill=pink!20!pink,mark indices=5}, 
     ] table[x=p/q,y=Acc]{\SCML}; \addlegendentry{SCML};
     
       \addplot [line width=0.05mm, 
     mark=asterisk, color=olive, mark size=1.5, mark options={mark indices=5}, 
     ] table[x=p/q,y=Acc]{\PMM}; \addlegendentry{PMM};
     
     
     \nextgroupplot[title=(d),xmin=1.98, xmax=3.02, ymin=0.3, ymax = 1,extra y ticks={0.3,0.5,0.7,0.9},xlabel={p/q}, ylabel={NMI}]
     
      \addplot [line width=0.05mm, 
     mark=o, color=red, mark size=1.5, mark options={fill=red!20!red,mark indices=5}, 
     ] table[x=p/q,y=NMI]{\EVM}; 
     
     \addplot [line width=0.05mm, 
     mark=square, color=blue, mark size=1.5, mark options={fill=blue!20!blue,mark indices=5}, 
     ] table[x=p/q,y=NMI]{\MA}; 
     
      \addplot [line width=0.05mm, 
     mark=otimes, color=brown, mark size=1.5, mark options={fill=brown!20!brown,mark indices=5}, 
     ] table[x=p/q,y=NMI]{\MAt}; 
     
       \addplot [line width=0.05mm, 
     mark=triangle, color=cyan, mark size=1.5, mark options={fill=cyan!20!cyan,mark indices=5}, 
     ] table[x=p/q,y=NMI]{\MVM}; 
     
       \addplot [line width=0.05mm, 
     mark=diamond, color=violet, mark size=1.5, mark options={fill=violet!20!violet,mark indices=5}, 
     ] table[x=p/q,y=NMI]{\MVMt}; 
     
       \addplot [line width=0.05mm, 
     mark=star, color=green, mark size=1.5, mark options={fill=green!20!green,mark indices=5}, 
     ] table[x=p/q,y=NMI]{\GL}; 
     
       \addplot [line width=0.05mm, 
     mark=x, color=orange, mark size=1.5, mark options={fill=orange!20!orange,mark indices=5}, 
     ] table[x=p/q,y=NMI]{\CoReg}; 
     
       \addplot [line width=0.05mm, 
     mark=10-pointed star, color=teal, mark size=1.5, mark options={fill=teal!20!teal,mark indices=5}, 
     ] table[x=p/q,y=NMI]{\AWP};  
     
       \addplot [line width=0.05mm, 
     mark=oplus, color=yellow, mark size=1.5, mark options={fill=yellow!20!yellow,mark indices=5}, 
     ] table[x=p/q,y=NMI]{\MCGC}; 
     
       \addplot [line width=0.05mm,
     mark=Mercedes star, color=magenta, mark size=1.5, mark options={fill=magenta!20!magenta,mark indices=5}, 
     ] table[x=p/q,y=NMI]{\PM}; 
     
       \addplot [line width=0.05mm,
     mark=pentagon, color=black, mark size=1.5, mark options={fill=black!20!black,mark indices=5}, 
     ] table[x=p/q,y=NMI]{\MT}; 
     
       \addplot [line width=0.05mm,
     mark=|, color=pink, mark size=1.5, mark options={fill=pink!20!pink,mark indices=5}, 
     ] table[x=p/q,y=NMI]{\SCML}; 
     
       \addplot [line width=0.05mm, 
     mark=asterisk, color=olive, mark size=1.5, mark options={fill=olive!20!olive,mark indices=5}, 
     ] table[x=p/q,y=NMI]{\PMM}; 

    \end{groupplot}
\end{tikzpicture}}
\caption{Average values of accuracy and NMI over 10 random networks sampled from SBM with equally distributed informative layers (2 layers (a)(b) and 3 layers (c)(d)) with four clusters of equal size, for $p=0.1$ and  $p/q \in \{2, 2.3, 2.5, 2.8, 3\}$.   }
\label{fig:plot1}
\end{figure*}

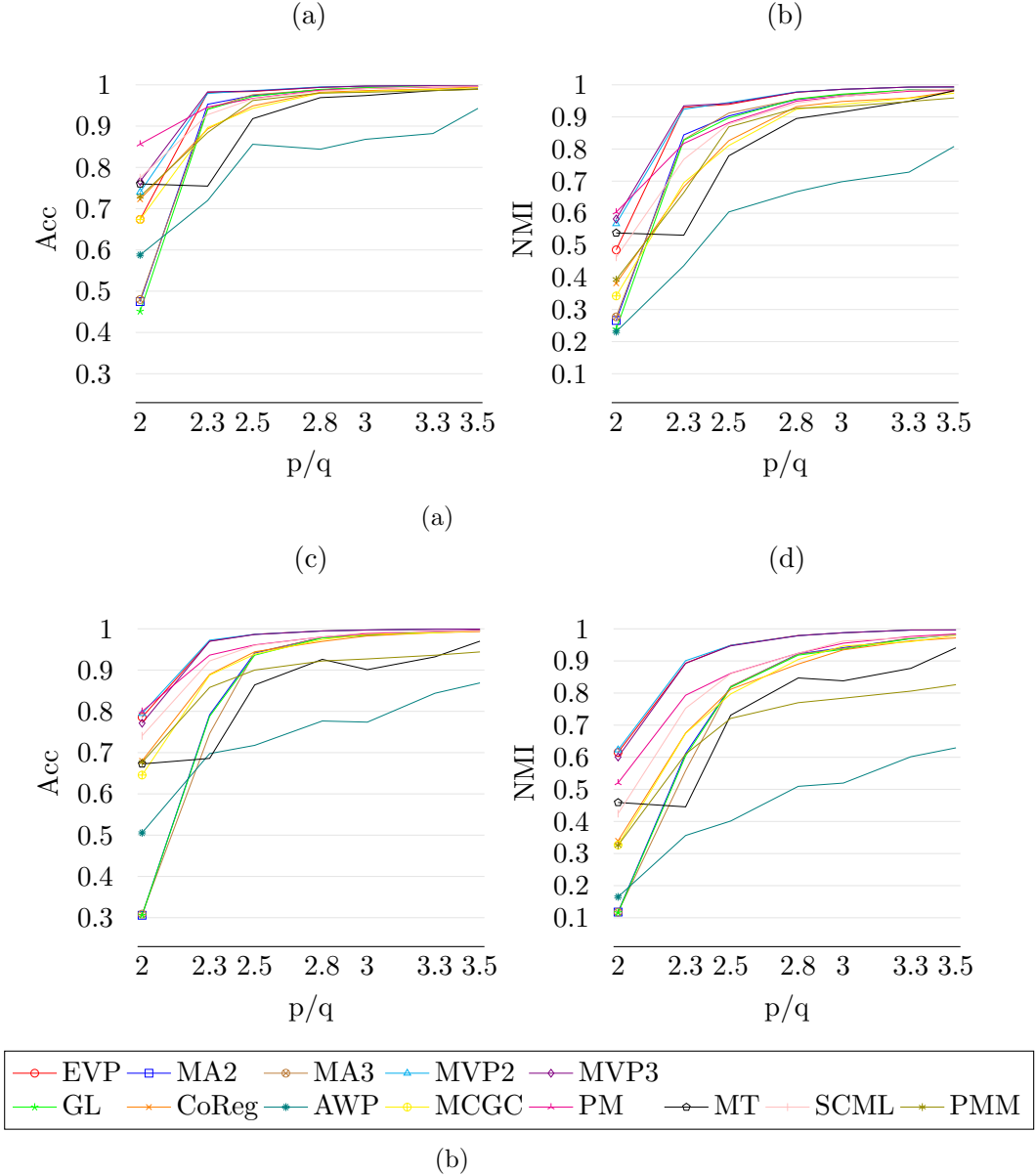
\begin{figure*}[p]
\centering
\captionsetup{oneside,margin={-0.53cm,0.6cm}}
    \hspace*{-1.9cm}
    \subfloat[]
    {
\pgfplotstableread{
Methods p/q Acc NMI 
AWP 3.5 0.943 0.808
AWP 3.3 0.882 0.728
AWP 3 0.868 0.698
AWP 2.8 0.844 0.667
AWP 2.5 0.856 0.604
AWP 2.3 0.7198 0.436490401
AWP 2 0.5878 0.230931294
}{\AWP}

\pgfplotstableread{
Methods p/q Acc NMI 
CoReg 3.5 0.995 0.98
CoReg 3.3 0.99 0.959
CoReg 3 0.986 0.948
CoReg 2.8 0.982 0.932
CoReg 2.5 0.949 0.826
CoReg 2.3 0.89304 0.682441872
CoReg 2 0.72284 0.381754362
}{\CoReg}

\pgfplotstableread{
Methods p/q Acc NMI 
EVP 3.5 0.998 0.993
EVP 3.3 0.998 0.993
EVP 3 0.997 0.986
EVP 2.8 0.994 0.977
EVP 2.5 0.984 0.939
EVP 2.3 0.98144 0.929802153
EVP 2 0.67392 0.485834157
}{\EVP}

\pgfplotstableread{
Methods p/q Acc NMI 
GL 3.5 0.996 0.984
GL 3.3 0.996 0.984
GL 3 0.993 0.971
GL 2.8 0.989 0.956
GL 2.5 0.972 0.897
GL 2.3 0.94052 0.827548766
GL 2 0.45132 0.239199845
    }{\GL}

\pgfplotstableread{
Methods p/q Acc NMI 
MA2 3.5 0.996 0.985
MA2 3.3 0.996 0.984
MA2 3 0.993 0.969
MA2 2.8 0.988 0.954
MA2 2.5 0.974 0.903
MA2 2.3 0.95288 0.843982803
MA2 2 0.47492 0.265945692
}{\MA}

\pgfplotstableread{
Methods p/q Acc NMI 
MA3 3.5 0.996 0.986
MA3 3.3 0.996 0.983
MA3 3 0.993 0.969
MA3 2.8 0.988 0.955
MA3 2.5 0.976 0.912
MA3 2.3 0.94236 0.82927328
MA3 2 0.47976 0.277121227
}{\MAt}

\pgfplotstableread{
Methods p/q Acc NMI 
MCGC 3.5 0.993 0.974
MCGC 3.3 0.989 0.958
MCGC 3 0.984 0.939
MCGC 2.8 0.98 0.924
MCGC 2.5 0.943 0.811
MCGC 2.3 0.8964 0.694846656
MCGC 2 0.6736 0.342345916
}{\MCGC}

\pgfplotstableread{
Methods p/q Acc NMI 
MT 3.5 0.99 0.982
MT 3.3 0.986 0.949
MT 3 0.974 0.915
MT 2.8 0.969 0.895
MT 2.5 0.918 0.779
MT 2.3 0.7546 0.531572474
MT 2 0.76 0.538637733
}{\MT}

\pgfplotstableread{
Methods p/q Acc NMI 
MVP2 3.5 0.998 0.993
MVP2 3.3 0.998 0.993
MVP2 3 0.997 0.986
MVP2 2.8 0.995 0.978
MVP2 2.5 0.986 0.945
MVP2 2.3 0.97924 0.92389341
MVP2 2 0.7402 0.567394194
}{\MVP}

\pgfplotstableread{
Methods p/q Acc NMI 
MVP3 3.5 0.998 0.994
MVP3 3.3 0.998 0.993
MVP3 3 0.997 0.986
MVP3 2.8 0.994 0.977
MVP3 2.5 0.985 0.942
MVP3 2.3 0.98304 0.934543868
MVP3 2 0.766 0.582102741
}{\MVPt}

\pgfplotstableread{
Methods p/q Acc NMI 
PM 3.5 0.996 0.982
PM 3.3 0.995 0.98
PM 3 0.991 0.964
PM 2.8 0.987 0.949
PM 2.5 0.967 0.882
PM 2.3 0.94628 0.817212865
PM 2 0.85624 0.603403862
}{\PM}

\pgfplotstableread{
Methods p/q Acc NMI 
PMM 3.5 0.989 0.959
PMM 3.3 0.986 0.948
PMM 3 0.982 0.932
PMM 2.8 0.98 0.927
PMM 2.5 0.962 0.869
PMM 2.3 0.8844 0.664095369
PMM 2 0.7294 0.393451259
}{\PMM}

\pgfplotstableread{
Methods p/q Acc NMI 
SCML 3.5 0.997 0.987
SCML 3.3 0.996 0.983
SCML 3 0.991 0.963
SCML 2.8 0.985 0.943
SCML 2.5 0.966 0.878
SCML 2.3 0.9276 0.767645045
SCML 2 0.775 0.462756859
}{\SCML}

\begin{tikzpicture}
    \begin{groupplot}[
        group style={
        group name=art_info, 
        group size= 2 by 1,
        horizontal sep=50pt
        },
        width=6.3cm,
        height=6.3cm,
        enlarge y limits,
        axis y line*=left,
        axis x line*=bottom, 
        xtick=data,
        ymajorgrids,
		grid style={line width=.2pt,draw=gray!20},
        y axis line style={draw=none},
        xtick style={draw=none},
		ytick style={draw=none},
        yticklabel style={xshift=-0.5em},
    ]
     
     \nextgroupplot[title=(a),xmin=1.98, xmax=3.52, ymin=0.3, ymax = 1,extra y ticks={0.3,0.5,0.7,0.9}, xlabel={p/q}, ylabel={Acc}]
     
      \addplot [line width=0.05mm, 
     mark=o, color=red, mark size=1.5, mark options={fill=red!20!red,mark indices=7}, 
     ] table[x=p/q,y=Acc]{\EVP}; 
     
     \addplot [line width=0.05mm, 
     mark=square, color=blue, mark size=1.5, mark options={fill=blue!20!blue,mark indices=7}, 
     ] table[x=p/q,y=Acc]{\MA}; 
     
      \addplot [line width=0.05mm, 
     mark=otimes, color=brown, mark size=1.5, mark options={fill=brown!20!brown,mark indices=7}, 
     ] table[x=p/q,y=Acc]{\MAt}; 
     
       \addplot [line width=0.05mm, 
     mark=triangle, color=cyan, mark size=1.5, mark options={fill=cyan!20!cyan,mark indices=7}, 
     ] table[x=p/q,y=Acc]{\MVP}; 
     
       \addplot [line width=0.05mm, 
     mark=diamond, color=violet, mark size=1.5, mark options={fill=violet!20!violet,mark indices=7}, 
     ] table[x=p/q,y=Acc]{\MVPt}; 
     
       \addplot [line width=0.05mm, 
     mark=star, color=green, mark size=1.5, mark options={fill=green!20!green,mark indices=7}, 
     ] table[x=p/q,y=Acc]{\GL}; 
     
       \addplot [line width=0.05mm, 
     mark=x, color=orange, mark size=1.5, mark options={fill=orange!20!orange,mark indices=7}, 
     ] table[x=p/q,y=Acc]{\CoReg}; 
     
       \addplot [line width=0.05mm, 
     mark=10-pointed star, color=teal, mark size=1.5, mark options={fill=teal!20!teal,mark indices=7}, 
     ] table[x=p/q,y=Acc]{\AWP};  
     
       \addplot [line width=0.05mm, 
     mark=oplus, color=yellow, mark size=1.5, mark options={fill=yellow!20!yellow,mark indices=7}, 
     ] table[x=p/q,y=Acc]{\MCGC}; 
     
       \addplot [line width=0.05mm,
     mark=Mercedes star, color=magenta, mark size=1.5, mark options={fill=magenta!20!magenta,mark indices=7}, 
     ] table[x=p/q,y=Acc]{\PM}; 
     
       \addplot [line width=0.05mm,
     mark=pentagon, color=black, mark size=1.5, mark options={fill=black!20!black,mark indices=7}, 
     ] table[x=p/q,y=Acc]{\MT}; 
     
       \addplot [line width=0.05mm,
     mark=|, color=pink, mark size=1.5, mark options={fill=pink!20!pink,mark indices=7}, 
     ] table[x=p/q,y=Acc]{\SCML}; 
     
       \addplot [line width=0.05mm, 
     mark=asterisk, color=olive, mark size=1.5, mark options={fill=olive!20!olive,mark indices=7}, 
     ] table[x=p/q,y=Acc]{\PMM};

     
     \nextgroupplot[title=(b),xmin=1.98, xmax=3.52, ymin=0.1, ymax = 1,extra y ticks={0.1,0.2,0.3,0.4,0.6,0.7,0.8,0.9},xlabel={p/q}, ylabel={NMI}]
     
        \addplot [line width=0.05mm, 
     mark=o, color=red, mark size=1.5, mark options={fill=red!20!red,mark indices=7}, 
     ] table[x=p/q,y=NMI]{\EVP}; 
     
     \addplot [line width=0.05mm, 
     mark=square, color=blue, mark size=1.5, mark options={fill=blue!20!blue,mark indices=7}, 
     ] table[x=p/q,y=NMI]{\MA}; 
     
      \addplot [line width=0.05mm, 
     mark=otimes, color=brown, mark size=1.5, mark options={fill=brown!20!brown,mark indices=7}, 
     ] table[x=p/q,y=NMI]{\MAt}; 
     
       \addplot [line width=0.05mm, 
     mark=triangle, color=cyan, mark size=1.5, mark options={fill=cyan!20!cyan,mark indices=7}, 
     ] table[x=p/q,y=NMI]{\MVP}; 
     
       \addplot [line width=0.05mm, 
     mark=diamond, color=violet, mark size=1.5, mark options={fill=violet!20!violet,mark indices=7}, 
     ] table[x=p/q,y=NMI]{\MVPt}; 
     
       \addplot [line width=0.05mm, 
     mark=star, color=green, mark size=1.5, mark options={fill=green!20!green,mark indices=7}, 
     ] table[x=p/q,y=NMI]{\GL}; 
     
       \addplot [line width=0.05mm, 
     mark=x, color=orange, mark size=1.5, mark options={fill=orange!20!orange,mark indices=7}, 
     ] table[x=p/q,y=NMI]{\CoReg}; 
     
       \addplot [line width=0.05mm, 
     mark=10-pointed star, color=teal, mark size=1.5, mark options={fill=teal!20!teal,mark indices=7}, 
     ] table[x=p/q,y=NMI]{\AWP};  
     
       \addplot [line width=0.05mm, 
     mark=oplus, color=yellow, mark size=1.5, mark options={fill=yellow!20!yellow,mark indices=7}, 
     ] table[x=p/q,y=NMI]{\MCGC}; 
     
       \addplot [line width=0.05mm,
     mark=Mercedes star, color=magenta, mark size=1.5, mark options={fill=magenta!20!magenta,mark indices=7}, 
     ] table[x=p/q,y=NMI]{\PM}; 
     
       \addplot [line width=0.05mm,
     mark=pentagon, color=black, mark size=1.5, mark options={fill=black!20!black,mark indices=7}, 
     ] table[x=p/q,y=NMI]{\MT}; 
     
       \addplot [line width=0.05mm,
     mark=|, color=pink, mark size=1.5, mark options={fill=pink!20!pink,mark indices=7}, 
     ] table[x=p/q,y=NMI]{\SCML}; 
     
       \addplot [line width=0.05mm, 
     mark=asterisk, color=olive, mark size=1.5, mark options={fill=olive!20!olive,mark indices=7}, 
     ] table[x=p/q,y=NMI]{\PMM}; 

    \end{groupplot}
\end{tikzpicture}}
    \hfill
    \hspace*{-1.5cm}
    \subfloat[]
    {
\pgfplotstableread{
Methods p/q Acc NMI 
AWP 3.5 0.868857143 0.629347559
AWP 3.3 0.8438 0.601651672
AWP 3 0.7738 0.519533113
AWP 2.8 0.7766 0.509345359
AWP 2.5 0.7176 0.401304204
AWP 2.3 0.6976 0.35632056
AWP 2 0.5056 0.165056672
}{\AWP}

\pgfplotstableread{
Methods p/q Acc NMI 
CoReg 3.5 0.993190476 0.971721541
CoReg 3.3 0.9906 0.962583255
CoReg 3 0.9828 0.93452979
CoReg 2.8 0.96938 0.890393359
CoReg 2.5 0.94436 0.811711315
CoReg 2.3 0.88944 0.675593885
CoReg 2 0.68008 0.339591007
}{\CoReg}

\pgfplotstableread{
Methods p/q Acc NMI 
EVP 3.5 0.999176191 0.996536518
EVP 3.3 0.99916 0.99646854
EVP 3 0.99702 0.987536923
EVP 2.8 0.99488 0.979353532
EVP 2.5 0.98674 0.948883536
EVP 2.3 0.9702 0.892720084
EVP 2 0.7858 0.615058476
}{\EVP}

\pgfplotstableread{
Methods p/q Acc NMI 
GL 3.5 0.995638095 0.98196894
GL 3.3 0.99256 0.969875005
GL 3 0.98374 0.936264865
GL 2.8 0.97786 0.918174074
GL 2.5 0.93614 0.817699229
GL 2.3 0.78796 0.604410578
GL 2 0.30604 0.115814903
}{\GL}

\pgfplotstableread{
Methods p/q Acc NMI 
MA2 3.5 0.995766667 0.982580916
MA2 3.3 0.99256 0.970045723
MA2 3 0.9853 0.941514624
MA2 2.8 0.97924 0.922028638
MA2 2.5 0.94128 0.819651564
MA2 2.3 0.79168 0.611869509
MA2 2 0.30588 0.117724381
}{\MA}

\pgfplotstableread{
Methods p/q Acc NMI 
MA3 3.5 0.995928571 0.983076511
MA3 3.3 0.99298 0.971277549
MA3 3 0.9847 0.939366101
MA3 2.8 0.97884 0.921000131
MA3 2.5 0.94248 0.821006398
MA3 2.3 0.74624 0.558262185
MA3 2 0.30864 0.117470037
}{\MAt}

\pgfplotstableread{
Methods p/q Acc NMI 
MCGC 3.5 0.995047619 0.979191952
MCGC 3.3 0.99 0.959988837
MCGC 3 0.9866 0.946601265
MCGC 2.8 0.9738 0.904502213
MCGC 2.5 0.9384 0.797174799
MCGC 2.3 0.8876 0.674122675
MCGC 2 0.6456 0.326629887
}{\MCGC}

\pgfplotstableread{
Methods p/q Acc NMI 
MT 3.5 0.97 0.941
MT 3.3 0.932 0.877
MT 3 0.901 0.838
MT 2.8 0.926 0.847
MT 2.5 0.864 0.731
MT 2.3 0.686 0.445506779
MT 2 0.6726 0.459120334
}{\MT}

\pgfplotstableread{
Methods p/q Acc NMI 
MVP2 3.5 0.999176191 0.996536478
MVP2 3.3 0.99908 0.996152079
MVP2 3 0.9971 0.987852214
MVP2 2.8 0.99474 0.978853016
MVP2 2.5 0.9866 0.948511151
MVP2 2.3 0.97252 0.900864446
MVP2 2 0.79472 0.622866898
}{\MVP}

\pgfplotstableread{
Methods p/q Acc NMI 
MVP3 3.5 0.999176191 0.996536478
MVP3 3.3 0.9991 0.996216168
MVP3 3 0.99716 0.988084185
MVP3 2.8 0.9947 0.978652303
MVP3 2.5 0.98614 0.947000393
MVP3 2.3 0.96968 0.892723536
MVP3 2 0.77132 0.60085497
}{\MVPt}

\pgfplotstableread{
Methods p/q Acc NMI 
PM 3.5 0.996190476 0.983988961
PM 3.3 0.99418 0.976371071
PM 3 0.9886 0.95526181
PM 2.8 0.98024 0.924131896
PM 2.5 0.96104 0.861275239
PM 2.3 0.93624 0.793150503
PM 2 0.79984 0.520060414
}{\PM}

\pgfplotstableread{
Methods p/q Acc NMI 
PMM 3.5 0.944 0.826
PMM 3.3 0.936 0.806
PMM 3 0.927 0.784
PMM 2.8 0.922 0.77
PMM 2.5 0.9 0.721
PMM 2.3 0.858 0.609993532
PMM 2 0.6764 0.325698402
}{\PMM}

\pgfplotstableread{
Methods p/q Acc NMI 
SCML 3.5 0.995 0.98
SCML 3.3 0.993 0.973
SCML 3 0.991 0.962
SCML 2.8 0.98 0.924
SCML 2.5 0.96 0.86
SCML 2.3 0.9222 0.752700638
SCML 2 0.7406 0.424481096
}{\SCML}

\begin{tikzpicture}
    \begin{groupplot}[
        group style={
        group name=art_info, 
        group size= 2 by 1,
        horizontal sep=50pt
        },
        width=6.3cm,
        height=6.3cm,
        enlarge y limits,
        axis y line*=left,
        axis x line*=bottom, 
        xtick=data,
        ymajorgrids,
		grid style={draw=gray!20},
        y axis line style={draw=none},
        xtick style={draw=none},
		ytick style={draw=none},
        yticklabel style={xshift=-0.5em},
        legend style={
        at={(1.1,-0.30)},
        legend columns=8,
        legend cell align=left,
        anchor=north},
        legend image post style={mark options={scale=1}}
    ]
     
     \nextgroupplot[title=(c),xmin=1.98, xmax=3.52, ymin=0.3, ymax = 1,extra y ticks={0.3,0.5,0.7,0.9},xlabel={p/q}, ylabel={Acc}]
     
      \addplot [line width=0.05mm, 
     mark=o, color=red, mark size=1.5, mark options={fill=red!20!red,mark indices=7}, 
     ] table[x=p/q,y=Acc]{\EVP}; \addlegendentry{EVP};
     
     \addplot [line width=0.05mm, 
     mark=square, color=blue, mark size=1.5, mark options={fill=blue!20!blue,mark indices=7}, 
     ] table[x=p/q,y=Acc]{\MA}; \addlegendentry{MA2};
     
      \addplot [line width=0.05mm, 
     mark=otimes, color=brown, mark size=1.5, mark options={fill=brown!20!brown,mark indices=7}, 
     ] table[x=p/q,y=Acc]{\MAt}; \addlegendentry{MA3};
     
       \addplot [line width=0.05mm, 
     mark=triangle, color=cyan, mark size=1.5, mark options={fill=cyan!20!cyan,mark indices=7}, 
     ] table[x=p/q,y=Acc]{\MVP}; \addlegendentry{MVP2};
     
       \addplot [line width=0.05mm, 
     mark=diamond, color=violet, mark size=1.5, mark options={fill=violet!20!violet,mark indices=7}, 
     ] table[x=p/q,y=Acc]{\MVPt}; \addlegendentry{MVP3};
     
      \addlegendimage{empty legend}
            \addlegendentry{}
            
     \addlegendimage{empty legend}
            \addlegendentry{}
            
     \addlegendimage{empty legend}
            \addlegendentry{}
            
       \addplot [line width=0.05mm, 
     mark=star, color=green, mark size=1.5, mark options={fill=green!20!green,mark indices=7}, 
     ] table[x=p/q,y=Acc]{\GL}; \addlegendentry{GL};
     
       \addplot [line width=0.05mm, 
     mark=x, color=orange, mark size=1.5, mark options={fill=orange!20!orange,mark indices=7}, 
     ] table[x=p/q,y=Acc]{\CoReg}; \addlegendentry{CoReg};
     
       \addplot [line width=0.05mm, 
     mark=10-pointed star, color=teal, mark size=1.5, mark options={fill=teal!20!teal,mark indices=7}, 
     ] table[x=p/q,y=Acc]{\AWP};  \addlegendentry{AWP};
     
       \addplot [line width=0.05mm, 
     mark=oplus, color=yellow, mark size=1.5, mark options={fill=yellow!20!yellow,mark indices=7}, 
     ] table[x=p/q,y=Acc]{\MCGC}; \addlegendentry{MCGC};
     
       \addplot [line width=0.05mm,
     mark=Mercedes star, color=magenta, mark size=1.5, mark options={fill=magenta!20!magenta,mark indices=7}, 
     ] table[x=p/q,y=Acc]{\PM}; \addlegendentry{PM};
     
       \addplot [line width=0.05mm,
     mark=pentagon, color=black, mark size=1.5, mark options={fill=black!20!black,mark indices=7}, 
     ] table[x=p/q,y=Acc]{\MT}; \addlegendentry{MT};
     
       \addplot [line width=0.05mm,
     mark=|, color=pink, mark size=1.5, mark options={fill=pink!20!pink,mark indices=7}, 
     ] table[x=p/q,y=Acc]{\SCML}; \addlegendentry{SCML};
     
       \addplot [line width=0.05mm, 
     mark=asterisk, color=olive, mark size=1.5, mark options={fill=olive!20!olive,mark indices=7}, 
     ] table[x=p/q,y=Acc]{\PMM}; \addlegendentry{PMM};
     
     
     \nextgroupplot[title=(d),xmin=1.98, xmax=3.52, ymin=0.1, ymax = 1,extra y ticks={0.1,0.2,0.3,0.4,0.6,0.7,0.8,0.9},xlabel={p/q}, ylabel={NMI}]
     
     \addplot [line width=0.05mm, 
     mark=o, color=red, mark size=1.5, mark options={fill=red!20!red,mark indices=7}, 
     ] table[x=p/q,y=NMI]{\EVP}; 
     
     \addplot [line width=0.05mm, 
     mark=square, color=blue, mark size=1.5, mark options={fill=blue!20!blue,mark indices=7}, 
     ] table[x=p/q,y=NMI]{\MA}; 
     
      \addplot [line width=0.05mm, 
     mark=otimes, color=brown, mark size=1.5, mark options={fill=brown!20!brown,mark indices=7}, 
     ] table[x=p/q,y=NMI]{\MAt}; 
     
       \addplot [line width=0.05mm, 
     mark=triangle, color=cyan, mark size=1.5, mark options={fill=cyan!20!cyan,mark indices=7}, 
     ] table[x=p/q,y=NMI]{\MVP}; 
     
       \addplot [line width=0.05mm, 
     mark=diamond, color=violet, mark size=1.5, mark options={fill=violet!20!violet,mark indices=7}, 
     ] table[x=p/q,y=NMI]{\MVPt}; 
     
       \addplot [line width=0.05mm, 
     mark=star, color=green, mark size=1.5, mark options={fill=green!20!green,mark indices=7}, 
     ] table[x=p/q,y=NMI]{\GL}; 
     
       \addplot [line width=0.05mm, 
     mark=x, color=orange, mark size=1.5, mark options={fill=orange!20!orange,mark indices=7}, 
     ] table[x=p/q,y=NMI]{\CoReg}; 
     
       \addplot [line width=0.05mm, 
     mark=10-pointed star, color=teal, mark size=1.5, mark options={fill=teal!20!teal,mark indices=7}, 
     ] table[x=p/q,y=NMI]{\AWP};  
     
       \addplot [line width=0.05mm, 
     mark=oplus, color=yellow, mark size=1.5, mark options={fill=yellow!20!yellow,mark indices=7}, 
     ] table[x=p/q,y=NMI]{\MCGC}; 
     
       \addplot [line width=0.05mm,
     mark=Mercedes star, color=magenta, mark size=1.5, mark options={fill=magenta!20!magenta,mark indices=7}, 
     ] table[x=p/q,y=NMI]{\PM}; 
     
       \addplot [line width=0.05mm,
     mark=pentagon, color=black, mark size=1.5, mark options={fill=black!20!black,mark indices=7}, 
     ] table[x=p/q,y=NMI]{\MT}; 
     
       \addplot [line width=0.05mm,
     mark=|, color=pink, mark size=1.5, mark options={fill=pink!20!pink,mark indices=7}, 
     ] table[x=p/q,y=NMI]{\SCML}; 
     
       \addplot [line width=0.05mm, 
     mark=asterisk, color=olive, mark size=1.5, mark options={fill=olive!20!olive,mark indices=7}, 
     ] table[x=p/q,y=NMI]{\PMM}; 

    \end{groupplot}
\end{tikzpicture}}
\caption{Average values of accuracy and NMI over 10 random networks sampled from SBM with both informative and noisy layers (two informative and one noisy in (a)(b);  two informative and two noisy in (c)(d)). The informative layers are equally distributed SBM graphs with four clusters of equal size, $p=0.1$ and $p/q \in \{2,2.3,2.5, 2.8, 3, 3.3, 3.5\}$. The noisy layers are SBM graphs with $p=q=0.1$. }
\label{fig:plot2}
\end{figure*}

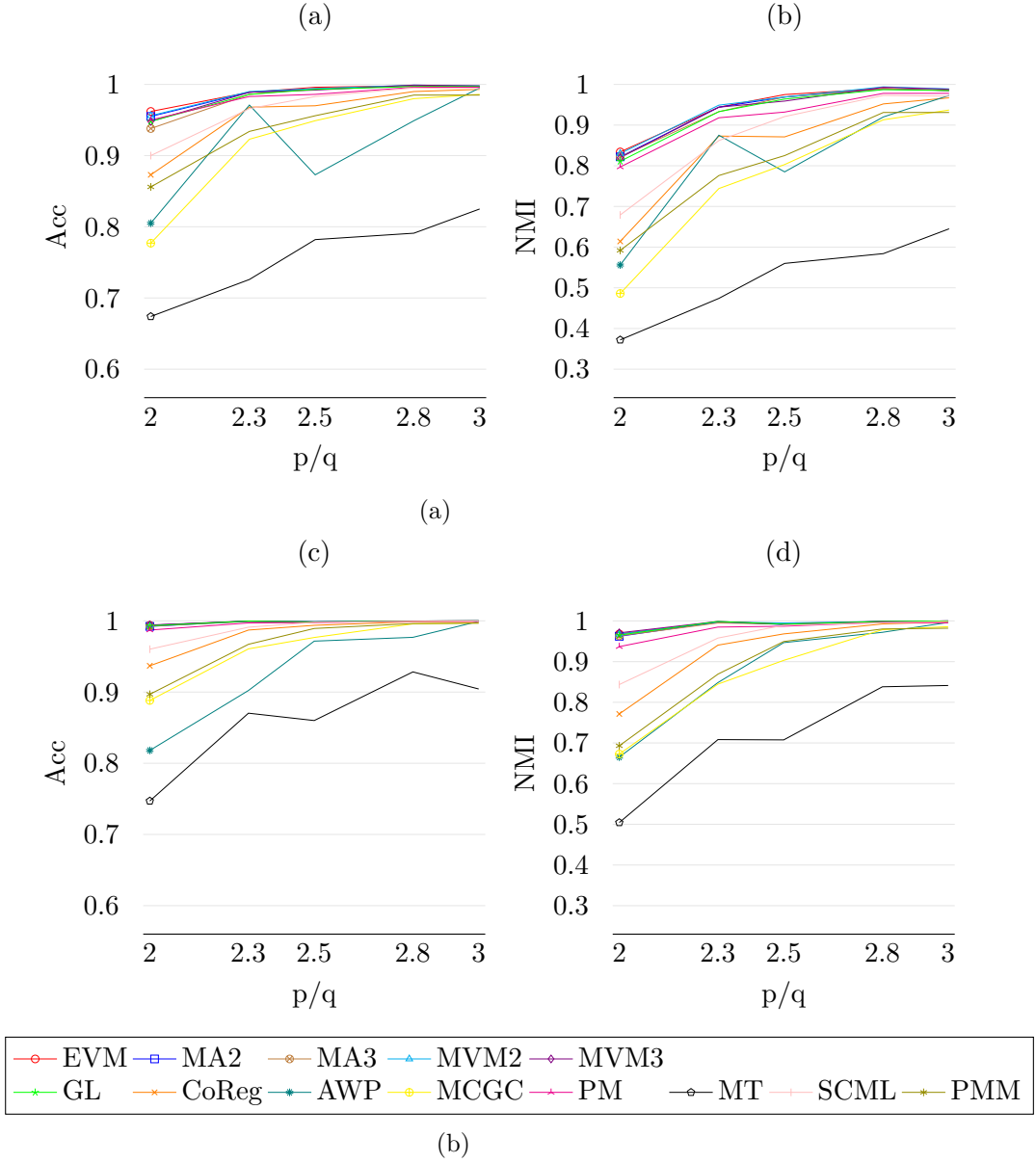
\begin{figure*}[p]
\centering
\captionsetup{oneside,margin={-0.53cm,0.6cm}}
    \hspace*{-2cm}
    \subfloat[]
    {
\pgfplotstableread{
Methods p/q Acc NMI 
AWP 3 0.995 0.973
AWP 2.8 0.949 0.92
AWP 2.5 0.873 0.785
AWP 2.3 0.971 0.875
AWP 2 0.805 0.556
}{\AWP}

\pgfplotstableread{
Methods p/q Acc NMI 
CoReg 3 0.993 0.967
CoReg 2.8 0.99 0.952
CoReg 2.5 0.97 0.871
CoReg 2.3 0.968 0.873
CoReg 2 0.873 0.614
}{\CoReg}

\pgfplotstableread{
Methods p/q Acc NMI 
EVM 3 0.997 0.985
EVM 2.8 0.998 0.99
EVM 2.5 0.996 0.975
EVM 2.3 0.989 0.944
EVM 2 0.962 0.834
}{\EVM}

\pgfplotstableread{
Methods p/q Acc NMI 
GL 3 0.997 0.985
GL 2.8 0.997 0.987
GL 2.5 0.993 0.964
GL 2.3 0.986 0.933
GL 2 0.948 0.811
}{\GL}

\pgfplotstableread{
Methods p/q Acc NMI 
MA2 3 0.997 0.985
MA2 2.8 0.998 0.989
MA2 2.5 0.994 0.969
MA2 2.3 0.989 0.944
MA2 2 0.955 0.822
}{\MA}

\pgfplotstableread{
Methods p/q Acc NMI 
MA3 3 0.997 0.985
MA3 2.8 0.998 0.989
MA3 2.5 0.994 0.969
MA3 2.3 0.986 0.933
MA3 2 0.938 0.822
}{\MAt}

\pgfplotstableread{
Methods p/q Acc NMI 
MCGC 3 0.986 0.937
MCGC 2.8 0.98 0.913
MCGC 2.5 0.949 0.803
MCGC 2.3 0.923 0.744
MCGC 2 0.777 0.486
}{\MCGC}

\pgfplotstableread{
Methods p/q Acc NMI 
MT 3 0.825 0.645
MT 2.8 0.791 0.584
MT 2.5 0.782 0.56
MT 2.3 0.726 0.474
MT 2 0.674 0.372
}{\MT}

\pgfplotstableread{
Methods p/q Acc NMI 
MVM2 3 0.998 0.989
MVM2 2.8 0.999 0.994
MVM2 2.5 0.994 0.969
MVM2 2.3 0.99 0.949
MVM2 2 0.956 0.83
}{\MVM}

\pgfplotstableread{
Methods p/q Acc NMI 
MVM3 3 0.998 0.988
MVM3 2.8 0.999 0.993
MVM3 2.5 0.992 0.959
MVM3 2.3 0.989 0.944
MVM3 2 0.948 0.82
}{\MVMt}

\pgfplotstableread{
Methods p/q Acc NMI 
PM 3 0.996 0.978
PM 2.8 0.996 0.978
PM 2.5 0.986 0.932
PM 2.3 0.983 0.918
PM 2 0.95 0.797
}{\PM}

\pgfplotstableread{
Methods p/q Acc NMI 
PMM 3 0.985 0.931
PMM 2.8 0.985 0.931
PMM 2.5 0.956 0.825
PMM 2.3 0.934 0.776
PMM 2 0.856 0.592
}{\PMM}

\pgfplotstableread{
Methods p/q Acc NMI 
SCML 3 0.994 0.971
SCML 2.8 0.995 0.973
SCML 2.5 0.983 0.921
SCML 2.3 0.966 0.862
SCML 2 0.9 0.679
}{\SCML}

\begin{tikzpicture}
    \begin{groupplot}[
        group style={
        group name=art_info, 
        group size= 2 by 1,
        horizontal sep=50pt
        },
        width=6.3cm,
        height=6.3cm,
        enlarge y limits,
        axis y line*=left,
        axis x line*=bottom, 
        xtick=data,
        ymajorgrids,
		grid style={line width=.2pt,draw=gray!20},
        y axis line style={draw=none},
        xtick style={draw=none},
		ytick style={draw=none},
        yticklabel style={xshift=-0.5em},
    ]
     
     \nextgroupplot[title=(a),xmin=1.98, xmax=3.02, ymin=0.6, ymax = 1,extra y ticks={0.7,0.9}, xlabel={p/q}, ylabel={Acc}]
     
       \addplot [line width=0.05mm, 
     mark=o, color=red, mark size=1.5, mark options={fill=red!20!red,mark indices=5}, 
     ] table[x=p/q,y=Acc]{\EVM}; 
     
     \addplot [line width=0.05mm, 
     mark=square, color=blue, mark size=1.5, mark options={fill=blue!20!blue,mark indices=5}, 
     ] table[x=p/q,y=Acc]{\MA}; 
     
      \addplot [line width=0.05mm, 
     mark=otimes, color=brown, mark size=1.5, mark options={fill=brown!20!brown,mark indices=5}, 
     ] table[x=p/q,y=Acc]{\MAt}; 
     
       \addplot [line width=0.05mm, 
     mark=triangle, color=cyan, mark size=1.5, mark options={fill=cyan!20!cyan,mark indices=5}, 
     ] table[x=p/q,y=Acc]{\MVM}; 
     
       \addplot [line width=0.05mm, 
     mark=diamond, color=violet, mark size=1.5, mark options={fill=violet!20!violet,mark indices=5}, 
     ] table[x=p/q,y=Acc]{\MVMt}; 
     
       \addplot [line width=0.05mm, 
     mark=star, color=green, mark size=1.5, mark options={fill=green!20!green,mark indices=5}, 
     ] table[x=p/q,y=Acc]{\GL}; 
     
       \addplot [line width=0.05mm, 
     mark=x, color=orange, mark size=1.5, mark options={fill=orange!20!orange,mark indices=5}, 
     ] table[x=p/q,y=Acc]{\CoReg}; 
     
       \addplot [line width=0.05mm, 
     mark=10-pointed star, color=teal, mark size=1.5, mark options={fill=teal!20!teal,mark indices=5}, 
     ] table[x=p/q,y=Acc]{\AWP};  
     
       \addplot [line width=0.05mm, 
     mark=oplus, color=yellow, mark size=1.5, mark options={fill=yellow!20!yellow,mark indices=5}, 
     ] table[x=p/q,y=Acc]{\MCGC}; 
     
       \addplot [line width=0.05mm,
     mark=Mercedes star, color=magenta, mark size=1.5, mark options={fill=magenta!20!magenta,mark indices=5}, 
     ] table[x=p/q,y=Acc]{\PM}; 
     
       \addplot [line width=0.05mm,
     mark=pentagon, color=black, mark size=1.5, mark options={fill=black!20!black,mark indices=5}, 
     ] table[x=p/q,y=Acc]{\MT}; 
     
       \addplot [line width=0.05mm,
     mark=|, color=pink, mark size=1.5, mark options={fill=pink!20!pink,mark indices=5}, 
     ] table[x=p/q,y=Acc]{\SCML}; 
     
       \addplot [line width=0.05mm, 
     mark=asterisk, color=olive, mark size=1.5, mark options={fill=olive!20!olive,mark indices=5}, 
     ] table[x=p/q,y=Acc]{\PMM}; 
     
     
     \nextgroupplot[title=(b),xmin=1.98, xmax=3.02, ymin=0.3, ymax = 1,extra y ticks={0.3,0.5,0.7,0.9},xlabel={p/q}, ylabel={NMI}]
     
       \addplot [line width=0.05mm, 
     mark=o, color=red, mark size=1.5, mark options={fill=red!20!red,mark indices=5}, 
     ] table[x=p/q,y=NMI]{\EVM}; 
     
     \addplot [line width=0.05mm, 
     mark=square, color=blue, mark size=1.5, mark options={fill=blue!20!blue,mark indices=5}, 
     ] table[x=p/q,y=NMI]{\MA}; 
     
      \addplot [line width=0.05mm, 
     mark=otimes, color=brown, mark size=1.5, mark options={fill=brown!20!brown,mark indices=5}, 
     ] table[x=p/q,y=NMI]{\MAt}; 
     
       \addplot [line width=0.05mm, 
     mark=triangle, color=cyan, mark size=1.5, mark options={fill=cyan!20!cyan,mark indices=5}, 
     ] table[x=p/q,y=NMI]{\MVM}; 
     
       \addplot [line width=0.05mm, 
     mark=diamond, color=violet, mark size=1.5, mark options={fill=violet!20!violet,mark indices=5}, 
     ] table[x=p/q,y=NMI]{\MVMt}; 
     
       \addplot [line width=0.05mm, 
     mark=star, color=green, mark size=1.5, mark options={fill=green!20!green,mark indices=5}, 
     ] table[x=p/q,y=NMI]{\GL}; 
     
       \addplot [line width=0.05mm, 
     mark=x, color=orange, mark size=1.5, mark options={fill=orange!20!orange,mark indices=5}, 
     ] table[x=p/q,y=NMI]{\CoReg}; 
     
       \addplot [line width=0.05mm, 
     mark=10-pointed star, color=teal, mark size=1.5, mark options={fill=teal!20!teal,mark indices=5}, 
     ] table[x=p/q,y=NMI]{\AWP};  
     
       \addplot [line width=0.05mm, 
     mark=oplus, color=yellow, mark size=1.5, mark options={fill=yellow!20!yellow,mark indices=5}, 
     ] table[x=p/q,y=NMI]{\MCGC}; 
     
       \addplot [line width=0.05mm,
     mark=Mercedes star, color=magenta, mark size=1.5, mark options={fill=magenta!20!magenta,mark indices=5}, 
     ] table[x=p/q,y=NMI]{\PM}; 
     
       \addplot [line width=0.05mm,
     mark=pentagon, color=black, mark size=1.5, mark options={fill=black!20!black,mark indices=5}, 
     ] table[x=p/q,y=NMI]{\MT}; 
     
       \addplot [line width=0.05mm,
     mark=|, color=pink, mark size=1.5, mark options={fill=pink!20!pink,mark indices=5}, 
     ] table[x=p/q,y=NMI]{\SCML}; 
     
       \addplot [line width=0.05mm, 
     mark=asterisk, color=olive, mark size=1.5, mark options={fill=olive!20!olive,mark indices=5}, 
     ] table[x=p/q,y=NMI]{\PMM}; 

    \end{groupplot}
\end{tikzpicture}}
    \hfill 
    \hspace*{-1.5cm}
    \subfloat[]
    {

\pgfplotstableread{
Methods p/q Acc NMI 
AWP 3 0.999333333 0.996318296
AWP 2.8 0.976888889 0.972020095
AWP 2.5 0.971555556 0.946711823
AWP 2.3 0.902222222 0.84844588
AWP 2 0.818 0.665182296
}{\AWP}

\pgfplotstableread{
Methods p/q Acc NMI 
CoReg 3 0.999555556 0.997535253
CoReg 2.8 0.998666667 0.99258049
CoReg 2.5 0.993777778 0.968116063
CoReg 2.3 0.987111111 0.940505969
CoReg 2 0.936888889 0.771350841
}{\CoReg}

\pgfplotstableread{
Methods p/q Acc NMI 
EVM 3 1 1
EVM 2.8 0.999777778 0.99871109
EVM 2.5 0.998888889 0.993786176
EVM 2.3 0.999555556 0.997463702
EVM 2 0.993777778 0.968018851
}{\EVM}

\pgfplotstableread{
Methods p/q Acc NMI 
GL 3 1 1
GL 2.8 0.999777778 0.99871109
GL 2.5 0.998444444 0.992008419
GL 2.3 0.999777778 0.99871109
GL 2 0.993111111 0.965110932
}{\GL}

\pgfplotstableread{
Methods p/q Acc NMI 
MA2 3 1 1
MA2 2.8 0.999777778 0.99871109
MA2 2.5 0.998444444 0.991293385
MA2 2.3 0.999555556 0.997464988
MA2 2 0.992222222 0.962208607
}{\MA}

\pgfplotstableread{
Methods p/q Acc NMI 
MA3 3 1 1
MA3 2.8 0.999777778 0.99871109
MA3 2.5 0.998666667 0.992508957
MA3 2.3 0.999111111 0.994970211
MA3 2 0.992666667 0.963352174
}{\MAt}

\pgfplotstableread{
Methods p/q Acc NMI 
MCGC 3 0.997333333 0.985897705
MCGC 2.8 0.995777778 0.978477342
MCGC 2.5 0.976666667 0.90307113
MCGC 2.3 0.960666667 0.844706575
MCGC 2 0.888222222 0.672754983
}{\MCGC}

\pgfplotstableread{
Methods p/q Acc NMI 
MT 3 0.904333333 0.841301491
MT 2.8 0.928333333 0.838102103
MT 2.5 0.86 0.707567781
MT 2.3 0.870333333 0.708293595
MT 2 0.747 0.504350346
}{\MT}

\pgfplotstableread{
Methods p/q Acc NMI 
MVM2 3 1 1
MVM2 2.8 0.999777778 0.99871109
MVM2 2.5 0.999111111 0.994999357
MVM2 2.3 0.999555556 0.99750651
MVM2 2 0.993777778 0.967749088
}{\MVM}

\pgfplotstableread{
Methods p/q Acc NMI 
MVM3 3 1 1
MVM3 2.8 0.999777778 0.99871109
MVM3 2.5 0.998444444 0.991580631
MVM3 2.3 0.999777778 0.998753898
MVM3 2 0.994222222 0.970705045
}{\MVMt}

\pgfplotstableread{
Methods p/q Acc NMI 
PM 3 0.999111111 0.995360799
PM 2.8 0.999333333 0.996288267
PM 2.5 0.997555556 0.986981305
PM 2.3 0.997111111 0.98490156
PM 2 0.987111111 0.93681657
}{\PM}

\pgfplotstableread{
Methods p/q Acc NMI 
PMM 3 0.996666667 0.982070922
PMM 2.8 0.996222222 0.980184925
PMM 2.5 0.989333333 0.949381936
PMM 2.3 0.967111111 0.869259451
PMM 2 0.896888889 0.693103459
}{\PMM}

\pgfplotstableread{
Methods p/q Acc NMI 
SCML 3 0.999777778 0.998783044
SCML 2.8 0.999111111 0.995395873
SCML 2.5 0.998 0.989284796
SCML 2.3 0.991111111 0.95790278
SCML 2 0.960222222 0.843622483
}{\SCML}

\begin{tikzpicture}
    \begin{groupplot}[
        group style={
        group name=art_info, 
        group size= 2 by 1,
        horizontal sep=50pt
        },
        width=6.3cm,
        height=6.3cm,
        enlarge y limits,
        axis y line*=left,
        axis x line*=bottom, 
        xtick=data,
        ymajorgrids,
		grid style={draw=gray!20},
        y axis line style={draw=none},
        xtick style={draw=none},
		ytick style={draw=none},
        yticklabel style={xshift=-0.5em},
        legend style={
        at={(1.1,-0.30)},
        legend columns=8,
        legend cell align=left,
        anchor=north},
        legend image post style={mark options={scale=1}}
    ]
     
     \nextgroupplot[title=(c),xmin=1.98, xmax=3.02, ymin=0.6, ymax = 1,extra y ticks={0.7,0.9},xlabel={p/q}, ylabel={Acc}]
     
       \addplot [line width=0.05mm, 
     mark=o, color=red, mark size=1.5, mark options={fill=red!20!red,mark indices=5}, 
     ] table[x=p/q,y=Acc]{\EVM}; \addlegendentry{EVM};
     
     \addplot [line width=0.05mm, 
     mark=square, color=blue, mark size=1.5, mark options={fill=blue!20!blue,mark indices=5}, 
     ] table[x=p/q,y=Acc]{\MA}; \addlegendentry{MA2};
     
      \addplot [line width=0.05mm, 
     mark=otimes, color=brown, mark size=1.5, mark options={fill=brown!20!brown,mark indices=5}, 
     ] table[x=p/q,y=Acc]{\MAt}; \addlegendentry{MA3};
     
       \addplot [line width=0.05mm, 
     mark=triangle, color=cyan, mark size=1.5, mark options={fill=cyan!20!cyan,mark indices=5}, 
     ] table[x=p/q,y=Acc]{\MVM}; \addlegendentry{MVM2};
     
       \addplot [line width=0.05mm, 
     mark=diamond, color=violet, mark size=1.5, mark options={fill=violet!20!violet,mark indices=5}, 
     ] table[x=p/q,y=Acc]{\MVMt}; \addlegendentry{MVM3};
     
      \addlegendimage{empty legend}
            \addlegendentry{}
            
     \addlegendimage{empty legend}
            \addlegendentry{}
            
     \addlegendimage{empty legend}
            \addlegendentry{}
            
       \addplot [line width=0.05mm, 
     mark=star, color=green, mark size=1.5, mark options={fill=green!20!green,mark indices=5}, 
     ] table[x=p/q,y=Acc]{\GL}; \addlegendentry{GL};
     
       \addplot [line width=0.05mm, 
     mark=x, color=orange, mark size=1.5, mark options={fill=orange!20!orange,mark indices=5}, 
     ] table[x=p/q,y=Acc]{\CoReg}; \addlegendentry{CoReg};
     
       \addplot [line width=0.05mm, 
     mark=10-pointed star, color=teal, mark size=1.5, mark options={fill=teal!20!teal,mark indices=5}, 
     ] table[x=p/q,y=Acc]{\AWP};  \addlegendentry{AWP};
     
       \addplot [line width=0.05mm, 
     mark=oplus, color=yellow, mark size=1.5, mark options={fill=yellow!20!yellow,mark indices=5}, 
     ] table[x=p/q,y=Acc]{\MCGC}; \addlegendentry{MCGC};
     
       \addplot [line width=0.05mm,
     mark=Mercedes star, color=magenta, mark size=1.5, mark options={fill=magenta!20!magenta,mark indices=5}, 
     ] table[x=p/q,y=Acc]{\PM}; \addlegendentry{PM};
     
       \addplot [line width=0.05mm,
     mark=pentagon, color=black, mark size=1.5, mark options={fill=black!20!black,mark indices=5}, 
     ] table[x=p/q,y=Acc]{\MT}; \addlegendentry{MT};
     
       \addplot [line width=0.05mm,
     mark=|, color=pink, mark size=1.5, mark options={fill=pink!20!pink,mark indices=5}, 
     ] table[x=p/q,y=Acc]{\SCML}; \addlegendentry{SCML};
     
       \addplot [line width=0.05mm, 
     mark=asterisk, color=olive, mark size=1.5, mark options={fill=olive!20!olive,mark indices=5}, 
     ] table[x=p/q,y=Acc]{\PMM}; \addlegendentry{PMM};
     
     
     \nextgroupplot[title=(d),xmin=1.98, xmax=3.02, ymin=0.3, ymax = 1,extra y ticks={0.3,0.5,0.7,0.9},xlabel={p/q}, ylabel={NMI}]
     
      \addplot [line width=0.05mm, 
     mark=o, color=red, mark size=1.5, mark options={fill=red!20!red,mark indices=5}, 
     ] table[x=p/q,y=NMI]{\EVM}; 
     
     \addplot [line width=0.05mm, 
     mark=square, color=blue, mark size=1.5, mark options={fill=blue!20!blue,mark indices=5}, 
     ] table[x=p/q,y=NMI]{\MA}; 
     
      \addplot [line width=0.05mm, 
     mark=otimes, color=brown, mark size=1.5, mark options={fill=brown!20!brown,mark indices=5}, 
     ] table[x=p/q,y=NMI]{\MAt}; 
     
       \addplot [line width=0.05mm, 
     mark=triangle, color=cyan, mark size=1.5, mark options={fill=cyan!20!cyan,mark indices=5}, 
     ] table[x=p/q,y=NMI]{\MVM}; 
     
       \addplot [line width=0.05mm, 
     mark=diamond, color=violet, mark size=1.5, mark options={fill=violet!20!violet,mark indices=5}, 
     ] table[x=p/q,y=NMI]{\MVMt}; 
     
       \addplot [line width=0.05mm, 
     mark=star, color=green, mark size=1.5, mark options={fill=green!20!green,mark indices=5}, 
     ] table[x=p/q,y=NMI]{\GL}; 
     
       \addplot [line width=0.05mm, 
     mark=x, color=orange, mark size=1.5, mark options={fill=orange!20!orange,mark indices=5}, 
     ] table[x=p/q,y=NMI]{\CoReg}; 
     
       \addplot [line width=0.05mm, 
     mark=10-pointed star, color=teal, mark size=1.5, mark options={fill=teal!20!teal,mark indices=5}, 
     ] table[x=p/q,y=NMI]{\AWP};  
     
       \addplot [line width=0.05mm, 
     mark=oplus, color=yellow, mark size=1.5, mark options={fill=yellow!20!yellow,mark indices=5}, 
     ] table[x=p/q,y=NMI]{\MCGC}; 
     
       \addplot [line width=0.05mm,
     mark=Mercedes star, color=magenta, mark size=1.5, mark options={fill=magenta!20!magenta,mark indices=5}, 
     ] table[x=p/q,y=NMI]{\PM}; 
     
       \addplot [line width=0.05mm,
     mark=pentagon, color=black, mark size=1.5, mark options={fill=black!20!black,mark indices=5}, 
     ] table[x=p/q,y=NMI]{\MT}; 
     
       \addplot [line width=0.05mm,
     mark=|, color=pink, mark size=1.5, mark options={fill=pink!20!pink,mark indices=5}, 
     ] table[x=p/q,y=NMI]{\SCML}; 
     
       \addplot [line width=0.05mm, 
     mark=asterisk, color=olive, mark size=1.5, mark options={fill=olive!20!olive,mark indices=5}, 
     ] table[x=p/q,y=NMI]{\PMM};  

    \end{groupplot}
\end{tikzpicture}}
\caption{Average values of accuracy and NMI over 10 random networks sampled from SBM with equally distributed informative layers (2 layers (a)(b) and 3 layers (c)(d)) with three clusters of 100, 150 and 200 nodes, respectively, for $p=0.1$ and  $p/q \in \{2, 2.3, 2.5, 2.8, 3\}$.   }
\label{fig:plot5}
\end{figure*}

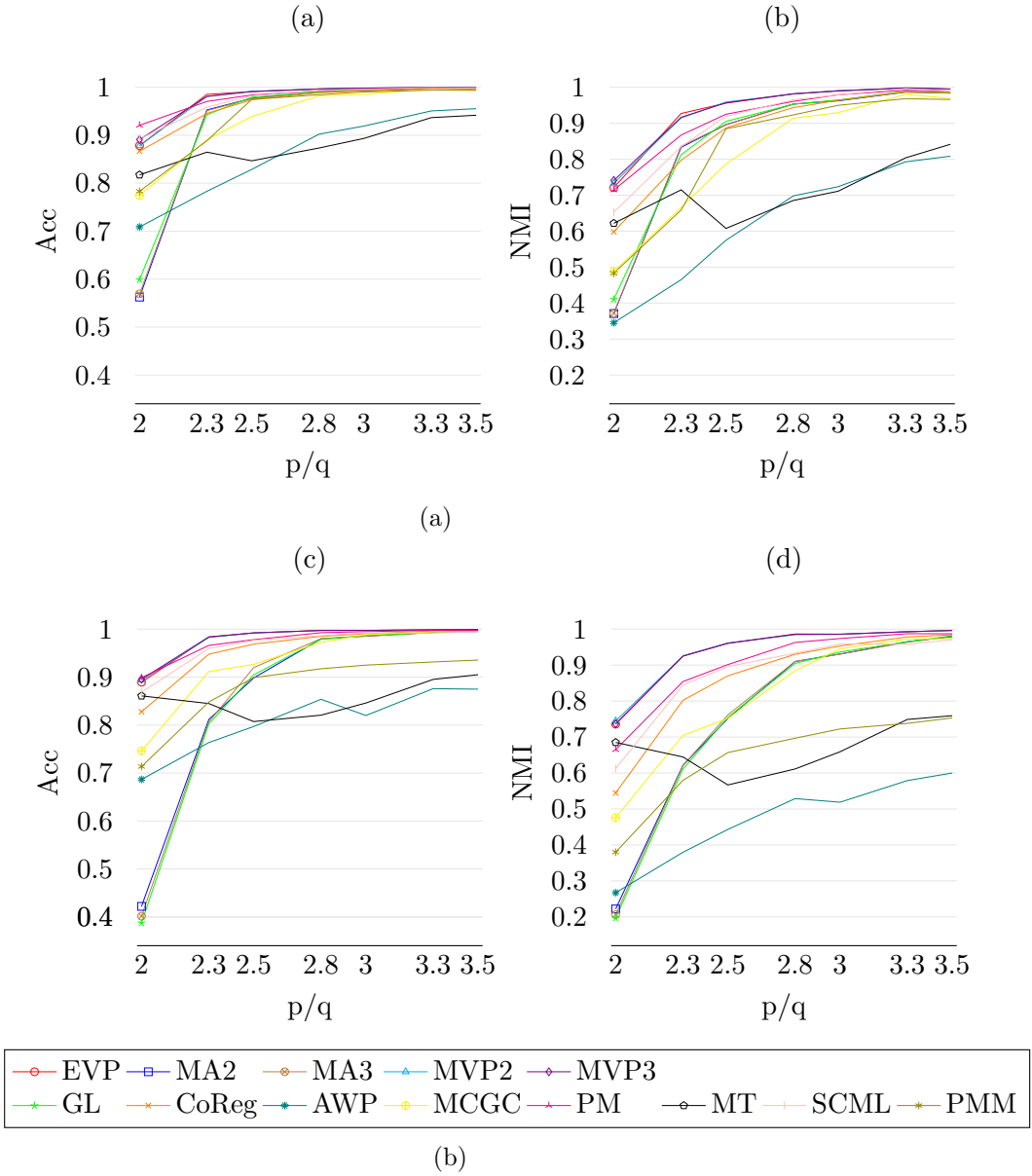
\begin{figure*}[p]
\centering
\captionsetup{oneside,margin={-0.53cm,0.6cm}}
    \hspace*{-1.9cm}
    \subfloat[]
    {
\pgfplotstableread{
Methods p/q Acc NMI 
AWP 3.5 0.955333333 0.808342424
AWP 3.3 0.950666667 0.792582379
AWP 3 0.919111111 0.723535677
AWP 2.8 0.902222222 0.697924792
AWP 2.5 0.828666667 0.574572336
AWP 2.3 0.783111111 0.464906355
AWP 2 0.708666667 0.345235725
}{\AWP}

\pgfplotstableread{
Methods p/q Acc NMI 
CoReg 3.5 0.996888889 0.984313396
CoReg 3.3 0.997555556 0.986690512
CoReg 3 0.992888889 0.963663255
CoReg 2.8 0.988444444 0.943385146
CoReg 2.5 0.973911111 0.88704811
CoReg 2.3 0.945422222 0.796825602
CoReg 2 0.866577778 0.598097323
}{\CoReg}

\pgfplotstableread{
Methods p/q Acc NMI 
EVP 3.5 0.999 0.9950632
EVP 3.3 0.999555556 0.997464988
EVP 3 0.9982 0.99064645
EVP 2.8 0.9964 0.98147404
EVP 2.5 0.991022222 0.955982532
EVP 2.3 0.985111111 0.927069313
EVP 2 0.878444444 0.720961584
}{\EVP}

\pgfplotstableread{
Methods p/q Acc NMI 
GL 3.5 0.997 0.98473596
GL 3.3 0.997911111 0.988481774
GL 3 0.9932 0.96409741
GL 2.8 0.9906 0.95290047
GL 2.5 0.979466667 0.905051824
GL 2.3 0.9428 0.812285762
GL 2 0.598844444 0.410364416
    }{\GL}

\pgfplotstableread{
Methods p/q Acc NMI 
MA2 3.5 0.997 0.98355102
MA2 3.3 0.997511111 0.98655636
MA2 3 0.9927 0.96221141
MA2 2.8 0.9908 0.95349713
MA2 2.5 0.976933333 0.895979842
MA2 2.3 0.952755556 0.833029862
MA2 2 0.5624 0.370845409
}{\MA}

\pgfplotstableread{
Methods p/q Acc NMI 
MA3 3.5 0.997 0.98574448
MA3 3.3 0.997822222 0.988171647
MA3 3 0.9932 0.96447131
MA3 2.8 0.9906 0.95269925
MA3 2.5 0.976533333 0.896162686
MA3 2.3 0.951111111 0.8361741
MA3 2 0.569155556 0.371599333
}{\MAt}

\pgfplotstableread{
Methods p/q Acc NMI 
MCGC 3.5 0.993777778 0.96929481
MCGC 3.3 0.996222222 0.980116072
MCGC 3 0.985111111 0.929317089
MCGC 2.8 0.981777778 0.913935143
MCGC 2.5 0.939111111 0.786730019
MCGC 2.3 0.890222222 0.66538732
MCGC 2 0.774444444 0.487393577
}{\MCGC}

\pgfplotstableread{
Methods p/q Acc NMI 
MT 3.5 0.941666667 0.841473925
MT 3.3 0.936666667 0.803660211
MT 3 0.893333333 0.711034632
MT 2.8 0.873333333 0.685221767
MT 2.5 0.846666667 0.607570822
MT 2.3 0.864666667 0.714771158
MT 2 0.817333333 0.621756992
}{\MT}

\pgfplotstableread{
Methods p/q Acc NMI 
MVP2 3.5 0.999 0.99481372
MVP2 3.3 0.999466667 0.996972658
MVP2 3 0.9982 0.99053965
MVP2 2.8 0.9964 0.9810895
MVP2 2.5 0.991688889 0.958980581
MVP2 2.3 0.980666667 0.914498908
MVP2 2 0.878266667 0.733230691
}{\MVP}

\pgfplotstableread{
Methods p/q Acc NMI 
MVP3 3.5 0.999 0.9950632
MVP3 3.3 0.999733333 0.998478993
MVP3 3 0.998 0.98936198
MVP3 2.8 0.9966 0.98203708
MVP3 2.5 0.991022222 0.956707804
MVP3 2.3 0.980977778 0.91654701
MVP3 2 0.890844444 0.740964672
}{\MVPt}

\pgfplotstableread{
Methods p/q Acc NMI 
PM 3.5 0.998 0.989187071
PM 3.3 0.998444444 0.991609515
PM 3 0.996 0.97919508
PM 2.8 0.992444444 0.960677354
PM 2.5 0.983866667 0.925039569
PM 2.3 0.970444444 0.867045565
PM 2 0.920666667 0.714200194
}{\PM}

\pgfplotstableread{
Methods p/q Acc NMI 
PMM 3.5 0.993555556 0.965920201
PMM 3.3 0.994 0.968257121
PMM 3 0.990222222 0.949839554
PMM 2.8 0.9844 0.923197321
PMM 2.5 0.974666667 0.884336592
PMM 2.3 0.888444444 0.658584749
PMM 2 0.783111111 0.483057208
}{\PMM}

\pgfplotstableread{
Methods p/q Acc NMI 
SCML 3.5 0.998 0.98931484
SCML 3.3 0.999333333 0.996321354
SCML 3 0.996 0.978712785
SCML 2.8 0.993111111 0.96510689
SCML 2.5 0.982222222 0.919470509
SCML 2.3 0.958444444 0.834987957
SCML 2 0.892444444 0.652425358
}{\SCML}

\begin{tikzpicture}
    \begin{groupplot}[
        group style={
        group name=art_info, 
        group size= 2 by 1,
        horizontal sep=50pt
        },
        width=6.3cm,
        height=6.3cm,
        enlarge y limits,
        axis y line*=left,
        axis x line*=bottom, 
        xtick=data,
        ymajorgrids,
		grid style={line width=.2pt,draw=gray!20},
        y axis line style={draw=none},
        xtick style={draw=none},
		ytick style={draw=none},
        yticklabel style={xshift=-0.5em},
    ]
     
     \nextgroupplot[title=(a),xmin=1.98, xmax=3.52, ymin=0.4, ymax = 1,extra y ticks={0.5,0.7,0.9}, xlabel={p/q}, ylabel={Acc}]
     
     \addplot [line width=0.05mm, 
     mark=o, color=red, mark size=1.5, mark options={fill=red!20!red,mark indices=7}, 
     ] table[x=p/q,y=Acc]{\EVP}; 
     
     \addplot [line width=0.05mm, 
     mark=square, color=blue, mark size=1.5, mark options={fill=blue!20!blue,mark indices=7}, 
     ] table[x=p/q,y=Acc]{\MA}; 
     
      \addplot [line width=0.05mm, 
     mark=otimes, color=brown, mark size=1.5, mark options={fill=brown!20!brown,mark indices=7}, 
     ] table[x=p/q,y=Acc]{\MAt}; 
     
       \addplot [line width=0.05mm, 
     mark=triangle, color=cyan, mark size=1.5, mark options={fill=cyan!20!cyan,mark indices=7}, 
     ] table[x=p/q,y=Acc]{\MVP}; 
     
       \addplot [line width=0.05mm, 
     mark=diamond, color=violet, mark size=1.5, mark options={fill=violet!20!violet,mark indices=7}, 
     ] table[x=p/q,y=Acc]{\MVPt}; 
     
       \addplot [line width=0.05mm, 
     mark=star, color=green, mark size=1.5, mark options={fill=green!20!green,mark indices=7}, 
     ] table[x=p/q,y=Acc]{\GL}; 
     
       \addplot [line width=0.05mm, 
     mark=x, color=orange, mark size=1.5, mark options={fill=orange!20!orange,mark indices=7}, 
     ] table[x=p/q,y=Acc]{\CoReg}; 
     
       \addplot [line width=0.05mm, 
     mark=10-pointed star, color=teal, mark size=1.5, mark options={fill=teal!20!teal,mark indices=7}, 
     ] table[x=p/q,y=Acc]{\AWP};  
     
       \addplot [line width=0.05mm, 
     mark=oplus, color=yellow, mark size=1.5, mark options={fill=yellow!20!yellow,mark indices=7}, 
     ] table[x=p/q,y=Acc]{\MCGC}; 
     
       \addplot [line width=0.05mm,
     mark=Mercedes star, color=magenta, mark size=1.5, mark options={fill=magenta!20!magenta,mark indices=7}, 
     ] table[x=p/q,y=Acc]{\PM}; 
     
       \addplot [line width=0.05mm,
     mark=pentagon, color=black, mark size=1.5, mark options={fill=black!20!black,mark indices=7}, 
     ] table[x=p/q,y=Acc]{\MT}; 
     
       \addplot [line width=0.05mm,
     mark=|, color=pink, mark size=1.5, mark options={fill=pink!20!pink,mark indices=7}, 
     ] table[x=p/q,y=Acc]{\SCML}; 
     
       \addplot [line width=0.05mm, 
     mark=asterisk, color=olive, mark size=1.5, mark options={fill=olive!20!olive,mark indices=7}, 
     ] table[x=p/q,y=Acc]{\PMM}; 
     
     
    \nextgroupplot[title=(b),xmin=1.98, xmax=3.52, ymin=0.2, ymax = 1,extra y ticks={0.3,0.5,0.7,0.9},xlabel={p/q}, ylabel={NMI}]
    
        \addplot [line width=0.05mm, 
     mark=o, color=red, mark size=1.5, mark options={fill=red!20!red,mark indices=7}, 
     ] table[x=p/q,y=NMI]{\EVP}; 
     
     \addplot [line width=0.05mm, 
     mark=square, color=blue, mark size=1.5, mark options={fill=blue!20!blue,mark indices=7}, 
     ] table[x=p/q,y=NMI]{\MA}; 
     
      \addplot [line width=0.05mm, 
     mark=otimes, color=brown, mark size=1.5, mark options={fill=brown!20!brown,mark indices=7}, 
     ] table[x=p/q,y=NMI]{\MAt}; 
     
       \addplot [line width=0.05mm, 
     mark=triangle, color=cyan, mark size=1.5, mark options={fill=cyan!20!cyan,mark indices=7}, 
     ] table[x=p/q,y=NMI]{\MVP}; 
     
       \addplot [line width=0.05mm, 
     mark=diamond, color=violet, mark size=1.5, mark options={fill=violet!20!violet,mark indices=7}, 
     ] table[x=p/q,y=NMI]{\MVPt}; 
     
       \addplot [line width=0.05mm, 
     mark=star, color=green, mark size=1.5, mark options={fill=green!20!green,mark indices=7}, 
     ] table[x=p/q,y=NMI]{\GL}; 
     
       \addplot [line width=0.05mm, 
     mark=x, color=orange, mark size=1.5, mark options={fill=orange!20!orange,mark indices=7}, 
     ] table[x=p/q,y=NMI]{\CoReg}; 
     
       \addplot [line width=0.05mm, 
     mark=10-pointed star, color=teal, mark size=1.5, mark options={fill=teal!20!teal,mark indices=7}, 
     ] table[x=p/q,y=NMI]{\AWP};  
     
       \addplot [line width=0.05mm, 
     mark=oplus, color=yellow, mark size=1.5, mark options={fill=yellow!20!yellow,mark indices=7}, 
     ] table[x=p/q,y=NMI]{\MCGC}; 
     
       \addplot [line width=0.05mm,
     mark=Mercedes star, color=magenta, mark size=1.5, mark options={fill=magenta!20!magenta,mark indices=7}, 
     ] table[x=p/q,y=NMI]{\PM}; 
     
       \addplot [line width=0.05mm,
     mark=pentagon, color=black, mark size=1.5, mark options={fill=black!20!black,mark indices=7}, 
     ] table[x=p/q,y=NMI]{\MT}; 
     
       \addplot [line width=0.05mm,
     mark=|, color=pink, mark size=1.5, mark options={fill=pink!20!pink,mark indices=7}, 
     ] table[x=p/q,y=NMI]{\SCML}; 
     
       \addplot [line width=0.05mm, 
     mark=asterisk, color=olive, mark size=1.5, mark options={fill=olive!20!olive,mark indices=7}, 
     ] table[x=p/q,y=NMI]{\PMM}; 

    \end{groupplot}
\end{tikzpicture}}
    \hfill
    \hspace*{-1.5cm}
    \subfloat[]
    {
\pgfplotstableread{
Methods p/q Acc NMI 
AWP 3.5 0.874888889 0.599988379
AWP 3.3 0.876 0.578949942
AWP 3 0.82 0.518791654
AWP 2.8 0.853777778 0.528901313
AWP 2.5 0.797111111 0.443131454
AWP 2.3 0.763555556 0.379635586
AWP 2 0.686666667 0.267093457
}{\AWP}

\pgfplotstableread{
Methods p/q Acc NMI 
CoReg 3.5 0.997111111 0.984198174
CoReg 3.3 0.996 0.978276514
CoReg 3 0.990666667 0.95314849
CoReg 2.8 0.985066667 0.930087658
CoReg 2.5 0.968888889 0.870098335
CoReg 2.3 0.947866667 0.802161201
CoReg 2 0.827733333 0.54459534
}{\CoReg}

\pgfplotstableread{
Methods p/q Acc NMI 
EVP 3.5 0.999333333 0.996320552
EVP 3.3 0.998711111 0.992779891
EVP 3 0.997422222 0.985999098
EVP 2.8 0.997466667 0.986368459
EVP 2.5 0.992222222 0.960412804
EVP 2.3 0.983822222 0.925021764
EVP 2 0.889422222 0.735779013
}{\EVP}

\pgfplotstableread{
Methods p/q Acc NMI 
GL 3.5 0.996622222 0.981496298
GL 3.3 0.993155556 0.964913899
GL 3 0.9868 0.936462105
GL 2.8 0.9788 0.904798305
GL 2.5 0.904844444 0.754780708
GL 2.3 0.802444444 0.611205494
GL 2 0.387511111 0.196882745
}{\GL}

\pgfplotstableread{
Methods p/q Acc NMI 
MA2 3.5 0.996133333 0.978979956
MA2 3.3 0.993244444 0.965146434
MA2 3 0.985511111 0.93085434
MA2 2.8 0.979777778 0.910069516
MA2 2.5 0.8984 0.751664373
MA2 2.3 0.811733333 0.621507433
MA2 2 0.421955556 0.222031997
}{\MA}

\pgfplotstableread{
Methods p/q Acc NMI 
MA3 3.5 0.996355556 0.980470233
MA3 3.3 0.992888889 0.963519042
MA3 3 0.9856 0.930503341
MA3 2.8 0.980311111 0.909990819
MA3 2.5 0.920044444 0.761688546
MA3 2.3 0.806666667 0.620884154
MA3 2 0.401511111 0.206273964
}{\MAt}

\pgfplotstableread{
Methods p/q Acc NMI 
MCGC 3.5 0.994888889 0.973243537
MCGC 3.3 0.994666667 0.972646018
MCGC 3 0.988666667 0.945394818
MCGC 2.8 0.972444444 0.884304798
MCGC 2.5 0.926666667 0.751917323
MCGC 2.3 0.911555556 0.704446888
MCGC 2 0.746 0.475691811
}{\MCGC}

\pgfplotstableread{
Methods p/q Acc NMI 
MT 3.5 0.905 0.759283197
MT 3.3 0.895333333 0.748923192
MT 3 0.846 0.659083644
MT 2.8 0.820666667 0.611283089
MT 2.5 0.807333333 0.566762059
MT 2.3 0.844888889 0.644637652
MT 2 0.860888889 0.684967057
}{\MT}

\pgfplotstableread{
Methods p/q Acc NMI 
MVP2 3.5 0.999288889 0.996071253
MVP2 3.3 0.998711111 0.99277953
MVP2 3 0.997288889 0.985373084
MVP2 2.8 0.997066667 0.984383225
MVP2 2.5 0.992488889 0.961399517
MVP2 2.3 0.982666667 0.925787625
MVP2 2 0.893955556 0.744603764
}{\MVP}

\pgfplotstableread{
Methods p/q Acc NMI 
MVP3 3.5 0.999333333 0.996326635
MVP3 3.3 0.998711111 0.992782263
MVP3 3 0.997422222 0.98587178
MVP3 2.8 0.997377778 0.985861462
MVP3 2.5 0.992177778 0.960880891
MVP3 2.3 0.983777778 0.925022597
MVP3 2 0.895822222 0.736635259
}{\MVPt}

\pgfplotstableread{
Methods p/q Acc NMI 
PM 3.5 0.997555556 0.987049924
PM 3.3 0.997555556 0.987244988
PM 3 0.994888889 0.973903917
PM 2.8 0.992666667 0.962920123
PM 2.5 0.978222222 0.901413541
PM 2.3 0.966222222 0.854338103
PM 2 0.898844444 0.664696135
}{\PM}

\pgfplotstableread{
Methods p/q Acc NMI 
PMM 3.5 0.935644444 0.753133682
PMM 3.3 0.931644444 0.738022601
PMM 3 0.925066667 0.722484784
PMM 2.8 0.917466667 0.696740985
PMM 2.5 0.899022222 0.656833808
PMM 2.3 0.848222222 0.57960935
PMM 2 0.713555556 0.379935997
}{\PMM}

\pgfplotstableread{
Methods p/q Acc NMI 
SCML 3.5 0.994444444 0.9706812
SCML 3.3 0.992 0.957899605
SCML 3 0.992 0.95860422
SCML 2.8 0.986444444 0.933093144
SCML 2.5 0.976888889 0.895838096
SCML 2.3 0.959777778 0.844589125
SCML 2 0.868444444 0.612336174
}{\SCML}

\begin{tikzpicture}
    \begin{groupplot}[
        group style={
        group name=art_info, 
        group size= 2 by 1,
        horizontal sep=50pt
        },
        width=6.3cm,
        height=6.3cm,
        enlarge y limits,
        axis y line*=left,
        axis x line*=bottom, 
        xtick=data,
        ymajorgrids,
		grid style={draw=gray!20},
        y axis line style={draw=none},
        xtick style={draw=none},
		ytick style={draw=none},
        yticklabel style={xshift=-0.5em},
        legend style={
        at={(1.1,-0.30)},
        legend columns=8,
        legend cell align=left,
        anchor=north},
        legend image post style={mark options={scale=1}}
    ]
     
     \nextgroupplot[title=(c),xmin=1.98, xmax=3.52, ymin=0.4, ymax = 1,extra y ticks={0.3,0.4,0.5,0.7,0.9},xlabel={p/q}, ylabel={Acc}]
     
       \addplot [line width=0.05mm, 
     mark=o, color=red, mark size=1.5, mark options={fill=red!20!red,mark indices=7}, 
     ] table[x=p/q,y=Acc]{\EVP}; \addlegendentry{EVP};
     
     \addplot [line width=0.05mm, 
     mark=square, color=blue, mark size=1.5, mark options={fill=blue!20!blue,mark indices=7}, 
     ] table[x=p/q,y=Acc]{\MA}; \addlegendentry{MA2};
     
      \addplot [line width=0.05mm, 
     mark=otimes, color=brown, mark size=1.5, mark options={fill=brown!20!brown,mark indices=7}, 
     ] table[x=p/q,y=Acc]{\MAt}; \addlegendentry{MA3};
     
       \addplot [line width=0.05mm, 
     mark=triangle, color=cyan, mark size=1.5, mark options={fill=cyan!20!cyan,mark indices=7}, 
     ] table[x=p/q,y=Acc]{\MVP}; \addlegendentry{MVP2};
     
       \addplot [line width=0.05mm, 
     mark=diamond, color=violet, mark size=1.5, mark options={fill=violet!20!violet,mark indices=7}, 
     ] table[x=p/q,y=Acc]{\MVPt}; \addlegendentry{MVP3};
     
      \addlegendimage{empty legend}
            \addlegendentry{}
            
     \addlegendimage{empty legend}
            \addlegendentry{}
            
     \addlegendimage{empty legend}
            \addlegendentry{}
            
       \addplot [line width=0.05mm, 
     mark=star, color=green, mark size=1.5, mark options={fill=green!20!green,mark indices=7}, 
     ] table[x=p/q,y=Acc]{\GL}; \addlegendentry{GL};
     
       \addplot [line width=0.05mm, 
     mark=x, color=orange, mark size=1.5, mark options={fill=orange!20!orange,mark indices=7}, 
     ] table[x=p/q,y=Acc]{\CoReg}; \addlegendentry{CoReg};
     
       \addplot [line width=0.05mm, 
     mark=10-pointed star, color=teal, mark size=1.5, mark options={fill=teal!20!teal,mark indices=7}, 
     ] table[x=p/q,y=Acc]{\AWP};  \addlegendentry{AWP};
     
       \addplot [line width=0.05mm, 
     mark=oplus, color=yellow, mark size=1.5, mark options={fill=yellow!20!yellow,mark indices=7}, 
     ] table[x=p/q,y=Acc]{\MCGC}; \addlegendentry{MCGC};
     
       \addplot [line width=0.05mm,
     mark=Mercedes star, color=magenta, mark size=1.5, mark options={fill=magenta!20!magenta,mark indices=7}, 
     ] table[x=p/q,y=Acc]{\PM}; \addlegendentry{PM};
     
       \addplot [line width=0.05mm,
     mark=pentagon, color=black, mark size=1.5, mark options={fill=black!20!black,mark indices=7}, 
     ] table[x=p/q,y=Acc]{\MT}; \addlegendentry{MT};
     
       \addplot [line width=0.05mm,
     mark=|, color=pink, mark size=1.5, mark options={fill=pink!20!pink,mark indices=7}, 
     ] table[x=p/q,y=Acc]{\SCML}; \addlegendentry{SCML};
     
       \addplot [line width=0.05mm, 
     mark=asterisk, color=olive, mark size=1.5, mark options={fill=olive!20!olive,mark indices=7}, 
     ] table[x=p/q,y=Acc]{\PMM}; \addlegendentry{PMM};
     
     
     \nextgroupplot[title=(d),xmin=1.98, xmax=3.52, ymin=0.2, ymax = 1,extra y ticks={0.3,0.5,0.7,0.9},xlabel={p/q}, ylabel={NMI}]
     
      \addplot [line width=0.05mm, 
     mark=o, color=red, mark size=1.5, mark options={fill=red!20!red,mark indices=7}, 
     ] table[x=p/q,y=NMI]{\EVP}; 
     
     \addplot [line width=0.05mm, 
     mark=square, color=blue, mark size=1.5, mark options={fill=blue!20!blue,mark indices=7}, 
     ] table[x=p/q,y=NMI]{\MA}; 
     
      \addplot [line width=0.05mm, 
     mark=otimes, color=brown, mark size=1.5, mark options={fill=brown!20!brown,mark indices=7}, 
     ] table[x=p/q,y=NMI]{\MAt}; 
     
       \addplot [line width=0.05mm, 
     mark=triangle, color=cyan, mark size=1.5, mark options={fill=cyan!20!cyan,mark indices=7}, 
     ] table[x=p/q,y=NMI]{\MVP}; 
     
       \addplot [line width=0.05mm, 
     mark=diamond, color=violet, mark size=1.5, mark options={fill=violet!20!violet,mark indices=7}, 
     ] table[x=p/q,y=NMI]{\MVPt}; 
     
       \addplot [line width=0.05mm, 
     mark=star, color=green, mark size=1.5, mark options={fill=green!20!green,mark indices=7}, 
     ] table[x=p/q,y=NMI]{\GL}; 
     
       \addplot [line width=0.05mm, 
     mark=x, color=orange, mark size=1.5, mark options={fill=orange!20!orange,mark indices=7}, 
     ] table[x=p/q,y=NMI]{\CoReg}; 
     
       \addplot [line width=0.05mm, 
     mark=10-pointed star, color=teal, mark size=1.5, mark options={fill=teal!20!teal,mark indices=7}, 
     ] table[x=p/q,y=NMI]{\AWP};  
     
       \addplot [line width=0.05mm, 
     mark=oplus, color=yellow, mark size=1.5, mark options={fill=yellow!20!yellow,mark indices=7}, 
     ] table[x=p/q,y=NMI]{\MCGC}; 
     
       \addplot [line width=0.05mm,
     mark=Mercedes star, color=magenta, mark size=1.5, mark options={fill=magenta!20!magenta,mark indices=7}, 
     ] table[x=p/q,y=NMI]{\PM}; 
     
       \addplot [line width=0.05mm,
     mark=pentagon, color=black, mark size=1.5, mark options={fill=black!20!black,mark indices=7}, 
     ] table[x=p/q,y=NMI]{\MT}; 
     
       \addplot [line width=0.05mm,
     mark=|, color=pink, mark size=1.5, mark options={fill=pink!20!pink,mark indices=7}, 
     ] table[x=p/q,y=NMI]{\SCML}; 
     
       \addplot [line width=0.05mm, 
     mark=asterisk, color=olive, mark size=1.5, mark options={fill=olive!20!olive,mark indices=7}, 
     ] table[x=p/q,y=NMI]{\PMM}; 
     
    \end{groupplot}
\end{tikzpicture}}
\caption{Average values of accuracy and NMI over 10 random networks sampled from SBM with both informative and noisy layers (two informative and one noisy in (a)(b);  two informative and two noisy in (c)(d)). The informative layers are equally distributed SBM graphs with three clusters of 100, 150 and 200 nodes, respectively, for $p=0.1$ and $p/q \in \{2,2.3,2.5, 2.8, 3, 3.3, 3.5\}$. The noisy layers are SBM graphs with $p=q=0.1$. }
\label{fig:plot6}
\end{figure*}

We set $p>q$ in order to generate synthetic informative layers, while we simply let $p=q$ for the noisy ones. 
More precisely, we created networks with 4 communities of 125 nodes each and with $k=2,3$ layers, by fixing $p=0.1$ and varying the ratio $p/q$ in the generation of the informative layers. In the noisy layers, we fixed $p=q=0.1$. 
For each value of the pair $(p,q)$ we sample 10 random instances and we report average scores. Results are shown in Figures \ref{fig:plot1} and \ref{fig:plot2},  where we consider the following four settings: \ref{fig:plot1}(a)(b) two informative layers; \ref{fig:plot1}(c)(d) three informative layers; \ref{fig:plot2}(a)(b) two informative layers and one noisy layer; \ref{fig:plot2}(c)(d) two informative layers and two noise ones. 
In general, our proposed approaches  show good performance in almost all parameter settings, as  compared to the baselines, and overall  the variance-based multiobjective approaches (\MVM{} and \MVP{}) perform best, reaching very high accuracy and NMI even in the presence of two noisy layers (Fig.\ \ref{fig:plot2}(b)). It is also interesting to notice that, while  the community detection problem becomes easier when the ratio $p/q$ grows, our proposed approaches still show performance advantages in that setting. 
In order to verify how the difference in the size of the communities can influence the performances, we tested the methods on networks with communities of different sizes.
In particular, we generated networks by the SBM with 3 communities of 100, 150 and 200 nodes, respectively. 
We keep the same values for $p$ and $p/q$, and we studied the same cases. 
Results are shown in Figures \ref{fig:plot5} and \ref{fig:plot6}: \ref{fig:plot5}(a)(b) two informative layers, \ref{fig:plot5}(c)(d) three informative layers, \ref{fig:plot6}(a)(b) two informative layers and one noisy layer, \ref{fig:plot6}(c)(d) two informative layers and two noise ones.
We do not report the results of IM, because it achieves very low results with respect to the others, mostly including all the nodes in just one community.
In general, we see that the performances of the methods are not really affected by the dimension of the communities.

\subsection{ Synthetic Networks via LFR}
\label{LFR}

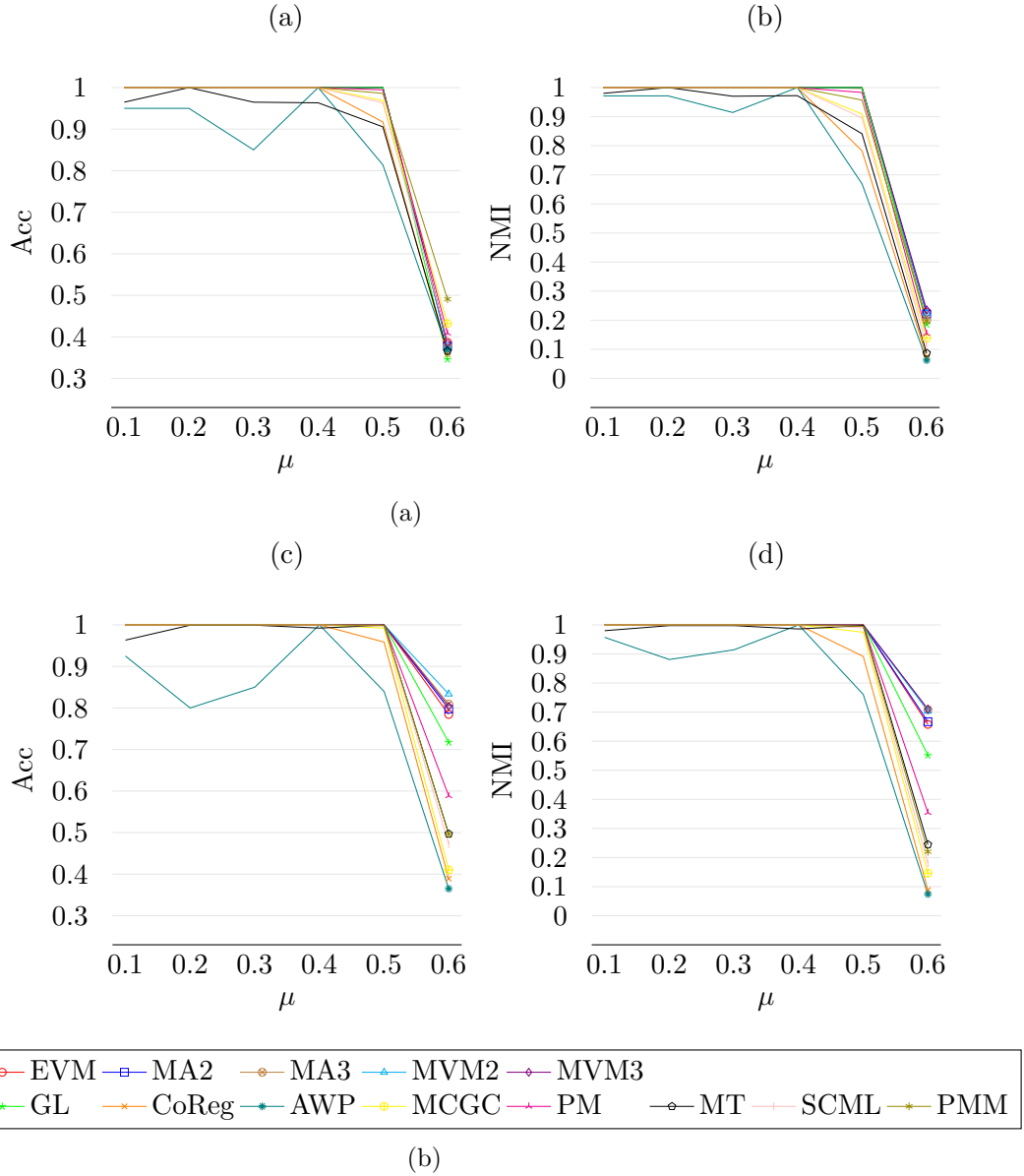
\begin{figure*}[p]
\centering
\captionsetup{oneside,margin={-0.53cm,0.6cm}}
    \hspace*{-2cm}
    \subfloat[]
    {
\pgfplotstableread{
Methods mu Acc NMI 
AWP 0.1 0.95 0.971428571
AWP 0.2 0.95 0.971428571
AWP 0.3 0.85 0.914285714
AWP 0.4 1 1
AWP 0.5 0.8140625 0.669195651
AWP 0.6 0.36875 0.062499548
}{\AWP}

\pgfplotstableread{
Methods mu Acc NMI
CoReg 0.1 1 1
CoReg 0.2 1 1
CoReg 0.3 1 1
CoReg 0.4 1 1
CoReg 0.5 0.9171875 0.782644055
CoReg 0.6 0.3578125 0.073767413
}{\CoReg}

\pgfplotstableread{
Methods mu Acc NMI
EVM 0.1 1 1
EVM 0.2 1 1
EVM 0.3 1 1
EVM 0.4 1 1
EVM 0.5 0.99921875 0.997483192
EVM 0.6 0.38671875 0.227639243
}{\EVM}

\pgfplotstableread{
Methods mu Acc NMI
GL 0.1 1 1
GL 0.2 1 1
GL 0.3 1 1
GL 0.4 1 1
GL 0.5 1 1
GL 0.6 0.34609375 0.185469813
}{\GL}

\pgfplotstableread{
Methods mu Acc NMI
IM 0.1 1 1
IM 0.2 1 1
IM 0.3 1 1
IM 0.4 1 1
IM 0.5 0.25 0
IM 0.6 0.25 0
}{\IM}

\pgfplotstableread{
Methods mu Acc NMI
MA2 0.1 1 1
MA2 0.2 1 1
MA2 0.3 1 1
MA2 0.4 1 1
MA2 0.5 1 1
MA2 0.6 0.37734375 0.221391384
}{\MA}

\pgfplotstableread{
Methods mu Acc NMI
MA3 0.1 1 1
MA3 0.2 1 1
MA3 0.3 1 1
MA3 0.4 1 1
MA3 0.5 1 1
MA3 0.6 0.37734375 0.200475503
}{\MAt}

\pgfplotstableread{
Methods mu Acc NMI 
MCGC 0.1 1 1
MCGC 0.2 1 1
MCGC 0.3 1 1
MCGC 0.4 1 1
MCGC 0.5 0.96875 0.908920484
MCGC 0.6 0.43203125 0.138767097
}{\MCGC}

\pgfplotstableread{
Methods mu Acc NMI 
MT 0.1 0.96484375 0.980136288
MT 0.2 1 1
MT 0.3 0.96484375 0.970224312
MT 0.4 0.96328125 0.972198573
MT 0.5 0.90546875 0.841307009
MT 0.6 0.365625 0.086823108
}{\MT}

\pgfplotstableread{
Methods mu Acc NMI
MVM2 0.1 1 1
MVM2 0.2 1 1
MVM2 0.3 1 1
MVM2 0.4 1 1
MVM2 0.5 1 1
MVM2 0.6 0.378125 0.224857091
}{\MVM}

\pgfplotstableread{
Methods mu Acc NMI
MVM3 0.1 1 1
MVM3 0.2 1 1
MVM3 0.3 1 1
MVM3 0.4 1 1
MVM3 0.5 1 1
MVM3 0.6 0.3859375 0.235467935
}{\MVMt}

\pgfplotstableread{
Methods mu Acc NMI
PM 0.1 1 1
PM 0.2 1 1
PM 0.3 1 1
PM 0.4 1 1
PM 0.5 0.99453125 0.9831719
PM 0.6 0.40859375 0.153922497
}{\PM}

\pgfplotstableread{
Methods mu Acc NMI
PMM 0.1 1 1
PMM 0.2 1 1
PMM 0.3 1 1
PMM 0.4 1 1
PMM 0.5 0.9859375 0.957070563
PMM 0.6 0.490625 0.196770811
}{\PMM}

\pgfplotstableread{
Methods mu Acc NMI
SCML 0.1 1 1
SCML 0.2 1 1
SCML 0.3 1 1
SCML 0.4 1 1
SCML 0.5 0.96328125 0.895447825
SCML 0.6 0.39375 0.112995124
}{\SCML}

\begin{tikzpicture}
    \begin{groupplot}[
        group style={
        group name=art_info, 
        group size= 2 by 1,
        horizontal sep=50pt
        },
        width=6.3cm,
        height=6.3cm,
        enlarge y limits,
        axis y line*=left,
        axis x line*=bottom, 
        xtick=data,
        ymajorgrids,
		grid style={line width=.2pt,draw=gray!20},
        y axis line style={draw=none},
        xtick style={draw=none},
		ytick style={draw=none},
        yticklabel style={xshift=-0.5em},
    ]
     
     \nextgroupplot[title=(a),xmin=0.08, xmax=0.62, ymin=0.3, ymax = 1,extra y ticks={0.3,0.5,0.7,0.9} , xlabel={$\mu$}, ylabel={Acc}]
     
      \addplot [line width=0.05mm, 
     mark=o, color=red, mark size=1.5, mark options={fill=red!20!red,mark indices=6}, 
     ] table[x=mu,y=Acc]{\EVM}; 
     
     \addplot [line width=0.05mm, 
     mark=square, color=blue, mark size=1.5, mark options={fill=blue!20!blue,mark indices=6}, 
     ] table[x=mu,y=Acc]{\MA}; 
     
      \addplot [line width=0.05mm, 
     mark=otimes, color=brown, mark size=1.5, mark options={fill=brown!20!brown,mark indices=6}, 
     ] table[x=mu,y=Acc]{\MAt}; 
     
       \addplot [line width=0.05mm, 
     mark=triangle, color=cyan, mark size=1.5, mark options={fill=cyan!20!cyan,mark indices=6}, 
     ] table[x=mu,y=Acc]{\MVM}; 
     
       \addplot [line width=0.05mm, 
     mark=diamond, color=violet, mark size=1.5, mark options={fill=violet!20!violet,mark indices=6}, 
     ] table[x=mu,y=Acc]{\MVMt}; 
     
       \addplot [line width=0.05mm, 
     mark=star, color=green, mark size=1.5, mark options={fill=green!20!green,mark indices=6}, 
     ] table[x=mu,y=Acc]{\GL}; 
     
       \addplot [line width=0.05mm, 
     mark=x, color=orange, mark size=1.5, mark options={fill=orange!20!orange,mark indices=6}, 
     ] table[x=mu,y=Acc]{\CoReg}; 
     
       \addplot [line width=0.05mm, 
     mark=10-pointed star, color=teal, mark size=1.5, mark options={fill=teal!20!teal,mark indices=6}, 
     ] table[x=mu,y=Acc]{\AWP};  
     
       \addplot [line width=0.05mm, 
     mark=oplus, color=yellow, mark size=1.5, mark options={fill=yellow!20!yellow,mark indices=6}, 
     ] table[x=mu,y=Acc]{\MCGC}; 
     
       \addplot [line width=0.05mm,
     mark=Mercedes star, color=magenta, mark size=1.5, mark options={fill=magenta!20!magenta,mark indices=6}, 
     ] table[x=mu,y=Acc]{\PM}; 
     
       \addplot [line width=0.05mm,
     mark=pentagon, color=black, mark size=1.5, mark options={fill=black!20!black,mark indices=6}, 
     ] table[x=mu,y=Acc]{\MT}; 
     
       \addplot [line width=0.05mm,
     mark=|, color=pink, mark size=1.5, mark options={fill=pink!20!pink,mark indices=6}, 
     ] table[x=mu,y=Acc]{\SCML}; 
     
       \addplot [line width=0.05mm, 
     mark=asterisk, color=olive, mark size=1.5, mark options={fill=olive!20!olive,mark indices=6}, 
     ] table[x=mu,y=Acc]{\PMM}; 
     
     
     \nextgroupplot[title=(b),xmin=0.08, xmax=0.62, ymin=0, ymax = 1,extra y ticks={0.1,0.2,0.3,0.4,0.6,0.7,0.8,0.9},xlabel={$\mu$}, ylabel={NMI}]
     
        \addplot [line width=0.05mm, 
     mark=o, color=red, mark size=1.5, mark options={fill=red!20!red,mark indices=6}, 
     ] table[x=mu,y=NMI]{\EVM}; 
     
     \addplot [line width=0.05mm, 
     mark=square, color=blue, mark size=1.5, mark options={fill=blue!20!blue,mark indices=6}, 
     ] table[x=mu,y=NMI]{\MA}; 
     
      \addplot [line width=0.05mm, 
     mark=otimes, color=brown, mark size=1.5, mark options={fill=brown!20!brown,mark indices=6}, 
     ] table[x=mu,y=NMI]{\MAt}; 
     
       \addplot [line width=0.05mm, 
     mark=triangle, color=cyan, mark size=1.5, mark options={fill=cyan!20!cyan,mark indices=6}, 
     ] table[x=mu,y=NMI]{\MVM}; 
     
       \addplot [line width=0.05mm, 
     mark=diamond, color=violet, mark size=1.5, mark options={fill=violet!20!violet,mark indices=6}, 
     ] table[x=mu,y=NMI]{\MVMt}; 
     
       \addplot [line width=0.05mm, 
     mark=star, color=green, mark size=1.5, mark options={fill=green!20!green,mark indices=6}, 
     ] table[x=mu,y=NMI]{\GL}; 
     
       \addplot [line width=0.05mm, 
     mark=x, color=orange, mark size=1.5, mark options={fill=orange!20!orange,mark indices=6}, 
     ] table[x=mu,y=NMI]{\CoReg}; 
     
       \addplot [line width=0.05mm, 
     mark=10-pointed star, color=teal, mark size=1.5, mark options={fill=teal!20!teal,mark indices=6}, 
     ] table[x=mu,y=NMI]{\AWP};  
     
       \addplot [line width=0.05mm, 
     mark=oplus, color=yellow, mark size=1.5, mark options={fill=yellow!20!yellow,mark indices=6}, 
     ] table[x=mu,y=NMI]{\MCGC}; 
     
       \addplot [line width=0.05mm,
     mark=Mercedes star, color=magenta, mark size=1.5, mark options={fill=magenta!20!magenta,mark indices=6}, 
     ] table[x=mu,y=NMI]{\PM}; 
     
       \addplot [line width=0.05mm,
     mark=pentagon, color=black, mark size=1.5, mark options={fill=black!20!black,mark indices=6}, 
     ] table[x=mu,y=NMI]{\MT}; 
     
       \addplot [line width=0.05mm,
     mark=|, color=pink, mark size=1.5, mark options={fill=pink!20!pink,mark indices=6}, 
     ] table[x=mu,y=NMI]{\SCML}; 
     
       \addplot [line width=0.05mm, 
     mark=asterisk, color=olive, mark size=1.5, mark options={fill=olive!20!olive,mark indices=6}, 
     ] table[x=mu,y=NMI]{\PMM}; 

    \end{groupplot}
\end{tikzpicture}}
    \hfill
    \hspace*{-1.5cm}
    \subfloat[]
    {
\pgfplotstableread{
Methods mu Acc NMI 
AWP 0.1 0.925 0.957142857
AWP 0.2 0.8 0.880952381
AWP 0.3 0.85 0.914285714
AWP 0.4 1 1
AWP 0.5 0.83984375 0.760799497
AWP 0.6 0.36484375 0.074204183
}{\AWP}

\pgfplotstableread{
Methods mu Acc NMI
CoReg 0.1 1 1
CoReg 0.2 1 1
CoReg 0.3 1 1
CoReg 0.4 1 1
CoReg 0.5 0.95859375 0.891516598
CoReg 0.6 0.3890625 0.088183295
}{\CoReg}

\pgfplotstableread{
Methods mu Acc NMI
EVM 0.1 1 1
EVM 0.2 1 1
EVM 0.3 1 1
EVM 0.4 1 1
EVM 0.5 1 1
EVM 0.6 0.784375 0.657849531
}{\EVM}

\pgfplotstableread{
Methods mu Acc NMI
GL 0.1 1 1
GL 0.2 1 1
GL 0.3 1 1
GL 0.4 1 1
GL 0.5 1 1
GL 0.6 0.7171875 0.551450028
}{\GL}

\pgfplotstableread{
Methods mu Acc NMI
IM 0.1 1 1
IM 0.2 1 1
IM 0.3 1 1
IM 0.4 0.996875 0.990731572
IM 0.5 0.25 0
IM 0.6 0.25 0
}{\IM}

\pgfplotstableread{
Methods mu Acc NMI
MA2 0.1 1 1
MA2 0.2 1 1
MA2 0.3 1 1
MA2 0.4 1 1
MA2 0.5 1 1
MA2 0.6 0.79765625 0.667114537
}{\MA}

\pgfplotstableread{
Methods mu Acc NMI
MA3 0.1 1 1
MA3 0.2 1 1
MA3 0.3 1 1
MA3 0.4 1 1
MA3 0.5 1 1
MA3 0.6 0.809375 0.708419482
}{\MAt}

\pgfplotstableread{
Methods mu Acc NMI 
MCGC 0.1 1 1
MCGC 0.2 1 1
MCGC 0.3 1 1
MCGC 0.4 1 1
MCGC 0.5 0.9921875 0.974830708
MCGC 0.6 0.41015625 0.146308096
}{\MCGC}

\pgfplotstableread{
Methods mu Acc NMI 
MT 0.1 0.96328125 0.980015041
MT 0.2 0.99921875 0.997483192
MT 0.3 0.99921875 0.997483192
MT 0.4 0.9921875 0.985962587
MT 0.5 0.99921875 0.997483192
MT 0.6 0.496875 0.244813096
}{\MT}

\pgfplotstableread{
Methods mu Acc NMI
MVM2 0.1 1 1
MVM2 0.2 1 1
MVM2 0.3 1 1
MVM2 0.4 1 1
MVM2 0.5 1 1
MVM2 0.6 0.83359375 0.703338639
}{\MVM}

\pgfplotstableread{
Methods mu Acc NMI
MVM3 0.1 1 1
MVM3 0.2 1 1
MVM3 0.3 1 1
MVM3 0.4 1 1
MVM3 0.5 1 1
MVM3 0.6 0.80390625 0.710303985
}{\MVMt}

\pgfplotstableread{
Methods mu Acc NMI
PM 0.1 1 1
PM 0.2 1 1
PM 0.3 1 1
PM 0.4 1 1
PM 0.5 1 1
PM 0.6 0.58828125 0.354303541
}{\PM}

\pgfplotstableread{
Methods mu Acc NMI
PMM 0.1 1 1
PMM 0.2 1 1
PMM 0.3 1 1
PMM 0.4 1 1
PMM 0.5 0.9984375 0.994965941
PMM 0.6 0.49609375 0.219988768
}{\PMM}

\pgfplotstableread{
Methods mu Acc NMI
SCML 0.1 1 1
SCML 0.2 1 1
SCML 0.3 1 1
SCML 0.4 1 1
SCML 0.5 0.996875 0.989932769
SCML 0.6 0.471875 0.179759354
}{\SCML}

\begin{tikzpicture}
    \begin{groupplot}[
        group style={
        group name=art_info, 
        group size= 2 by 1,
        horizontal sep=50pt
        },
        width=6.3cm,
        height=6.3cm,
        enlarge y limits,
        axis y line*=left,
        axis x line*=bottom, 
        xtick=data,
        ymajorgrids,
		grid style={draw=gray!20},
        y axis line style={draw=none},
        xtick style={draw=none},
		ytick style={draw=none},
        yticklabel style={xshift=-0.5em},
        legend style={
        at={(1.1,-0.30)},
        legend columns=8,
        legend cell align=left,
        anchor=north},
        legend image post style={mark options={scale=1}}
    ]
     
     \nextgroupplot[title=(c),xmin=0.08, xmax=0.62, ymin=0.3, ymax = 1,extra y ticks={0.3,0.5,0.7,0.9}, xlabel={$\mu$}, ylabel={Acc}]
     
      \addplot [line width=0.05mm, 
     mark=o, color=red, mark size=1.5, mark options={fill=red!20!red,mark indices=6}, 
     ] table[x=mu,y=Acc]{\EVM}; \addlegendentry{EVM};
     
     \addplot [line width=0.05mm, 
     mark=square, color=blue, mark size=1.5, mark options={fill=blue!20!blue,mark indices=6}, 
     ] table[x=mu,y=Acc]{\MA}; \addlegendentry{MA2};
     
      \addplot [line width=0.05mm, 
     mark=otimes, color=brown, mark size=1.5, mark options={fill=brown!20!brown,mark indices=6}, 
     ] table[x=mu,y=Acc]{\MAt}; \addlegendentry{MA3};
     
       \addplot [line width=0.05mm, 
     mark=triangle, color=cyan, mark size=1.5, mark options={fill=cyan!20!cyan,mark indices=6}, 
     ] table[x=mu,y=Acc]{\MVM}; \addlegendentry{MVM2};
     
       \addplot [line width=0.05mm, 
     mark=diamond, color=violet, mark size=1.5, mark options={fill=violet!20!violet,mark indices=6}, 
     ] table[x=mu,y=Acc]{\MVMt}; \addlegendentry{MVM3};
     
      \addlegendimage{empty legend}
            \addlegendentry{}
            
     \addlegendimage{empty legend}
            \addlegendentry{}
            
     \addlegendimage{empty legend}
            \addlegendentry{}
            
       \addplot [line width=0.05mm, 
     mark=star, color=green, mark size=1.5, mark options={fill=green!20!green,mark indices=6}, 
     ] table[x=mu,y=Acc]{\GL}; \addlegendentry{GL};
     
       \addplot [line width=0.05mm, 
     mark=x, color=orange, mark size=1.5, mark options={fill=orange!20!orange,mark indices=6}, 
     ] table[x=mu,y=Acc]{\CoReg}; \addlegendentry{CoReg};
     
       \addplot [line width=0.05mm, 
     mark=10-pointed star, color=teal, mark size=1.5, mark options={fill=teal!20!teal,mark indices=6}, 
     ] table[x=mu,y=Acc]{\AWP};  \addlegendentry{AWP};
     
       \addplot [line width=0.05mm, 
     mark=oplus, color=yellow, mark size=1.5, mark options={fill=yellow!20!yellow,mark indices=6}, 
     ] table[x=mu,y=Acc]{\MCGC}; \addlegendentry{MCGC};
     
       \addplot [line width=0.05mm,
     mark=Mercedes star, color=magenta, mark size=1.5, mark options={fill=magenta!20!magenta,mark indices=6}, 
     ] table[x=mu,y=Acc]{\PM}; \addlegendentry{PM};
     
       \addplot [line width=0.05mm,
     mark=pentagon, color=black, mark size=1.5, mark options={fill=black!20!black,mark indices=6}, 
     ] table[x=mu,y=Acc]{\MT}; \addlegendentry{MT};
     
       \addplot [line width=0.05mm,
     mark=|, color=pink, mark size=1.5, mark options={fill=pink!20!pink,mark indices=6}, 
     ] table[x=mu,y=Acc]{\SCML}; \addlegendentry{SCML};
     
       \addplot [line width=0.05mm, 
     mark=asterisk, color=olive, mark size=1.5, mark options={fill=olive!20!olive,mark indices=6}, 
     ] table[x=mu,y=Acc]{\PMM}; \addlegendentry{PMM};
     
     
     \nextgroupplot[title=(d),xmin=0.08, xmax=0.62, ymin=0, ymax = 1,extra y ticks={0.1,0.2,0.3,0.4,0.6,0.7,0.8,0.9},xlabel={$\mu$}, ylabel={NMI}]
     
      \addplot [line width=0.05mm, 
     mark=o, color=red, mark size=1.5, mark options={fill=red!20!red,mark indices=6}, 
     ] table[x=mu,y=NMI]{\EVM}; 
     
     \addplot [line width=0.05mm, 
     mark=square, color=blue, mark size=1.5, mark options={fill=blue!20!blue,mark indices=6}, 
     ] table[x=mu,y=NMI]{\MA}; 
     
      \addplot [line width=0.05mm, 
     mark=otimes, color=brown, mark size=1.5, mark options={fill=brown!20!brown,mark indices=6}, 
     ] table[x=mu,y=NMI]{\MAt}; 
     
       \addplot [line width=0.05mm, 
     mark=triangle, color=cyan, mark size=1.5, mark options={fill=cyan!20!cyan,mark indices=6}, 
     ] table[x=mu,y=NMI]{\MVM}; 
     
       \addplot [line width=0.05mm, 
     mark=diamond, color=violet, mark size=1.5, mark options={fill=violet!20!violet,mark indices=6}, 
     ] table[x=mu,y=NMI]{\MVMt}; 
     
       \addplot [line width=0.05mm, 
     mark=star, color=green, mark size=1.5, mark options={fill=green!20!green,mark indices=6}, 
     ] table[x=mu,y=NMI]{\GL}; 
     
       \addplot [line width=0.05mm, 
     mark=x, color=orange, mark size=1.5, mark options={fill=orange!20!orange,mark indices=6}, 
     ] table[x=mu,y=NMI]{\CoReg}; 
     
       \addplot [line width=0.05mm, 
     mark=10-pointed star, color=teal, mark size=1.5, mark options={fill=teal!20!teal,mark indices=6}, 
     ] table[x=mu,y=NMI]{\AWP};  
     
       \addplot [line width=0.05mm, 
     mark=oplus, color=yellow, mark size=1.5, mark options={fill=yellow!20!yellow,mark indices=6}, 
     ] table[x=mu,y=NMI]{\MCGC}; 
     
       \addplot [line width=0.05mm,
     mark=Mercedes star, color=magenta, mark size=1.5, mark options={fill=magenta!20!magenta,mark indices=6}, 
     ] table[x=mu,y=NMI]{\PM}; 
     
       \addplot [line width=0.05mm,
     mark=pentagon, color=black, mark size=1.5, mark options={fill=black!20!black,mark indices=6}, 
     ] table[x=mu,y=NMI]{\MT}; 
     
       \addplot [line width=0.05mm,
     mark=|, color=pink, mark size=1.5, mark options={fill=pink!20!pink,mark indices=6}, 
     ] table[x=mu,y=NMI]{\SCML}; 
     
       \addplot [line width=0.05mm, 
     mark=asterisk, color=olive, mark size=1.5, mark options={fill=olive!20!olive,mark indices=6}, 
     ] table[x=mu,y=NMI]{\PMM};

    \end{groupplot}
\end{tikzpicture}}
\caption{Average values of accuracy and NMI over 10 random networks sampled from LFR with  equally distributed informative layers (2 layers (a)(b) and 3 layers (c)(d)), with four clusters and $\mu \in \{0.1, 0.2, 0.3, 0.4, 0.5, 0.6\}$.}
\label{fig:plot3}
\end{figure*}

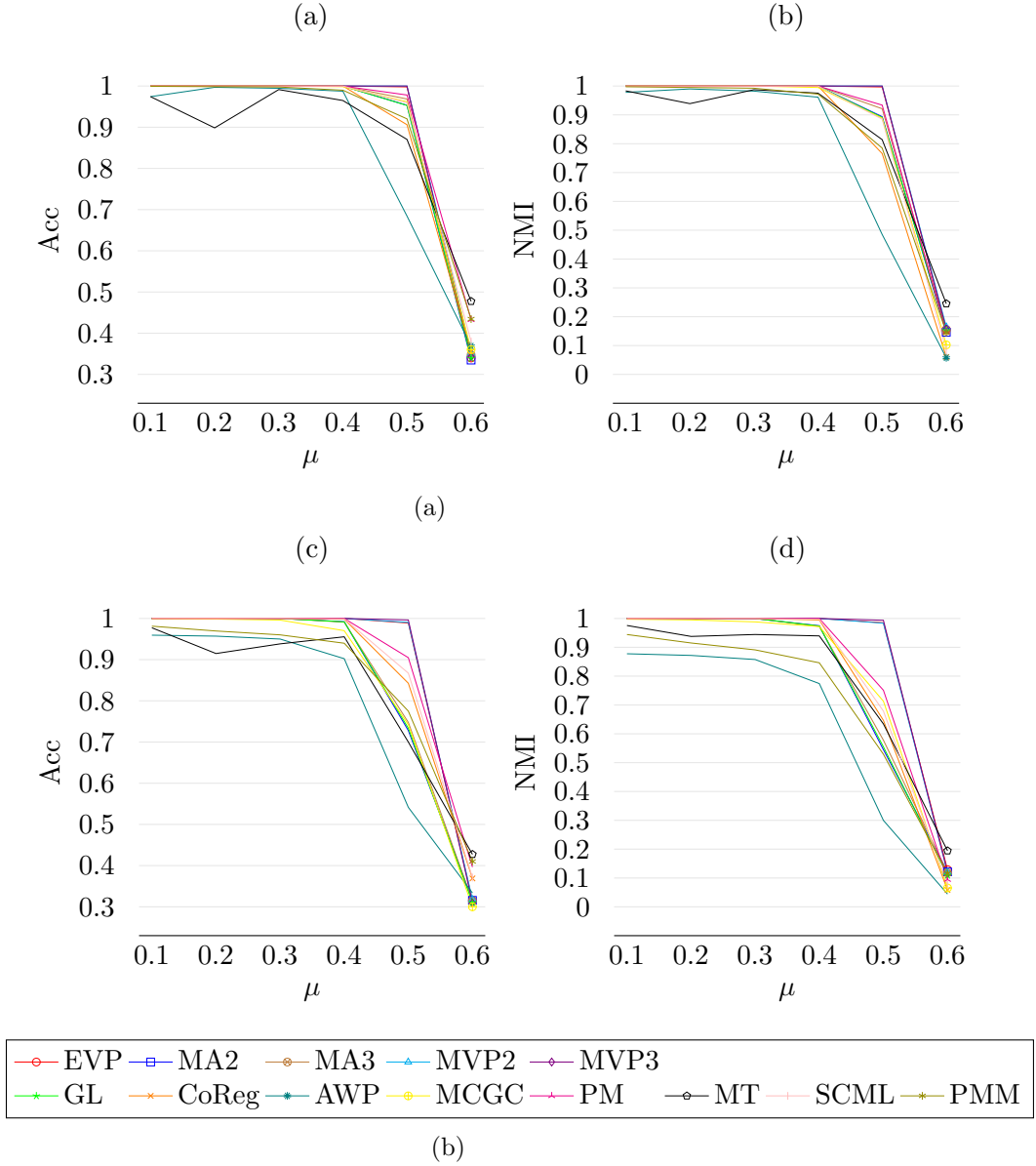
\begin{figure*}[p]
\centering
\captionsetup{oneside,margin={-0.53cm,0.6cm}}
    \hspace*{-2cm}
    \subfloat[]
    {
\pgfplotstableread{
Methods mu Acc NMI 
AWP 0.1 0.97421875 0.977995565
AWP 0.2 0.996875 0.989932769
AWP 0.3 0.99453125 0.982390711
AWP 0.4 0.9875 0.961305302
AWP 0.5 0.684375 0.48512492
AWP 0.6 0.36875 0.057460917
}{\AWP}

\pgfplotstableread{
Methods mu Acc NMI 
CoReg 0.1 1 1
CoReg 0.2 1 1
CoReg 0.3 1 1
CoReg 0.4 1 1
CoReg 0.5 0.90609375 0.765752707
CoReg 0.6 0.354609375 0.060332525
}{\CoReg}

\pgfplotstableread{
Methods mu Acc NMI  
EVP 0.1 1 1
EVP 0.2 1 1
EVP 0.3 1 1
EVP 0.4 1 1
EVP 0.5 0.99796875 0.996855084
EVP 0.6 0.33984375 0.156094833
}{\EVP}

\pgfplotstableread{
Methods mu Acc NMI  
GL 0.1 1 1
GL 0.2 1 1
GL 0.3 1 1
GL 0.4 1 1
GL 0.5 0.95296875 0.890943666
GL 0.6 0.33890625 0.152082515
    }{\GL}

\pgfplotstableread{
Methods mu Acc NMI     
    IM 0.1 1 1
IM 0.2 1 1
IM 0.3 1 1
IM 0.4 1 1
IM 0.5 0.25 0
IM 0.6 0.25 0
}{\IM}

\pgfplotstableread{
Methods mu Acc NMI 
MA2 0.1 1 1
MA2 0.2 1 1
MA2 0.3 1 1
MA2 0.4 1 1
MA2 0.5 0.953203125 0.893637511
MA2 0.6 0.334765625 0.14614946
}{\MA}

\pgfplotstableread{
Methods mu Acc NMI 
MA3 0.1 1 1
MA3 0.2 1 1
MA3 0.3 1 1
MA3 0.4 1 1
MA3 0.5 0.968515625 0.921221063
MA3 0.6 0.347578125 0.151198203
}{\MAt}

\pgfplotstableread{
Methods mu Acc NMI 
MCGC 0.1 1 1
MCGC 0.2 1 1
MCGC 0.3 1 1
MCGC 0.4 0.9984375 0.994966385
MCGC 0.5 0.96171875 0.888641009
MCGC 0.6 0.36015625 0.102419966
}{\MCGC}

\pgfplotstableread{
Methods mu Acc NMI 
MT 0.1 0.97421875 0.983197478
MT 0.2 0.8984375 0.938953209
MT 0.3 0.99140625 0.98716842
MT 0.4 0.96484375 0.97529059
MT 0.5 0.8703125 0.814040969
MT 0.6 0.47734375 0.245301752
}{\MT}

\pgfplotstableread{
Methods mu Acc NMI 
MVP2 0.1 1 1
MVP2 0.2 1 1
MVP2 0.3 1 1
MVP2 0.4 1 1
MVP2 0.5 1 1
MVP2 0.6 0.351796875 0.164641642
}{\MVP}

\pgfplotstableread{
Methods mu Acc NMI 
MVP3 0.1 1 1
MVP3 0.2 1 1
MVP3 0.3 1 1
MVP3 0.4 1 1
MVP3 0.5 1 1
MVP3 0.6 0.342890625 0.157583673
}{\MVPt}

\pgfplotstableread{
Methods mu Acc NMI 
PM 0.1 1 1
PM 0.2 1 1
PM 0.3 1 1
PM 0.4 1 1
PM 0.5 0.978203125 0.9337983
PM 0.6 0.432421875 0.146752763
}{\PM}

\pgfplotstableread{
Methods mu Acc NMI 
PMM 0.1 0.99921875 0.997483192
PMM 0.2 0.9984375 0.994966385
PMM 0.3 0.99765625 0.992449577
PMM 0.4 0.98984375 0.972004755
PMM 0.5 0.9203125 0.784839555
PMM 0.6 0.434375 0.145338169
}{\PMM}

\pgfplotstableread{
Methods mu Acc NMI  
SCML 0.1 1 1
SCML 0.2 1 1
SCML 0.3 1 1
SCML 0.4 1 1
SCML 0.5 0.959375 0.885247359
SCML 0.6 0.3796875 0.083436308
}{\SCML}

\begin{tikzpicture}
    \begin{groupplot}[
        group style={
        group name=art_info, 
        group size= 2 by 1,
        horizontal sep=50pt
        },
        width=6.3cm,
        height=6.3cm,
        enlarge y limits,
        axis y line*=left,
        axis x line*=bottom, 
        xtick=data,
        ymajorgrids,
		grid style={line width=.2pt,draw=gray!20},
        y axis line style={draw=none},
        xtick style={draw=none},
		ytick style={draw=none},
        yticklabel style={xshift=-0.5em},
    ]
     
     \nextgroupplot[title=(a),xmin=0.08, xmax=0.62, ymin=0.3, ymax = 1,extra y ticks={0.3,0.5,0.7,0.9}, xlabel={$\mu$}, ylabel={Acc}]
     
       \addplot [line width=0.05mm, 
     mark=o, color=red, mark size=1.5, mark options={fill=red!20!red,mark indices=6}, 
     ] table[x=mu,y=Acc]{\EVP}; 
     
     \addplot [line width=0.05mm, 
     mark=square, color=blue, mark size=1.5, mark options={fill=blue!20!blue,mark indices=6}, 
     ] table[x=mu,y=Acc]{\MA}; 
     
      \addplot [line width=0.05mm, 
     mark=otimes, color=brown, mark size=1.5, mark options={fill=brown!20!brown,mark indices=6}, 
     ] table[x=mu,y=Acc]{\MAt}; 
     
       \addplot [line width=0.05mm, 
     mark=triangle, color=cyan, mark size=1.5, mark options={fill=cyan!20!cyan,mark indices=6}, 
     ] table[x=mu,y=Acc]{\MVP}; 
     
       \addplot [line width=0.05mm, 
     mark=diamond, color=violet, mark size=1.5, mark options={fill=violet!20!violet,mark indices=6}, 
     ] table[x=mu,y=Acc]{\MVPt}; 
     
       \addplot [line width=0.05mm, 
     mark=star, color=green, mark size=1.5, mark options={fill=green!20!green,mark indices=6}, 
     ] table[x=mu,y=Acc]{\GL}; 
     
       \addplot [line width=0.05mm, 
     mark=x, color=orange, mark size=1.5, mark options={fill=orange!20!orange,mark indices=6}, 
     ] table[x=mu,y=Acc]{\CoReg}; 
     
       \addplot [line width=0.05mm, 
     mark=10-pointed star, color=teal, mark size=1.5, mark options={fill=teal!20!teal,mark indices=6}, 
     ] table[x=mu,y=Acc]{\AWP};  
     
       \addplot [line width=0.05mm, 
     mark=oplus, color=yellow, mark size=1.5, mark options={fill=yellow!20!yellow,mark indices=6}, 
     ] table[x=mu,y=Acc]{\MCGC}; 
     
       \addplot [line width=0.05mm,
     mark=Mercedes star, color=magenta, mark size=1.5, mark options={fill=magenta!20!magenta,mark indices=6}, 
     ] table[x=mu,y=Acc]{\PM}; 
     
       \addplot [line width=0.05mm,
     mark=pentagon, color=black, mark size=1.5, mark options={fill=black!20!black,mark indices=6}, 
     ] table[x=mu,y=Acc]{\MT}; 
     
       \addplot [line width=0.05mm,
     mark=|, color=pink, mark size=1.5, mark options={fill=pink!20!pink,mark indices=6}, 
     ] table[x=mu,y=Acc]{\SCML}; 
     
       \addplot [line width=0.05mm, 
     mark=asterisk, color=olive, mark size=1.5, mark options={fill=olive!20!olive,mark indices=6}, 
     ] table[x=mu,y=Acc]{\PMM};

     
     \nextgroupplot[title=(b),xmin=0.08, xmax=0.62, ymin=0, ymax = 1,extra y ticks={0.1,0.2,0.3,0.4,0.6,0.7,0.8,0.9},xlabel={$\mu$}, ylabel={NMI}]
     
       \addplot [line width=0.05mm, 
     mark=o, color=red, mark size=1.5, mark options={fill=red!20!red,mark indices=6}, 
     ] table[x=mu,y=NMI]{\EVP}; 
     
     \addplot [line width=0.05mm, 
     mark=square, color=blue, mark size=1.5, mark options={fill=blue!20!blue,mark indices=6}, 
     ] table[x=mu,y=NMI]{\MA}; 
     
      \addplot [line width=0.05mm, 
     mark=otimes, color=brown, mark size=1.5, mark options={fill=brown!20!brown,mark indices=6}, 
     ] table[x=mu,y=NMI]{\MAt}; 
     
       \addplot [line width=0.05mm, 
     mark=triangle, color=cyan, mark size=1.5, mark options={fill=cyan!20!cyan,mark indices=6}, 
     ] table[x=mu,y=NMI]{\MVP}; 
     
       \addplot [line width=0.05mm, 
     mark=diamond, color=violet, mark size=1.5, mark options={fill=violet!20!violet,mark indices=6}, 
     ] table[x=mu,y=NMI]{\MVPt}; 
     
       \addplot [line width=0.05mm, 
     mark=star, color=green, mark size=1.5, mark options={fill=green!20!green,mark indices=6}, 
     ] table[x=mu,y=NMI]{\GL}; 
     
       \addplot [line width=0.05mm, 
     mark=x, color=orange, mark size=1.5, mark options={fill=orange!20!orange,mark indices=6}, 
     ] table[x=mu,y=NMI]{\CoReg}; 
     
       \addplot [line width=0.05mm, 
     mark=10-pointed star, color=teal, mark size=1.5, mark options={fill=teal!20!teal,mark indices=6}, 
     ] table[x=mu,y=NMI]{\AWP};  
     
       \addplot [line width=0.05mm, 
     mark=oplus, color=yellow, mark size=1.5, mark options={fill=yellow!20!yellow,mark indices=6}, 
     ] table[x=mu,y=NMI]{\MCGC}; 
     
       \addplot [line width=0.05mm,
     mark=Mercedes star, color=magenta, mark size=1.5, mark options={fill=magenta!20!magenta,mark indices=6}, 
     ] table[x=mu,y=NMI]{\PM}; 
     
       \addplot [line width=0.05mm,
     mark=pentagon, color=black, mark size=1.5, mark options={fill=black!20!black,mark indices=6}, 
     ] table[x=mu,y=NMI]{\MT}; 
     
       \addplot [line width=0.05mm,
     mark=|, color=pink, mark size=1.5, mark options={fill=pink!20!pink,mark indices=6}, 
     ] table[x=mu,y=NMI]{\SCML}; 
     
       \addplot [line width=0.05mm, 
     mark=asterisk, color=olive, mark size=1.5, mark options={fill=olive!20!olive,mark indices=6}, 
     ] table[x=mu,y=NMI]{\PMM}; 
     
    \end{groupplot}
\end{tikzpicture}}
    \hfill
    \hspace*{-1.5cm}
    \subfloat[]
    {
\pgfplotstableread{
Methods mu Acc NMI 
AWP 0.1 0.959375 0.877595541
AWP 0.2 0.95703125 0.871863512
AWP 0.3 0.95 0.857644314
AWP 0.4 0.90234375 0.77419143
AWP 0.5 0.54140625 0.299710349
AWP 0.6 0.33359375 0.044314874
}{\AWP}

\pgfplotstableread{
Methods mu Acc NMI 
CoReg 0.1 1 1
CoReg 0.2 1 1
CoReg 0.3 1 1
CoReg 0.4 0.9984375 0.994966385
CoReg 0.5 0.84265625 0.646647471
CoReg 0.6 0.36890625 0.058520173
}{\CoReg}

\pgfplotstableread{
Methods mu Acc NMI 
EVP 0.1 1 1
EVP 0.2 1 1
EVP 0.3 1 1
EVP 0.4 1 1
EVP 0.5 0.988984375 0.98502458
EVP 0.6 0.31515625 0.128778307
}{\EVP}

\pgfplotstableread{
Methods mu Acc NMI 
GL 0.1 1 1
GL 0.2 1 1
GL 0.3 1 1
GL 0.4 0.991640625 0.973781156
GL 0.5 0.733515625 0.554834986
GL 0.6 0.309921875 0.109272097
}{\GL}

\pgfplotstableread{
Methods mu Acc NMI 
IM 0.1 1 1
IM 0.2 1 1
IM 0.3 1 1
IM 0.4 0.996875 0.989932161
IM 0.5 0.25 0
IM 0.6 0.25 0
}{\IM}

\pgfplotstableread{
Methods mu Acc NMI 
MA2 0.1 1 1
MA2 0.2 1 1
MA2 0.3 0.99984375 0.999496638
MA2 0.4 0.992109375 0.974785776
MA2 0.5 0.727890625 0.545088304
MA2 0.6 0.315625 0.121288779
}{\MA}

\pgfplotstableread{
Methods mu Acc NMI 
MA3 0.1 1 1
MA3 0.2 1 1
MA3 0.3 1 1
MA3 0.4 0.991171875 0.972467885
MA3 0.5 0.747578125 0.577893888
MA3 0.6 0.31421875 0.117974549
}{\MAt}

\pgfplotstableread{
Methods mu Acc NMI 
MCGC 0.1 0.99921875 0.997483192
MCGC 0.2 0.9984375 0.994966385
MCGC 0.3 0.99609375 0.988177466
MCGC 0.4 0.9703125 0.971909885
MCGC 0.5 0.7421875 0.713094234
MCGC 0.6 0.3 0.065484026
}{\MCGC}

\pgfplotstableread{
Methods mu Acc NMI 
MT 0.1 0.97734375 0.975663585
MT 0.2 0.91484375 0.937725919
MT 0.3 0.93828125 0.944399629
MT 0.4 0.95546875 0.939527903
MT 0.5 0.7 0.634266601
MT 0.6 0.42734375 0.194365142
}{\MT}

\pgfplotstableread{
Methods mu Acc NMI 
MVP2 0.1 1 1
MVP2 0.2 1 1
MVP2 0.3 1 1
MVP2 0.4 1 1
MVP2 0.5 0.99046875 0.983583562
MVP2 0.6 0.31765625 0.119591971
}{\MVP}

\pgfplotstableread{
Methods mu Acc NMI 
MVP3 0.1 1 1
MVP3 0.2 1 1
MVP3 0.3 1 1
MVP3 0.4 1 1
MVP3 0.5 0.996328125 0.99372798
MVP3 0.6 0.309765625 0.119715843
}{\MVPt}

\pgfplotstableread{
Methods mu Acc NMI 
PM 0.1 1 1
PM 0.2 1 1
PM 0.3 1 1
PM 0.4 1 1
PM 0.5 0.904765625 0.751027068
PM 0.6 0.396484375 0.094481047
}{\PM}

\pgfplotstableread{
Methods mu Acc NMI 
PMM 0.1 0.98125 0.94436466
PMM 0.2 0.96953125 0.914991833
PMM 0.3 0.96015625 0.891077521
PMM 0.4 0.93984375 0.84622773
PMM 0.5 0.775 0.529230858
PMM 0.6 0.41015625 0.115749804
}{\PMM}

\pgfplotstableread{
Methods mu Acc NMI 
SCML 0.1 1 1
SCML 0.2 1 1
SCML 0.3 1 1
SCML 0.4 0.99921875 0.997483192
SCML 0.5 0.8671875 0.678746888
SCML 0.6 0.37109375 0.076727525
}{\SCML}

\begin{tikzpicture}
    \begin{groupplot}[
        group style={
        group name=art_info, 
        group size= 2 by 1,
        horizontal sep=50pt
        },
        width=6.3cm,
        height=6.3cm,
        enlarge y limits,
        axis y line*=left,
        axis x line*=bottom, 
        xtick=data,
        ymajorgrids,
		grid style={draw=gray!20},
        y axis line style={draw=none},
        xtick style={draw=none},
		ytick style={draw=none},
        yticklabel style={xshift=-0.5em},
        legend style={
        at={(1.1,-0.30)},
        legend columns=8,
        legend cell align=left,
        anchor=north},
        legend image post style={mark options={scale=1}}
    ]
     
     \nextgroupplot[title=(c),xmin=0.08, xmax=0.62, ymin=0.3, ymax = 1,extra y ticks={0.3,0.5,0.7,0.9}, xlabel={$\mu$}, ylabel={Acc}]
     
       \addplot [line width=0.05mm, 
     mark=o, color=red, mark size=1.5, mark options={fill=red!20!red,mark indices=6}, 
     ] table[x=mu,y=Acc]{\EVP}; \addlegendentry{EVP};
     
     \addplot [line width=0.05mm, 
     mark=square, color=blue, mark size=1.5, mark options={fill=blue!20!blue,mark indices=6}, 
     ] table[x=mu,y=Acc]{\MA}; \addlegendentry{MA2};
     
      \addplot [line width=0.05mm, 
     mark=otimes, color=brown, mark size=1.5, mark options={fill=brown!20!brown,mark indices=6}, 
     ] table[x=mu,y=Acc]{\MAt}; \addlegendentry{MA3};
     
       \addplot [line width=0.05mm, 
     mark=triangle, color=cyan, mark size=1.5, mark options={fill=cyan!20!cyan,mark indices=6}, 
     ] table[x=mu,y=Acc]{\MVP}; \addlegendentry{MVP2};
     
       \addplot [line width=0.05mm, 
     mark=diamond, color=violet, mark size=1.5, mark options={fill=violet!20!violet,mark indices=6}, 
     ] table[x=mu,y=Acc]{\MVPt}; \addlegendentry{MVP3};
     
      \addlegendimage{empty legend}
            \addlegendentry{}
            
     \addlegendimage{empty legend}
            \addlegendentry{}
            
     \addlegendimage{empty legend}
            \addlegendentry{}
            
       \addplot [line width=0.05mm, 
     mark=star, color=green, mark size=1.5, mark options={fill=green!20!green,mark indices=6}, 
     ] table[x=mu,y=Acc]{\GL}; \addlegendentry{GL};
     
       \addplot [line width=0.05mm, 
     mark=x, color=orange, mark size=1.5, mark options={fill=orange!20!orange,mark indices=6}, 
     ] table[x=mu,y=Acc]{\CoReg}; \addlegendentry{CoReg};
     
       \addplot [line width=0.05mm, 
     mark=10-pointed star, color=teal, mark size=1.5, mark options={fill=teal!20!teal,mark indices=7}, 
     ] table[x=mu,y=Acc]{\AWP};  \addlegendentry{AWP};
     
       \addplot [line width=0.05mm, 
     mark=oplus, color=yellow, mark size=1.5, mark options={fill=yellow!20!yellow,mark indices=6}, 
     ] table[x=mu,y=Acc]{\MCGC}; \addlegendentry{MCGC};
     
       \addplot [line width=0.05mm,
     mark=Mercedes star, color=magenta, mark size=1.5, mark options={fill=magenta!20!magenta,mark indices=7}, 
     ] table[x=mu,y=Acc]{\PM}; \addlegendentry{PM};
     
       \addplot [line width=0.05mm,
     mark=pentagon, color=black, mark size=1.5, mark options={fill=black!20!black,mark indices=6}, 
     ] table[x=mu,y=Acc]{\MT}; \addlegendentry{MT};
     
       \addplot [line width=0.05mm,
     mark=|, color=pink, mark size=1.5, mark options={fill=pink!20!pink,mark indices=6}, 
     ] table[x=mu,y=Acc]{\SCML}; \addlegendentry{SCML};
     
       \addplot [line width=0.05mm, 
     mark=asterisk, color=olive, mark size=1.5, mark options={fill=olive!20!olive,mark indices=6}, 
     ] table[x=mu,y=Acc]{\PMM}; \addlegendentry{PMM};
     
     
     \nextgroupplot[title=(d),xmin=0.08, xmax=0.62, ymin=0, ymax = 1,extra y ticks={0.1,0.2,0.3,0.4,0.6,0.7,0.8,0.9},xlabel={$\mu$}, ylabel={NMI}]
     
     \addplot [line width=0.05mm, 
     mark=o, color=red, mark size=1.5, mark options={fill=red!20!red,mark indices=6}, 
     ] table[x=mu,y=NMI]{\EVP}; 
     
     \addplot [line width=0.05mm, 
     mark=square, color=blue, mark size=1.5, mark options={fill=blue!20!blue,mark indices=6}, 
     ] table[x=mu,y=NMI]{\MA}; 
     
      \addplot [line width=0.05mm, 
     mark=otimes, color=brown, mark size=1.5, mark options={fill=brown!20!brown,mark indices=6}, 
     ] table[x=mu,y=NMI]{\MAt}; 
     
       \addplot [line width=0.05mm, 
     mark=triangle, color=cyan, mark size=1.5, mark options={fill=cyan!20!cyan,mark indices=6}, 
     ] table[x=mu,y=NMI]{\MVP}; 
     
       \addplot [line width=0.05mm, 
     mark=diamond, color=violet, mark size=1.5, mark options={fill=violet!20!violet,mark indices=7}, 
     ] table[x=mu,y=NMI]{\MVPt}; 
     
       \addplot [line width=0.05mm, 
     mark=star, color=green, mark size=1.5, mark options={fill=green!20!green,mark indices=6}, 
     ] table[x=mu,y=NMI]{\GL}; 
     
       \addplot [line width=0.05mm, 
     mark=x, color=orange, mark size=1.5, mark options={fill=orange!20!orange,mark indices=6}, 
     ] table[x=mu,y=NMI]{\CoReg}; 
     
       \addplot [line width=0.05mm, 
     mark=10-pointed star, color=teal, mark size=1.5, mark options={fill=teal!20!teal,mark indices=7}, 
     ] table[x=mu,y=NMI]{\AWP};  
     
       \addplot [line width=0.05mm, 
     mark=oplus, color=yellow, mark size=1.5, mark options={fill=yellow!20!yellow,mark indices=6}, 
     ] table[x=mu,y=NMI]{\MCGC}; 
     
       \addplot [line width=0.05mm,
     mark=Mercedes star, color=magenta, mark size=1.5, mark options={fill=magenta!20!magenta,mark indices=6}, 
     ] table[x=mu,y=NMI]{\PM}; 
     
       \addplot [line width=0.05mm,
     mark=pentagon, color=black, mark size=1.5, mark options={fill=black!20!black,mark indices=6}, 
     ] table[x=mu,y=NMI]{\MT}; 
     
       \addplot [line width=0.05mm,
     mark=|, color=pink, mark size=1.5, mark options={fill=pink!20!pink,mark indices=6}, 
     ] table[x=mu,y=NMI]{\SCML}; 
     
       \addplot [line width=0.05mm, 
     mark=asterisk, color=olive, mark size=1.5, mark options={fill=olive!20!olive,mark indices=6}, 
     ] table[x=mu,y=NMI]{\PMM}; 
    \end{groupplot}
\end{tikzpicture}}
\caption{Average values of accuracy and NMI over 10 random networks sampled from LFR with both informative and noisy layers (two informative and one noisy in (a)(b); two informative and two noisy in (c)(d)). The informative layers are equally distributed LFR graphs with four clusters and $\mu\in \{0.1, 0.2, 0.3, 0.4, 0.5, 0.6\}$. The noisy layers are LFR graphs with one community and $\mu=0$.}
\label{fig:plot4}
\end{figure*}

Our second test setting is on synthetic networks generated via the Lancichinetti-Fortunato-Radicchi (LFR) benchmark~\cite{lancichinetti2008benchmark}, which  allows us to model networks with  more heterogeneous node degrees and community sizes than the SBM.
We extended this benchmark to the multi-layer case, generating an independent network for each layer, using the same parameters. In particular, following \cite{de2015identifying}, we considered graphs with 128 nodes and 4 communities, each with 32 nodes with an average degree of 16 and maximum degree 32. We let the fraction of inter-community links $\mu$ vary. For the noisy layers, we forced the network to have just one community and we fixed $\mu=0$, as suggested by the authors.
As for the SBM, in our experiments we consider different combinations of informative and noisy layers.
Results are shown in Figures \ref{fig:plot3} and \ref{fig:plot4}, where the different pairs of panels correspond to accuracy and NMI for the settings: \ref{fig:plot3}(a)(b) two informative layers; \ref{fig:plot3}(c)(d) three informative layers; \ref{fig:plot4}(a)(b) two informative layers and one noisy layer; \ref{fig:plot4}(c)(d) two informative layers and two noise ones.   We can observe that the proposed methods give  high values of both accuracy and NMI overall and are very competitive with respect to the baseline approaches. It is important to highlight that when $\mu = 0.6$, each node has more neighbors in other communities than in its own community, thus communities are no longer well-defined. This is the reason why all the methods struggle to find a good solution in that case.

\clearpage

\subsection{Real World Networks}
\label{real}

We consider five real-world datasets frequently used for evaluation in multilayer graph clustering, \cite{mercado2018}:
\begin{itemize}[noitemsep]
    \item \textit{3sources} is a text dataset of articles from three online news sources (BBC, Reuters, and The Guardian), one for each layer, which have been manually assigned to one of six topical labels: business, entertainment, health, politics, sport and technology~\cite{BBCSports,3sources}. 
    \item \textit{BBCSport} is a news dataset of sports articles with five annotated topic labels. The two layers are created splitting each  document into segments and assigning them randomly to layers~\cite{BBCSports}. 
    \item \textit{Cora} is a citation dataset of research papers labeled with seven classes. The first layer is a citation network, whereas the second one is 
    built on documents features~\cite{mccallum2000}.
    \item \textit{UCI} is a dataset of features of handwritten digits (0-9). These digits are represented in terms of six different feature sets, forming the layers:
    Fourier coefficients of the character shapes, profile correlations, Karhunen-Love coefficients, pixel averages, Zernike moments and morphological features~\cite{Dua:2019, 3sources}. 
    \item \textit{Wikipedia} is a dataset of Wikipedia articles classified in ten categories: art \& architecture, biology, geography, history, literature \& theatre, media, music, royalty \& nobility, sport \& recreation, Warfare. These categories are assigned to both the text and image components of each article, corresponding to the two layers \cite{rasiwasia2010}. 
\end{itemize}
All the layers built from feature sets are formed by means of a symmetrized  $k$-nearest neighbor graph with $k = 10$, based on the Pearson linear correlation between nodes, i.e., the higher the correlation the smaller the nodes' distance. Thus, if $N_k(u)$ denotes the set of $k$ nodes that have highest correlation with node $u$, to each node $u$ we connect all nodes in the set 
\[
N_k(u)\cup \{v : u\in  N_k(v)\}.
\]
The main properties  of the various multilayer networks are reported in Table~\ref{tab:info-real-datasets}.

\setlength\tabcolsep{2pt}
\begin{table*}[t]
\caption{Basic statistics for the real-world datasets. For each dataset it shows the number of nodes $N$,  the number of layers $k$, the number of communities $c$, the size of each community and, for each layer, the number of edges $|E|$, the edge density $\delta$ and the average  and  standard deviation of the nodes' degrees, $\langle \text{deg}\rangle$ and $\sigma$, respectively.}
\label{tab:info-real-datasets}
\centering 

\setlength\tabcolsep{2pt}
\def \fns {\footnotesize}

\begin{tabular}{l cccc cccc cccc cccc cccc}
\toprule
&\multicolumn{4}{c}{\textbf{3sources}}&\multicolumn{4}{c}{\textbf{BBCSport}}&\multicolumn{4}{c}{\textbf{cora}} &\multicolumn{4}{c}{\textbf{UCI}}&\multicolumn{4}{c}{\textbf{Wikipedia}}\\
\midrule
$N$ & \multicolumn{4}{c}{\fns 169} & \multicolumn{4}{c}{\fns 544} & \multicolumn{4}{c}{\fns 2708} & \multicolumn{4}{c}{\fns 2000} & \multicolumn{4}{c}{\fns 693} \\
$k$ & \multicolumn{4}{c}{\fns 3} & \multicolumn{4}{c}{\fns 2} & \multicolumn{4}{c}{\fns 2} & \multicolumn{4}{c}{\fns 6} & \multicolumn{4}{c}{\fns 2}\\
$c$ & \multicolumn{4}{c}{\fns 6} & \multicolumn{4}{c}{\fns 5} & \multicolumn{4}{c}{\fns 7} & \multicolumn{4}{c}{\fns 10} & \multicolumn{4}{c}{\fns 10}\\
$|C_i|$ & \multicolumn{4}{c}{\begin{tabular}{c}\fns 56, 21, 11,\\\fns 18, 51, 12\end{tabular}} & \multicolumn{4}{c}{\begin{tabular}{c}\fns 62, 104, 193,\\\fns 124, 61\end{tabular}} & \multicolumn{4}{c}{\begin{tabular}{c}\fns 298, \fns 418, 818,  426, \\ \fns 217, 180, 351\end{tabular}} & \multicolumn{4}{c}{\fns 200 each} & \multicolumn{4}{c}{\begin{tabular}{c}\fns 34, \fns 88,  96,  85,  65, \\ \fns 58, 51, 41, 71, 104\end{tabular}} \\
\cmidrule(lr){2-5} \cmidrule(lr){6-9} \cmidrule(lr){10-13} \cmidrule(lr){14-17}  \cmidrule(lr){18-21} 
& \fns $|E|$ & \fns $\delta$ & \fns $\langle \text{deg} \rangle$ & \fns $\sigma$ &
  \fns $|E|$ & \fns $\delta$ & \fns $\langle \text{deg} \rangle$ & \fns $\sigma$ &
  \fns $|E|$ & \fns $\delta$ & \fns $\langle \text{deg} \rangle$ & \fns $\sigma$ &
  \fns $|E|$ & \fns $\delta$ & \fns $\langle \text{deg} \rangle$ & \fns $\sigma$ &
  \fns $|E|$ & \fns $\delta$ & \fns $\langle \text{deg} \rangle$ & \fns $\sigma$ \\
\cmidrule(lr){2-5} \cmidrule(lr){6-9} \cmidrule(lr){10-13} \cmidrule(lr){14-17}  \cmidrule(lr){18-21} 
L1 & \fns 1168 & \fns 0.04 &\fns 12.82 &\fns 3.37 &\fns 4075 &\fns 0.01 &\fns 13.98 &\fns 5.22 &\fns 5278 &\fns 7e-4 &\fns 3.90 &\fns 5.23 & \fns 14447 & \fns 3e-3& \fns 13.45& \fns 3.54 & \fns5606 & \fns 0.01 & \fns 15.18 & \fns 5.37\\
L2 & \fns 1223& \fns 0.04& \fns 13.47 & \fns 4.55  & \fns 4127 & \fns 0.01 & \fns 14.17 & \fns 6.42  & \fns 21273 & \fns 3e-3 & \fns 14.71 & \fns 9.52  & \fns 14600 & \fns 3e-3 & \fns 13.60 & \fns 3.7  & \fns 5385 & \fns 0.01 & \fns 28.83 & \fns 5.37\\
L3  & \fns 1272 & \fns 0.04 & \fns 14.05 & \fns 5.13  & \fns - & \fns - & \fns - & \fns -  & \fns - & \fns - & \fns - & \fns -  & \fns 14498 & \fns 3e-3 & \fns 13.50 & \fns 3.48  & \fns - & \fns - & \fns - & \fns -\\
L4  & \fns - & \fns - & \fns - & \fns -  & \fns - & \fns - & \fns - & \fns -  & \fns - & \fns - & \fns - & \fns - & \fns 12729 & \fns 3e-3 & \fns 11.73 & \fns 1.67  & \fns - & \fns - & \fns - & \fns -\\
L5  & \fns - & \fns - & \fns - & \fns -  & \fns - & \fns - & \fns - & \fns -  & \fns - & \fns - & \fns - & \fns - & \fns 14561 & \fns 3e-3 & \fns 13.56 & \fns 3.73  & \fns - & \fns - & \fns - & \fns -\\
L6  & \fns - & \fns - & \fns - & \fns -  & \fns - & \fns - & \fns - & \fns -  & \fns - & \fns - & \fns - & \fns - & \fns 14421 & \fns 3e-3 & \fns 13.42 & \fns 3.43  & \fns - & \fns - & \fns - & \fns -\\
\bottomrule
\end{tabular}

\end{table*}
 
\setlength\tabcolsep{10pt}

Given the ground truth community structure of the graphs, we analyzed both the informative and the noisy case.
For the noisy case, we study two settings. In the first one, we added a noisy layer to the informative layers, whereas in the second setting we considered networks with 2 layers, where the first one is the graph obtained aggregating all the layers and the other one is just noise.
The noisy layers are generated via uniform (Erd\H{o}s–R\'enyi) random graphs with edge probability $p \in \{0.01, 0.03, 0.05\}$. For each value of $p$, and each dataset, we generated 10 random instances and for each instance, we run our methods 10 times with different random initial community orderings. 

In  Tables \ref{tab:1}--\ref{tab:3} we  report the average accuracy and NMI scores over the samples and the random initializations. The best and second best values are highlighted with a gray box, with the best values having a bold font.  
We further consider the average performance ratio score values $\rho_{\mathrm{Acc}}$ and $\rho_{\mathrm{NMI}}$, quantified as follows: for a given measure $M_{a,d}$
(where $M_{a,d}$ is either accuracy or NMI obtained by algorithm $a$ over the dataset~$d$),
the \emph{performance ratio} is  $r_{a,d}=M_{a,d}/\max\{M_{a,d}\text{ over all }   a\}$. The average performance ratios of each algorithm 
$\rho_{\mathrm{Acc}}$ and $\rho_{\mathrm{NMI}}$ (for accuracy and NMI, respectively) are then obtained averaging $r_{a,d}$ over all the datasets $d$.
For any algorithm, the closer the average performance ratio to 1, the better the overall performance.

We can see that in many cases the proposed methods overcome the baselines. In particular, the methods that consider the variance in addition to the average  of the modularity across the layers usually work better, with the multiobjective approaches \MVM\ and \MVP\ achieving the best results in almost all cases. This is in agreement with the more sophisticated Pareto-based strategy and is consistent with what is observed in the experiments with synthetic data. Moreover, the last two tables highlight the robustness of the proposed methods with respect to noise. In fact, in the presence of noisy layers, the proposed methods --- in particular the multiobjective ones ---  achieve very high performance ratios as well as overall high values of accuracy and NMI.

\begin{figure*}[!t]
\captionsetup[subfigure]{labelformat=empty}
\centering
\captionsetup{oneside,margin={-0.53cm,0.6cm}}
    \hspace*{-2cm}
    \subfloat[]
    {
\pgfplotstableread{
Methods noisy_layers time
EVP/M 0 0.048
EVP/M 1 0.064
EVP/M 2 0.077
EVP/M 3 0.082
EVP/M 4 0.089
EVP/M 5 0.100
}{\EVPM}

\pgfplotstableread{
Methods noisy_layers time
MA2 0 0.296
MA2 1 0.078
MA2 2 0.992
MA2 3 1.127
MA2 4 1.365
MA2 5 1.685
}{\MA}

\pgfplotstableread{
Methods noisy_layers time
MA3 0 0.450
MA3 1 0.257
MA3 2 1.534
MA3 3 1.907
MA3 4 2.351
MA3 5 3.036
}{\MAt}

\pgfplotstableread{
Methods noisy_layers time
MVP/M2 0 0.354
MVP/M2 1 0.667
MVP/M2 2 0.705
MVP/M2 3 0.755
MVP/M2 4 0.852
MVP/M2 5 0.963
}{\MVPM}

\pgfplotstableread{
Methods noisy_layers time 
MVP/M3 0 0.602
MVP/M3 1 1.020
MVP/M3 2 1.139
MVP/M3 3 1.217
MVP/M3 4 1.297
MVP/M3 5 1.552
}{\MVPMt}

\pgfplotstableread{
Methods noisy_layers time 
GL 0 0.062
GL 1 0.075
GL 2 0.119
GL 3 0.129
GL 4 0.143
GL 5 0.169
}{\GL}

\pgfplotstableread{
Methods noisy_layers time
CoReg 0 0.078
CoReg 1 0.098
CoReg 2 0.111
CoReg 3 0.163
CoReg 4 0.305
CoReg 5 0.295
}{\CoReg}

\pgfplotstableread{
Methods noisy_layers time 
AWP 0 0.017
AWP 1 0.038
AWP 2 0.076
AWP 3 0.106
AWP 4 0.146
AWP 5 0.185
}{\AWP}

\pgfplotstableread{
Methods noisy_layers time
MCGC 0 0.142
MCGC 1 0.106
MCGC 2 0.151
MCGC 3 0.158
MCGC 4 0.187
MCGC 5 0.230
}{\MCGC}

\pgfplotstableread{
Methods noisy_layers time
PM 0 0.053
PM 1 0.024
PM 2 0.032
PM 3 0.033
PM 4 0.037
PM 5 0.044
}{\PM}

\pgfplotstableread{
Methods noisy_layers time
MT 0 8.592
MT 1 7.692
MT 2 14.253
MT 3 16.862
MT 4 22.442
MT 5 28.767
}{\MT}

\pgfplotstableread{
Methods noisy_layers time 
SCML 0 0.488
SCML 1 0.416
SCML 2 0.491
SCML 3 0.495
SCML 4 0.558
SCML 5 0.672
}{\SCML}

\pgfplotstableread{
Methods noisy_layers time 
PMM 0 0.029
PMM 1 0.019
PMM 2 0.028
PMM 3 0.031
PMM 4 0.037
PMM 5 0.047
}{\PMM}

\pgfplotstableread{
Methods noisy_layers time 
IM 0 0.845
IM 1 0.738
IM 2 0.991
IM 3 0.948
IM 4 0.971
IM 5 1.056
}{\IM}

\begin{tikzpicture}
    \begin{groupplot}[
        group style={
        group name=3sources_time, 
        group size= 2 by 1,
        horizontal sep=50pt
        },
        width=6.3cm,
        height=6.3cm,
        enlarge y limits,
        axis y line*=left,
        axis x line*=bottom, 
        xtick=data,
        ymajorgrids,
		grid style={line width=.2pt,draw=gray!20},
        y axis line style={draw=none},
        xtick style={draw=none},
		ytick style={draw=none},
        yticklabel style={xshift=-0.5em},
        legend style={
        at={(1.1,-0.30)},
        legend columns=8,
        legend cell align=left,
        anchor=north},
        legend image post style={mark options={scale=1}}
    ]
     
     \nextgroupplot[title=(a), xlabel={added noisy layers}, ylabel={Time}]
     
      \addplot [line width=0.05mm, 
     mark=o, color=red, mark size=1.5, mark options={fill=red!20!red,mark indices=6}, 
     ] table[x=noisy_layers,y=time]{\EVPM}; \addlegendentry{EVM/P};
     
     \addplot [line width=0.05mm, 
     mark=square, color=blue, mark size=1.5, mark options={fill=blue!20!blue,mark indices=6}, 
     ] table[x=noisy_layers,y=time]{\MA}; \addlegendentry{MA2};
     
      \addplot [line width=0.05mm, 
     mark=otimes, color=brown, mark size=1.5, mark options={fill=brown!20!brown,mark indices=6}, 
     ] table[x=noisy_layers,y=time]{\MAt}; \addlegendentry{MA3};
     
       \addplot [line width=0.05mm, 
     mark=triangle, color=cyan, mark size=1.5, mark options={fill=cyan!20!cyan,mark indices=6}, 
     ] table[x=noisy_layers,y=time]{\MVPM}; \addlegendentry{MVM/P2};
     
       \addplot [line width=0.05mm, 
     mark=diamond, color=violet, mark size=1.5, mark options={fill=violet!20!violet,mark indices=6}, 
     ] table[x=noisy_layers,y=time]{\MVPMt}; \addlegendentry{MVM/P3};
     
      \addlegendimage{empty legend}
            \addlegendentry{}
            
     \addlegendimage{empty legend}
            \addlegendentry{}
            
     \addlegendimage{empty legend}
            \addlegendentry{}
            
       \addplot [line width=0.05mm, 
     mark=star, color=green, mark size=1.5, mark options={fill=green!20!green,mark indices=6}, 
     ] table[x=noisy_layers,y=time]{\GL}; \addlegendentry{GL};
     
       \addplot [line width=0.05mm, 
     mark=x, color=orange, mark size=1.5, mark options={fill=orange!20!orange,mark indices=6}, 
     ] table[x=noisy_layers,y=time]{\CoReg}; \addlegendentry{CoReg};
     
       \addplot [line width=0.05mm, 
     mark=10-pointed star, color=teal, mark size=1.5, mark options={fill=teal!20!teal,mark indices=6}, 
     ] table[x=noisy_layers,y=time]{\AWP};  \addlegendentry{AWP};
     
       \addplot [line width=0.05mm, 
     mark=oplus, color=yellow, mark size=1.5, mark options={fill=yellow!20!yellow,mark indices=6}, 
     ] table[x=noisy_layers,y=time]{\MCGC}; \addlegendentry{MCGC};
     
       \addplot [line width=0.05mm,
     mark=Mercedes star, color=magenta, mark size=1.5, mark options={fill=magenta!20!magenta,mark indices=6}, 
     ] table[x=noisy_layers,y=time]{\PM}; \addlegendentry{PM};
     
       \addplot [line width=0.05mm,
     mark=pentagon, color=black, mark size=1.5, mark options={fill=black!20!black,mark indices=6}, 
     ] table[x=noisy_layers,y=time]{\MT}; \addlegendentry{MT};
     
       \addplot [line width=0.05mm,
     mark=|, color=pink, mark size=1.5, mark options={fill=pink!20!pink,mark indices=6}, 
     ] table[x=noisy_layers,y=time]{\SCML}; \addlegendentry{SCML};
     
       \addplot [line width=0.05mm, 
     mark=asterisk, color=olive, mark size=1.5, mark options={fill=olive!20!olive,mark indices=6}, 
     ] table[x=noisy_layers,y=time]{\PMM}; \addlegendentry{PMM};

     
     \nextgroupplot[title=(b), xlabel={added noisy layers}, ylabel={Time}]

      \addplot [line width=0.05mm, 
     mark=o, color=red, mark size=1.5, mark options={fill=red!20!red,mark indices=6}, 
     ] table[x=noisy_layers,y=time]{\EVPM}; 
     
     \addplot [line width=0.05mm, 
     mark=square, color=blue, mark size=1.5, mark options={fill=blue!20!blue,mark indices=6}, 
     ] table[x=noisy_layers,y=time]{\MA}; 
     
      \addplot [line width=0.05mm, 
     mark=otimes, color=brown, mark size=1.5, mark options={fill=brown!20!brown,mark indices=6}, 
     ] table[x=noisy_layers,y=time]{\MAt}; 
     
       \addplot [line width=0.05mm, 
     mark=triangle, color=cyan, mark size=1.5, mark options={fill=cyan!20!cyan,mark indices=6}, 
     ] table[x=noisy_layers,y=time]{\MVPM}; 
     
       \addplot [line width=0.05mm, 
     mark=diamond, color=violet, mark size=1.5, mark options={fill=violet!20!violet,mark indices=6}, 
     ] table[x=noisy_layers,y=time]{\MVPMt}; 
            
       \addplot [line width=0.05mm, 
     mark=star, color=green, mark size=1.5, mark options={fill=green!20!green,mark indices=6}, 
     ] table[x=noisy_layers,y=time]{\GL};
     
       \addplot [line width=0.05mm, 
     mark=x, color=orange, mark size=1.5, mark options={fill=orange!20!orange,mark indices=6}, 
     ] table[x=noisy_layers,y=time]{\CoReg}; 
     
       \addplot [line width=0.05mm, 
     mark=10-pointed star, color=teal, mark size=1.5, mark options={fill=teal!20!teal,mark indices=6}, 
     ] table[x=noisy_layers,y=time]{\AWP}; 
     
       \addplot [line width=0.05mm, 
     mark=oplus, color=yellow, mark size=1.5, mark options={fill=yellow!20!yellow,mark indices=6}, 
     ] table[x=noisy_layers,y=time]{\MCGC}; 
     
       \addplot [line width=0.05mm,
     mark=Mercedes star, color=magenta, mark size=1.5, mark options={fill=magenta!20!magenta,mark indices=6}, 
     ] table[x=noisy_layers,y=time]{\PM};
     
       \addplot [line width=0.05mm,
     mark=|, color=pink, mark size=1.5, mark options={fill=pink!20!pink,mark indices=6}, 
     ] table[x=noisy_layers,y=time]{\SCML}; 
     
       \addplot [line width=0.05mm, 
     mark=asterisk, color=olive, mark size=1.5, mark options={fill=olive!20!olive,mark indices=6}, 
     ] table[x=noisy_layers,y=time]{\PMM}; 

    \end{groupplot}
\end{tikzpicture}}
\caption{Computational time of different methods on the 3-sources dataset, with an increasing number of noisy layers (from 0 to 5).  In (a) all methods' runtimes are shown, in (b) excluding MT.}
\label{fig:3sources_time}
\end{figure*}
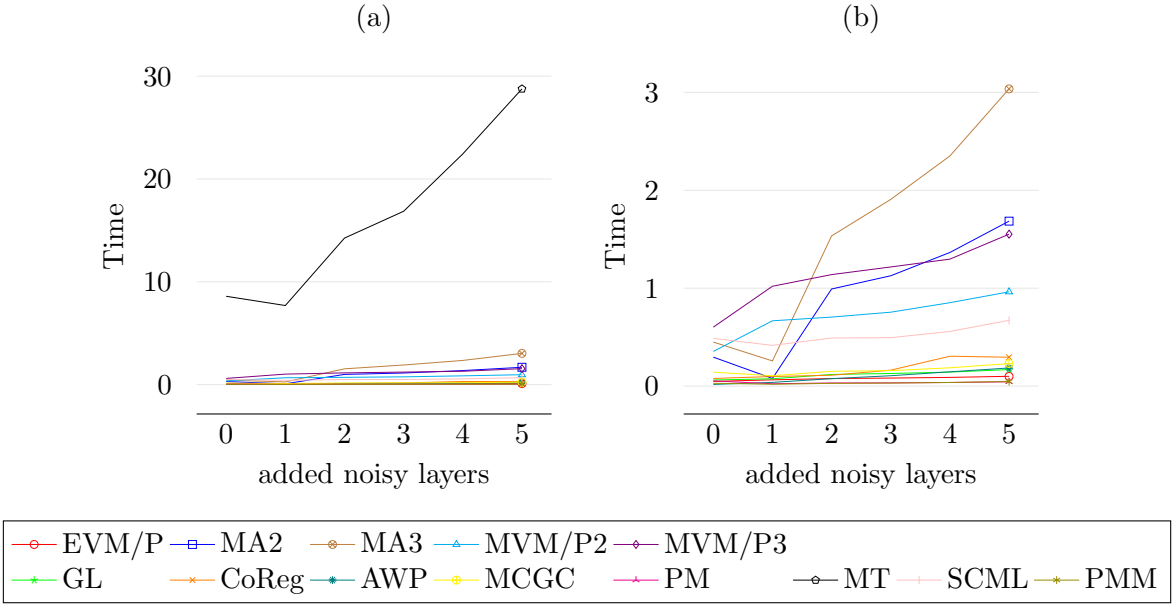

Finally, we aim to analyze the behavior of the methods subject to the addition of a larger number of noisy layers. To this end, we compare the performance of the different methods on  the \textit{3sources} dataset adding to it 
up to 5 noisy layers. 
In Table \ref{tab:3sources}, we report the average accuracy and NMI scores over the samples and the random initializations for the noisy cases. For the sake of comparison, we also report the values for the informative case, studied in Table \ref{tab:1}, which correspond to the addition of 0 noisy layers. 
We can see that, in all cases, the proposed methods overcome the baselines. In particular, the multiobjective approaches MVM and MVP achieve the best results in almost all cases. Note that we use MVM for the informative setting (as in Table \ref{tab:1}) and MVP for all the settings with a different number of noisy layers. This is because we want to enforce a larger variance across the layers in each of those cases.
It is interesting to notice that, unlike the competing baselines, the proposed multiobjective MVP approaches are almost insensitive to the number of noisy layers. The accuracy, for these methods, seems to be only marginally affected when moving from 1 to 5 layers of noise and, in some cases, more noise seems to yield better accuracy. This is probably due to the ability of the methods to exploit a high modularity variance as beneficial, rather than being negatively affected by it.\\ 
In Figure \ref{fig:3sources_time}, we report the computational time in seconds. In \ref{fig:3sources_time}(a) we report the computational time for all the methods, whereas in \ref{fig:3sources_time}(b) we remove the values for the most time-consuming method (MT), to have a clearer comparison. 
We can see the proposed methods are comparable to the baselines in terms of time efficiency.

\section{Conclusions}
\label{Conclusions}
In this paper, we presented a new method for community detection in multiplex graphs that  extends  the Louvain heuristic method by introducing a variance-aware quality function and by performing a vector-valued modularity ascending scheme based on a tailored Pareto  search.

We considered different versions of this method to better analyze two situations: the informative case, where each layer shows the same community structure, and the noisy case, where some layers present a community structure and all the others contain only noise. 
We provided extensive experiments comparing  with nine baselines borrowed from both the network science and the machine learning communities. We tested the performance of the proposed methods on synthetic networks, using the LFR and the stochastic block models, as well as five real-world multilayer datasets (i.e., 3sources, BBCSport, cora, UCI, Wikipedia). In both cases, we studied informative and noisy settings. 
The experimental results demonstrate that the proposed method is competitive with the baselines. In particular, the multiobjective approach combined with the modularity variance shows the best performance in almost all cases.

\def\first#1{\fcolorbox{black}{black!15}{\textbf{#1}}}
\def\second#1{\fcolorbox{black!50}{black!5}{#1}}
\renewcommand{\arraystretch}{1}

\setlength\tabcolsep{2pt}
\begin{table*}[h!]
\caption{Real-world dataset setting one: no noisy layers. Average accuracy, NMI and performance ratio scores  over 10 random initializations. All layers are informative. Best and second best values are highlighted with gray boxes.}
\label{tab:1}
\centering 
\resizebox{\textwidth}{!}{
\begin{tabular}{l cc cc cc cc cc | cc}

\toprule
&\multicolumn{2}{c}{\textbf{3sources}}&\multicolumn{2}{c}{\textbf{BBCSport}}&\multicolumn{2}{c}{\textbf{cora}}&\multicolumn{2}{c}{\textbf{UCI}}&\multicolumn{2}{c}{\textbf{Wikipedia}}&\multicolumn{2}{c}{\textbf{Perf. Ratios}}\\
\cmidrule(lr){2-3} \cmidrule(lr){4-5} \cmidrule(lr){6-7} \cmidrule(lr){8-9} \cmidrule(lr){10-11} \cmidrule(lr){12-13}
      & Acc      & NMI   & Acc      & NMI   & Acc   & NMI   & Acc   & NMI   & Acc       & NMI   &  $\rho_{\mathrm{Acc}}$ & $\rho_{\mathrm{NMI}}$\\
\midrule 
EVM   & \second{0.876} & \second{0.789} & 0.833 & 0.798 & \first{0.617} & \first{0.537} & 0.882 & \second{0.921} & 0.548 & 0.520 & 0.951 & \first{0.966} \\
MA2    & 0.858 & 0.749 & 0.899 & \second{0.825} & 0.407 & 0.434 & 0.753 & 0.862 & 0.525 & 0.521 &0.863&0.912\\
MA3   & \second{0.876} & \second{0.789} & 0.596 & 0.731 & 0.425 & 0.428 & 0.876 & 0.910 & 0.558 & \second{0.546} & 0.838 & 0.920\\
MVM2  & \first{0.888} & \first{0.812} & 0.844 & 0.784 & 0.597 & 0.514 & 0.883 & \first{0.925} & 0.544 & 0.508 & \second{0.952} & 0.959\\
MVM3 & \first{0.888} & \first{0.812} & \second{0.915} & \first{0.851} & 0.603 & 0.502 & 0.883 & \first{0.925} & 0.530 & 0.504 & \first{0.966} & \second{0.965} \\
\hline
GL    & 0.858 & 0.749 & 0.748 & 0.753 & 0.523 & \second{0.520} & 0.877 & 0.913 & 0.556 & 0.544 & 0.904 & 0.947\\
CoReg & 0.651 & 0.658 & 0.858 & 0.617 & 0.530 & 0.380 & \first{0.958} & 0.911 & 0.522 & 0.445 & 0.905 & 0.840\\
AWP   & 0.686 & 0.662 & 0.616 & 0.722 & 0.534 & 0.293 & 0.869 & 0.891 & 0.462 & 0.332 & 0.843 & 0.758\\
MCGC  & 0.544    & 0.595 & \first{0.919}    & 0.795 & 0.273 & 0.034 & \second{0.898} & 0.855 & 0.221     & 0.135 & 0.676 & 0.580\\
PM    & 0.734 & 0.707 & 0.778 & 0.690 & 0.551 & 0.456 & 0.876 & 0.879 &\first{0.569} & \first{0.560} & 0.896 & 0.892\\
MT   & 0.651 & 0.610  & 0.748 & 0.656 & 0.453 & 0.289 & 0.553 & 0.666 & 0.342 & 0.229 & 0.692 & 0.638\\
SCML   & 0.686 & 0.661 & 0.864 & 0.767 & \second{0.616} & 0.447 & 0.862 & 0.872 & \second{0.560} & 0.535 & 0.919 & 0.889\\
PMM   & 0.692 & 0.666 & 0.518 & 0.514 & 0.336 & 0.238 & 0.638 & 0.662 & 0.417& 0.302 & 0.658 & 0.625\\
IM   & 0.539 & 0.624 & 0.531 & 0.401 & 0.431 & 0.477 & 0.721 & 0.761 & 0.123 & 0.11 & 0.570 & 0.630\\
\bottomrule
\end{tabular}
}
\end{table*}

\begin{table*}[h!]
\caption{Real-world dataset setting two: informative layers plus one noisy layer. Average accuracy, NMI and performance ratio scores  over 10 random initializations and 10 random edge probabilities $p \in \{0.01,0.03,0.05\}$ for the noisy layer. Best and second best values are highlighted with gray boxes.}
\label{tab:2}
\centering 
\resizebox{\textwidth}{!}{
\begin{tabular}{l cc cc cc cc cc | cc}
\toprule
&\multicolumn{2}{c}{\textbf{3sources}}&\multicolumn{2}{c}{\textbf{BBCSport}}&\multicolumn{2}{c}{\textbf{cora}}&\multicolumn{2}{c}{\textbf{UCI}}&\multicolumn{2}{c}{\textbf{Wikipedia}}&\multicolumn{2}{c}{\textbf{Perf. Ratios}}\\
\cmidrule(lr){2-3} \cmidrule(lr){4-5} \cmidrule(lr){6-7} \cmidrule(lr){8-9} \cmidrule(lr){10-11} \cmidrule(lr){12-13}
      & Acc      & NMI   & Acc      & NMI   & Acc   & NMI   & Acc   & NMI   & Acc       & NMI   &   $\rho_{\mathrm{Acc}}$&     $\rho_{\mathrm{NMI}}$\\
\midrule 
EVP   & 0.703    & 0.649 & 0.825    & \first{0.797} & 0.541 & 0.517 & 0.880 & 0.916 & 0.577     & \first{0.556}&0.964& 0.988\\
MA2   & 0.692    & 0.609 & 0.790    & 0.761 & \second{0.551} & 0.519 & 0.881 & 0.920 & 0.558     & 0.518 &0.950&0.955\\
MA3   & 0.683    & 0.612 & 0.789    & 0.758 & 0.549 & 0.519 & 0.881 & \second{0.921} & 0.559     & 0.518 &0.947&0.955\\
MVP2  & \second{0.717}   & \second{0.668} & \second{0.828}    & \first{0.797} & 0.543 & 0.518 & 0.881 & 0.920 & \second{0.578}     & \second{0.555}&\second{0.970}&\second{0.994}\\
MVP3  & \first{0.730}    & \first{0.677} & 0.817    & \second{0.792} & 0.543 & \second{0.520} & 0.881 & 0.919 & \first{0.579}     & \first{0.556} &\first{0.971}&\first{0.996}\\
\hline
GL    & 0.678    & 0.607 & 0.777    & 0.754 & \first{0.555} & \first{0.521} & 0.881 & 0.920 & 0.561     & 0.520&0.945&0.953 \\
CoReg & 0.652    & 0.650 & \first{0.849}    & 0.753 & 0.407 & 0.191 & \second{0.957} & 0.912 & 0.435     & 0.324 &0.874&0.767\\
AWP   & 0.658    & 0.602 & 0.737    & 0.593 & 0.411 & 0.123 & 0.924 & 0.897 & 0.410     & 0.266 &0.835&0.662\\
MCGC  & 0.546    & 0.585 & 0.812    & 0.694 & 0.303 & 0.005 & 0.804 & 0.816 & 0.201     & 0.107 &0.686&0.563\\
PM    & 0.714    & 0.658 & 0.730    & 0.645 & 0.548 & 0.444 & 0.876 & 0.880 & 0.568     & \first{0.556}&0.943&0.916\\
MT    & 0.624 & 0.627 & 0.519 & 0.355 & 0.187 & 0.011 & 0.660 & 0.723  & 0.170  & 0.051 & 0.556 & 0.453 \\
SCML     & 0.639 & 0.593 & 0.772 & 0.604 & 0.222 & 0.028 & \first{0.964} & \first{0.930} & 0.185 & 0.057 & 0.701 & 0.558 \\
PMM    & 0.538 & 0.508 & 0.387 & 0.167 & 0.255 & 0.060 & 0.667 & 0.677  & 0.172  & 0.047 & 0.528 & 0.377 \\
IM    & 0.538 & 0.624 & 0.531 & 0.401 & 0.431 & 0.477 & 0.721 & 0.761  & 0.123  & 0.110 & 0.620 & 0.671 \\
\bottomrule
\end{tabular}
}
\end{table*}

\begin{table*}[h!]
\caption{Real-world dataset setting three: one aggregated informative layer plus one noisy layer. Average accuracy, NMI, and performance ratio scores  over 10 random initializations and 10 random edge probabilities $p \in \{0.01,0.03,0.05\}$ for the noisy layer. Best and second best values are highlighted with gray boxes.}
\label{tab:3}
\centering 
\resizebox{\textwidth}{!}{
\begin{tabular}{l cc cc cc cc cc | cc}
\toprule
&\multicolumn{2}{c}{\textbf{3sources}}&\multicolumn{2}{c}{\textbf{BBCSport}}&\multicolumn{2}{c}{\textbf{cora}}&\multicolumn{2}{c}{\textbf{UCI}}&\multicolumn{2}{c}{\textbf{Wikipedia}}&\multicolumn{2}{c}{\textbf{Perf. Ratios}}\\
\cmidrule(lr){2-3} \cmidrule(lr){4-5} \cmidrule(lr){6-7} \cmidrule(lr){8-9} \cmidrule(lr){10-11} \cmidrule(lr){12-13}
      & Acc      & NMI   & Acc      & NMI   & Acc   & NMI   & Acc   & NMI   & Acc       & NMI  &$\rho_{\mathrm{Acc}}$ & $\rho_{\mathrm{NMI}}$ \\
\midrule 
EVP   & 0.655    & 0.575 & \second{0.886}    & 0.799 & 0.550 & \first{0.432} & 0.869 & 0.903 & 0.556     & 0.526 &\second{0.938}&0.943\\
MA2  & 0.348    & 0.217 & 0.664    & 0.555 & 0.541 & 0.418 & 0.853 & 0.890 & 0.431     & 0.361 &0.761&0.717\\
MA3    & 0.332    & 0.207 & 0.659    & 0.550 & 0.537 & 0.416 & 0.858 & 0.893 & 0.427     & 0.359 &0.754&0.711\\
MVP2  & 0.744    & 0.675 & \first{0.914}    & \second{0.826} & 0.546 & 0.430 & \second{0.872} & \second{0.905} & \second{0.564}     & \second{0.538} &\first{0.969}&\second{0.980}\\
MVP3  & \second{0.754}    & \second{0.689} & \first{0.914}    & \first{0.828} & 0.544 & \second{0.431} & \first{0.873} & \first{0.906} & \first{0.566}     & \first{0.540} &\first{0.969}&\first{0.984}\\
\hline
GL    & 0.327    & 0.203 & 0.664    & 0.549 & 0.537 & 0.418 & 0.866 & 0.898 & 0.427     & 0.360 &0.756&0.713\\
CoReg & 0.566    & 0.436 & 0.608    & 0.338 & 0.435 & 0.190 & 0.761 & 0.642 & 0.446     & 0.305 &0.753&0.542\\
AWP   & 0.549    & 0.416 & 0.644    & 0.376 & 0.444 & 0.179 & 0.750 & 0.622 & 0.439 & 0.292 &0.754&0.531\\
MCGC  & 0.512    & 0.479 & 0.682    & 0.480 & 0.276 & 0.020 & 0.702 & 0.691 & 0.347     & 0.291 & 0.654&0.514\\
PM    & 0.512    & 0.363 & 0.729    & 0.658 & \second{0.569} & 0.414 & 0.834 & 0.828 & 0.457     & 0.360 & 0.831 & 0.764 \\
MT  & 0.693 & 0.614 & 0.729 & 0.614 & 0.272 & 0.113 & 0.634 & 0.699 & 0.406 & 0.306 &0.714&0.534   \\
SCML    & 0.514 & 0.410 & 0.797 & 0.661 & \first{0.600} & 0.416 & 0.846 & 0.835  & 0.479  & 0.365 & 0.867 & 0.783 \\
PMM    & 0.440 & 0.304 & 0.424 & 0.206 & 0.311 & 0.108 &0.641 & 0.535  & 0.386  & 0.249 & 0.591 & 0.392 \\
IM   & \first{0.793} & \first{0.742} & 0.730 & 0.751 & 0.323 & 0.376 & 0.199 & 0.338  & 0.496  & 0.532 & 0.687 & 0.827 \\
\bottomrule
\end{tabular}
}
\end{table*}

\begin{table*}[h!]
\centering
\caption{Real-world dataset setting four: 3-sources dataset with an increasing number of noisy layers (from 0 to 5). Average accuracy and NMI over 10 random initializations (for the noisy cases) and 10 random edge probabilities $p \in \{0.01,0.03,0.05\}$ for the noisy layers. Best and second best values are highlighted with gray boxes.
}
\label{tab:3sources}
\centering 
\setlength\tabcolsep{0pt}
\def \fns {\tiny}

\begin{tabular}{l cc cc cc cc cc cc cc cc}
\toprule
&\multicolumn{2}{c}{\textbf{0}}&\multicolumn{2}{c}{\textbf{1}}&\multicolumn{2}{c}{\textbf{2}}&\multicolumn{2}{c}{\textbf{3}}&\multicolumn{2}{c}{\textbf{4}}&\multicolumn{2}{c}{\textbf{5}}\\

\cmidrule(lr){2-3} \cmidrule(lr){4-5} \cmidrule(lr){6-7} \cmidrule(lr){8-9} \cmidrule(lr){10-11} \cmidrule(lr){12-13}
      & Acc      & NMI   & Acc      & NMI   & Acc   & NMI & Acc   & NMI & Acc   & NMI & Acc   & NMI  \\
\midrule 
EVM/P & \second{0.876} &\second{0.789}& 0.703 & 0.649 & 0.729 & 0.677 & 0.735 & 0.688 & 0.767 & 0.708 & 0.777 & 0.717 \\ 
MA2 & 0.858 &0.749& 0.692 & 0.609 & 0.615 & 0.538 & 0.577 & 0.495 & 0.526 & 0.442 & 0.503 & 0.426 \\ 
MA3 & \second{0.876} &\second{0.789}& 0.683 & 0.612 & 0.602 & 0.529 & 0.576 & 0.494 & 0.535 & 0.449 & 0.503 & 0.420 \\ 
MVM/P2 & \first{0.888} &\first{0.812}& \second{0.717} & \second{0.668} & \second{0.748} & \first{0.693} & \first{0.752} & \first{0.700} & \second{0.778} & \second{0.718} & \second{0.787} & \first{0.721} \\ 
MVM/P3 & \first{0.888} &\first{0.812}& \first{0.730} & \first{0.677} & \first{0.751} & \second{0.692} & \second{0.738} & \second{0.699} & \first{0.779} & \first{0.722} & \first{0.790} & \second{0.720} \\ \hline
GL & 0.858 &0.749& 0.678 & 0.607 & 0.615 & 0.541 & 0.571 & 0.492 & 0.508 & 0.443 & 0.506 & 0.424 \\ 
CoReg & 0.651 &0.658& 0.652 & 0.650 & 0.650 & 0.645 & 0.654 & 0.645 & 0.650 & 0.637 & 0.645 & 0.631 \\ 
AWP & 0.686 &0.662& 0.658 & 0.602 & 0.664 & 0.580 & 0.632 & 0.539 & 0.622 & 0.521 & 0.599 & 0.462 \\ 
MCGC & 0.544 &0.595& 0.546 & 0.585 & 0.541 & 0.577 & 0.550 & 0.575 & 0.538 & 0.551 & 0.486 & 0.503 \\ 
PM & 0.734 &0.707& 0.714 & 0.658 & 0.671 & 0.609 & 0.675 & 0.592 & 0.671 & 0.583 & 0.640 & 0.543 \\ 
MT & 0.651 &0.610& 0.624 & 0.627 & 0.629 & 0.619 & 0.650 & 0.654 & 0.642 & 0.632 & 0.665 & 0.633 \\
SCML & 0.686 &0.661& 0.639 & 0.593 & 0.658 & 0.641 & 0.666 & 0.632 & 0.652 & 0.620 & 0.649 & 0.596 \\ 
PMM & 0.692 &0.666& 0.538 & 0.508 & 0.649 & 0.608 & 0.637 & 0.603 & 0.581 & 0.555 & 0.570 & 0.563 \\
IM & 0.539 &0.624& 0.538 & 0.624 & 0.538 & 0.650 & 0.574 & 0.635 & 0.580 & 0.617 & 0.527 & 0.610 \\ 
\bottomrule
\end{tabular}
\end{table*}

\clearpage

\printbibliography

\end{document}